\documentclass[a4paper,12pt]{article}%
\usepackage[a4paper]{geometry}
\makeatletter
\def\@makefnmark{\hbox{\@textsuperscript{\normalfont\@thefnmark}}}
\renewcommand\@makefntext[1]%
    {\noindent\makebox[0pt][r]{\textsuperscript{\@thefnmark}\,}#1}
\makeatother

\usepackage{scrextend}
\deffootnote{0em}{1.6em}{\thefootnotemark\ }

\usepackage{setspace}
\usepackage{etoolbox}

\makeatletter
\patchcmd{\@makecaption}
  {\parbox}
  {\advance\@tempdima-\fontdimen2} 
  {}{}
\makeatother 
\usepackage{graphicx}

\usepackage{amssymb,amsmath,amsthm,amsfonts,bbm,bm,mathrsfs,xcolor,array}
\usepackage{mhequ}
\usepackage{accents,url}
\usepackage[colorinlistoftodos,prependcaption,textsize=tiny]{todonotes}
\usepackage{float}  
\usepackage{subfigure}
\usepackage[utf8]{inputenc} 
\usepackage[T1]{fontenc}    
\usepackage{hyperref}       
\usepackage{url}            
\usepackage{booktabs}       
\usepackage{nicefrac}       
\usepackage{microtype}      
\usepackage{lipsum}		
 \usepackage{fixltx2e}
 \usepackage{appendix}
 \usepackage{algorithmicx}
 \usepackage{multirow}
  \usepackage{natbib}
  \usepackage[ruled,vlined]{algorithm2e}

\DeclareMathOperator*{\argmin}{arg\,min}
\DeclareMathOperator*{\argmax}{arg\,max}

\def\m{\mathcal}
\def\mb{\mathbb}

\def\ms{\mathscr}

\newcommand{\be}{\begin{equs}}
\newcommand{\ee}{\end{equs}}

\providecommand{\keywords}[1]
{
 
  \textbf{\textit{Keywords---}} #1
}

\theoremstyle{plain}
\newtheorem{theorem}{Theorem}

\newtheorem{corollary}{Corollary}
\newtheorem{lemma}{Lemma}

\newtheorem*{definition*}{Definition}

\title{Statistical Inference for Bayesian Risk Minimization via Exponentially Tilted Empirical Likelihood}

\author{Rong Tang and Yun Yang} 

\date{University of Illinois Urbana-Champaign}

\begin{document}
\maketitle
\begin{abstract}
The celebrated Bernstein von-Mises theorem ensures that credible regions from Bayesian posterior are well-calibrated when the model is correctly-specified, in the frequentist sense that their coverage probabilities tend to the nominal values as data accrue. However, this conventional Bayesian framework is known to lack robustness when the model is misspecified or only partly specified, such as in quantile regression, risk minimization based supervised/unsupervised learning and robust estimation. To overcome this difficulty, we propose a new Bayesian inferential approach that substitutes the (misspecified or partly specified) likelihoods with proper exponentially tilted empirical likelihoods plus a regularization term.  Our surrogate empirical likelihood is carefully constructed by using the first order optimality condition of the empirical risk minimization as the moment condition. We show that the Bayesian posterior obtained by combining this surrogate empirical likelihood and the prior is asymptotically close to a normal distribution centering at the empirical risk minimizer with covariance matrix taking an appropriate sandwiched form. Consequently, the resulting Bayesian credible regions are automatically calibrated to deliver valid uncertainty quantification.  Computationally, the proposed method can be easily implemented by Markov Chain Monte Carlo sampling algorithms. Our numerical results show that the proposed method tends to be more accurate than existing state-of-the-art competitors.
\end{abstract}

\keywords{Bayesian inference; Risk minimization;  Exponentially tilted empirical likelihood; Gibbs posterior; Misspecified model; Robust estimation}

\deffootnote{0em}{1.6em}{\thefootnotemark.\,}
\renewcommand*{\thefootnote}{\arabic{footnote}}
\setcounter{footnote}{0}%

\section{Introduction}
We consider Bayesian approaches for estimation and probabilistic inference on model parameter that is defined by a loss function. Specifically, given a loss function $\ell:  \mathcal{X}\times \Theta \to \mb R$ of random variable $X\in\m X$ and parameter $\theta\in\Theta\subset \mb R^d$, we aim to estimate the global minimizer $\theta^*$ of the population risk function $\mathcal{R}(\theta)=\mathbb{E}[\ell(X,\theta)]$ where the expectation is taken with respect to the underlying data generating distribution $\mathcal{P}^*$ that generates $X$. Many statistical problems can be formulated as a risk minimization problem.  For example, for any $\tau\in(0,1)$, the $\tau$-th quantile of a random variable $X$ solves the population risk $\m R(\theta)$ with the check loss function $\ell(x,\,\theta) = (x-\theta) \big(\tau - \textbf{1}(x<\theta)\big)$ where $\textbf{1}(\cdot)$ stands for the indicator function. More generally, quantile regression~\citep{YU2001437,10.1214/13-BA817}, a widely used data analysis technique in statistics and econometrics, can be fitted via minimizing the empirical check loss on the residuals, where the true parameter (regression coefficients) minimizes the corresponding popular level risk (see Section~\ref{Sec:quantile_sim} for a concrete example).
In many high-dimensional problems, the model parameters can also be interpreted as the minimizer of the expectation of a loss function under some low-dimensional structural constraint. For example, sparse high-dimensional regression aims to estimate a regression coefficient vector $\theta^\ast$ that is at most $s$-sparse (an $s$-sparse vector is a vector with $s$ non-zero components). In this example, $\theta^\ast$ can also be defined as the minimizer of the expected squared loss on the residual vector subject to the constraint that it is at most $s$-sparse (see for example,~\cite{pmlr-v48-yangc16} and~\cite{Martin_2017}).

In statistical applications, distribution $\mathcal{P}^*$ is not directly observable, but instead a set of i.i.d samples $\{X_1,\cdots,X_n\}$ from $\mathcal{P}^*$ is available. Based on the formulation of the  population risk minimization problem, a natural strategy to estimate $\theta^*$ is the empirical risk minimization (ERM) approach~\citep{10.5555/2986916.2987018}, which uses any minimizer $\hat{\theta}\in\argmin_{\theta\in\Theta} \m R_n(\theta)$ of the empirical risk function $\mathcal{R}_n(\theta)=n^{-1}\sum_{i=1}^n \ell(X_i,\theta)$ as the estimator. Beyond point estimation, Bayesian approaches allow natural uncertainty quantification, or more broadly probabilistic inference, on the unknown parameter via the posterior distribution. However, the major challenge for the Bayesian or other likelihood-based inference is the requirement of assuming a distribution family that contains the underlying data generating distribution, even though the latter is not our primary objective of study. For example, in Bayesian quantile regression~\citep{YU2001437,10.1214/13-BA817}, although estimating and predicting certain quantile of the response in the presence of covariate variables is of the primary interest, a Bayesian procedure still needs to fully specify the error distribution, such as the asymmetric Laplace distribution, to mimic the check loss minimization~\citep{koenker1978regression,koenker_2005} method in the frequentist paradigm. When the error distribution is misspecified, consistency of point estimation remains valid~\citep{10.1214/13-BA817}. However, credible intervals derived from the Bayesian posterior no longer honestly reflect the estimation uncertainty (c.f.~our numerical results in Sections~\ref{Sec:quantile_sim} and~\ref{Sec:quantile_realdata}). More generally,~\cite{kleijn2012} establishes a Bernstein-von Mises theorem for Bayesian posterior under model misspecification, showing that Bayesian credible sets may not be valid confidence sets. Examples where models are not necessary to be fully specified, or only partly specified, are ubiquitous in various problems, including quantile regression, risk minimization based supervised and unsupervised learning~\citep{10.5555/2986916.2987018,NIPS1992_c06d06da,sun2019model,barlow1989unsupervised,buhmann1998empirical}, and robust estimation~\citep{huber1992robust,wilcox2011introduction,rousseeuw1984robust,rousseeuw2005robust}. Therefore, alternatives to the conventional likelihood based Bayesian approaches that do not require full model specification and are robust to model-misspecification are imperative.
 
A popular model-free surrogate to the conventional Bayesian posterior is the Gibbs posterior~\citep{10.1145/307400.307435,bhattacharya2020gibbs}, whose density function is defined as $\pi_G(\theta|X_{1:n})\propto \exp\{-n\beta\, \mathcal{R}_n(\theta)\}\,\pi(\theta)$, where $\pi$ denotes the prior density, $X_{1:n}=\{X_i\}_{i=1}^n$ is the sample of size $n$ and $\beta>0$ is a learning rate (sometimes called inverse temperature) parameter for balancing between the empirical risk $\m R_n$ and prior $\pi$. The Gibbs posterior uses the empirical risk to exponentially penalize a ``loss'' of parameter $\theta$ incurred on the data, thus avoiding full specification of a statistical model. Theoretically, it is shown in~\cite{bhattacharya2020gibbs,guedj2019primer,syring2020gibbs} that such a Gibbs posterior has good generalization ability --- it concentrates on parameter values with small population risk $\m R$. On the other side, uncertainty quantification remains problematic since the Gibbs posterior is generally incapable of honestly capturing the estimation variability ~\citep{kleijn2012, grunwald2017}.
In the one-dimensional case where $\theta\in\mb R$, the Gibbs posterior can be calibrated via empirically tuning the learning rate $\beta$ (e.g.~via bootstrapping as in
~\cite{Syring_2018,grunwald2017}) so that the frequentist coverage probability of its highest posterior region asymptotically agrees with its credible level. 
In practice, the performance of these calibration methods is highly sensitive to the choice of $\beta$~\citep{bhattacharya2020gibbs}. 
Furthermore, in the multivariate case where $d>1$, a single tuning parameter $\beta$ inflates all entries of the limiting covariance matrix of the Gibbs posterior by the same multiplicative factor. As a consequence, the dependence in the Gibbs posterior among different coordinates in $\theta$ cannot be adjusted.
Formally,~\cite{bhattacharya2020gibbs} shows that for general dimensions, the Gibbs posterior can only be calibrated via tuning the learning rate if a generalized information equality~\citep{CHERNOZHUKOV2003293} holds. This generalized information equality almost requires the empirical risk function $\m R_n$ to be proportional to the negative log-likelihood function in a local neighborhood of true parameter value $\theta^\ast$. 

 Another popular approach of statistical inference without full model specification is via the empirical likelihood (EL)~\citep{owen1990, 10.2307/20441164,10.2307/20441448, 10.2307/30042042}. In a nutshell, EL is an attractive nonparametric analogue of the conventional likelihood that only requires partial model specification through moment conditions~\citep{doi:10.1080/01621459.2017.1358172}.  Specifically, a moment condition takes the form of $\mathbb{E} [g(X,\theta)]=0$~\citep{broniatowski2012divergences} where $g$ is a known vector-valued function of $X$ and $\theta$, called the moment function. Statistical models satisfying moment conditions are called moment condition models. In the Bayesian paradigm, the conventional likelihood function can also be replaced by the EL, leading to the Bayesian EL posterior, whose frequentist properties from posterior inference have been shown to be valid in~\cite{10.2307/30042042, https://doi.org/10.1111/j.1467-9868.2010.00747.x, https://doi.org/10.1111/rssb.12342} for parameters defined through unbiased estimating functions.  As a popular variant of EL,
the exponentially tilted empirical likelihood (ETEL),  has been shown in ~\cite{10.2307/20441164} that  is closely related to EL but also admits a well-defined probabilistic interpretation arising from a Bayesian nonparametric perspective.  Moreover, a Bayesian posterior defined as being proportional to the product of the ETEL and the prior is guaranteed to admit the ``correct'' covariance structure, under the assumption that the moment function $g$ defining the ETEL is sufficiently smooth~\citep{doi:10.1080/01621459.2017.1358172}. Here, the correctness means that the asymptotic posterior covariance matches that of the frequentist sampling distribution of the posterior mean, or the maximum empirical likelihood estimator.

While statistical inference based on Bayesian ETEL for moment condition models enjoys appealing asymptotic properties, it is not clear how it can be applied to problems whose parameter of interest is defined as the minimizer of a population risk function $\m R$, such as quantile regression, risk minimization based statistical learning and robust estimation. One natural idea is to turn the $M$-estimation problem~\citep{geer2000empirical} of empirical risk minimization into the $Z$-estimation problem (or generalized methods of moments) of solving the moment condition equation $\nabla \m R_n(\theta)=0$ arising from its first order optimality condition. Unfortunately, this idea has an obvious limitation: not every first order stationary point that solves this equation is a global minimizer of $\m R_n$, unless restrictive assumptions such as strong convexity of $\m R_n$ are imposed. More rigorously, we show in this paper (c.f.~Theorem~\ref{th1}) that if equation $\nabla \m R_n(\theta)=0$ (or $\nabla R(\theta)=0$) admits multiple solutions, then the naive Bayesian ETEL posterior constructed with this equation as the moment condition is close to a Gaussian mixture distribution, where each mixture component corresponds to one solution with a non-vanishing mixture weight. This leads to estimation inconsistency. Computationally, due to the multi-modality any local move based sampling algorithm for simulating from the Bayesian ETEL posterior may suffer from slow-mixing as the algorithm may get stuck in the local modes.

 In this article, we propose a new ETEL-based Bayesian approach for risk minimization that enjoys good properties from both: it enjoys the estimation consistency as the Gibbs posterior and captures the estimation variability by exhibiting the ``correct'' asymptotic covariance as the Bayesian ETEL posterior. We call the resulting posterior distribution as Bayesian penalized exponentially tilted empirical likelihood (PETEL) posterior.  Unlike Bayesian inference with Gibbs posteriors, our approach is calibration free and circumvents the need of any restrictive assumption such as the generalized information equality, and is thus more broadly applicable. Unlike the aforementioned naive application of Bayesian ETEL that results in a multi-modal posterior, our proposed posterior concentrates on a shrinking neighborhood of the target $\theta^*$, and thus can be used to form consistent point estimators of the parameter. Our proposed methodology also provides an attractive Bayesian alternative to the bootstrapping for uncertainty quantification in empirical risk minimization with several advantages: 1.~Bayesian PETEL allows for a direct incorporation of prior information which embraces complicated hierarchical structures and promotes shrinkage estimation; 2.~conventional gradient based optimization algorithms for minimizing the empirical risk function tend to get stuck into first order stationary points, while the asymptotic uni-modality of the Bayesian PETEL posterior enhances the sampling efficiency of MCMC algorithms; 3.~Bayesian PETEL exhibits superior performance in our numerical studies and tends to be more accurate in terms of coverage probabilities than bootstrapping especially when the risk function is non-convex (c.f.~Section~\ref{Sec:robust_sim}). Our proposed Bayesian PETEL method can also be generalized in several ways. First, it applies to non-smooth loss functions by using any sub-gradient of the empirical risk function to substitute the gradient in smooth cases. This improves the results in~\cite{doi:10.1080/01621459.2017.1358172} where the validity of their Bayesian ETEL method requires moment function $g$ to be at least twice differentiable,  excluding many important examples such as quantile regression, soft-margin support vector machines (SVM) for classification and Huber loss based robust estimation. Specifically, we show in Section~\ref{Sec:BPETELnsmooth} that by replacing the gradient with any subgradient in the Bayesian PETEL posterior, the resulting Bayesian credible region remains well-calibrated~\citep{10.2307/41000330}.  Second, Bayesian PETEL can be extended to high-dimensional models under sparsity constraints by incorporating sparsity inducing priors. For the high-dimensional extension, we show that under proper conditions: (1) those unimportant parameters shrink to zero in the posterior; (2) the joint posterior distribution of those important non-zero parameters is well-approximated by a normal distribution as if working with the low-dimensional (true) model. 
 
 The rest of the paper is organized as follows. In Section 2, we  summarize the notation and give a background introduction to the Bayesian exponentially tilted empirical likelihood (ETEL) and Gibbs posterior. Our proposed Bayesian PETEL posterior is introduced in Section 3.1 and  its extensions to non-smooth loss functions and high-dimensional problems are introduced in Section 3.2. The non-asymptotic properties on the Bayesian ETEL/PETEL posterior  are provided in Section 4 for both smooth and non-smooth loss functions.  Numerical comparisons of our proposed method with calibrated Gibbs posteriors~\citep{Syring_2018} and bootstrapping are provided in Sections 5 and 6.  In Appendix A, we discuss in detail computational aspects of our method, which can be easily implemented via MCMC algorithms. As two representative examples, we apply our theory to quantile regression  and soft-margin SVM in  Appendix B. Proofs of main results and technical results are deferred to the Appendix C and D respectively.

\section{Background and Problem Formulation}
In this section, we begin with the problem setup and summarize some necessary notations. After that, we review two candidate approaches, namely, Gibbs posterior and Bayesian ETEL posterior, for Bayesian inference in risk minimization, and discuss their limitations. As we will see, the Gibbs posterior approach is consistent for parameter estimation, but does not capture the dependence structure and leads to incorrect uncertainty quantification; in contrast, the Bayesian ETEL posterior captures the local covariance, but is susceptible to spurious local minima (a spurious local minimum is a local minimum that is not global) and leads to inconsistent estimation. Both of these will serve as the motivation to our proposed method to be described in the next section. 

Recall from the beginning of the introduction section that in the risk minimization problem, we observe i.i.d.~copies of a random variable $X$ from an unknown underlying distribution $\m P^\ast$, and our goal is to estimate a parameter $\theta^\ast$ as the evaluation at $\m P^\ast$ of a functional $\theta:\,\ms P(\m X) \to \Theta$, where functional output $\theta(\m P)$ at an input distribution $\m P\in\ms P(\m X)$ is implicitly defined through the following population risk minimization problem 
 \begin{align*}
\theta(\m P) \in \argmin_{\theta\in\Theta} \m R(\theta; \m P), \quad\mbox{with} \ \   \m R(\theta;\m P):\,= \mb E_{\m P}\big[\ell(X,\theta)\big],
 \end{align*}
where $\mb E_{\m P}$ denotes the expectation with respect to $\m P$, and recall that $\ell:\,\m X\times\Theta\to\mb R$ is the loss function. When no ambiguity arises, we will omit the $\m P^\ast$ in the expectation $\mb E_{\m P}$ and the population risk function $\m R(\,\cdot\,,\m P)$ when $\m P=\m P^\ast$ in the rest of the paper.  We use $\m H_{\theta}$ to denote the Hessian of population risk function $\mathcal{R}(\theta)$ at $\theta$, and $\Delta_{\theta}=\mathbb{E}\left( \nabla_{\theta} \ell(X,\theta) \nabla_{\theta} \ell(X,\theta)^T\right)$ the covariance matrix of the ``score'' vector $\nabla_{\theta} \ell(X,\theta)$ at $\theta$. 

\subsection{Notation}

We use $\|\cdot\|_p$ to denote the vector $\ell_p$ norm
and $\textbf{1}_{A}$ the indicator function of a set $A$ so that $\textbf{1}_A(x)=1$ if $x\in A$ and zero otherwise. For a vector $\theta\in\mb R^d$, we use $\m S(\theta)$ to denote the support of vector $\theta$, the set of all indices from $1$ to $d$ corresponding to non-zero components of $\theta$. For any set $S=\{s_1,\cdots,s_p\}\subseteq \{1,\cdots,d\}$, let $|S|$ denote its cardinality, $\theta_S=(\theta_{s_1}, \cdots, \theta_{s_p})^T\in\mb R^{|S|}$, and $\Theta_S= \{\theta_S\,|\, \theta\in \Theta\}$ the $S$-section of $\Theta$.
When no ambiguity arises, we may also use the density function, for example $\pi$, to refer an absolutely continuous probability measure $\Pi$. For a set $\Omega$, we use $\Omega^\circ$ to denotes its interior and $\ms P(\Omega)$ to denote the space of all probability distributions over $\Omega$.
Let $d_{\rm TV}(\mu,\nu)$ the total variation distance between two probability measures $\mu$ and $\nu$.   For two discrete probability measures $p=(p_{1},\cdots,p_{n})$ and $p^*=(p_{1}^*,\cdots,p_{n}^*)$, the ``forward''  Kullback–Leibler (KL) divergence  between $p$ and $p^*$ is defined as $\sum_{i=1}^n p_{i}^*\log (p_{i}^*/p_{i})$; the ``backward''  Kullback–Leibler (KL) divergence  between $p$ and $p^*$ is defined as $\sum_{i=1}^n p_{i}\log (p_{i}/p_{i}^*)$~\citep{kullback1997information}.
For any function $f:\,\m X\times\Theta \to \mb R$, we use $ \nabla_{\theta}f(x,\theta)$ to denote the gradient of $f(x,\theta)$ respect to $\theta$ for $x\in\m X$ and $\theta\in\Theta$. 
 For a sample $X_{1:n}=\{X_1,\ldots,X_n\}$ of size $n$ and any measurable function on $\m X$, we use $\m P_n$ to denote its empirical distribution which assigns probability mass $n^{-1}$ to each observation. We use $[d]$ to denote the set $\{1,2,\ldots,d\}$ for any $d\in \mb N_+$. For two sequences $\{a_n\}$ and $\{b_n\}$, we use the notation $a_n \lesssim b_n$ and $a_n \gtrsim b_n$ to mean $a_n \leq Cb_n$ and $a_n \geq C b_n$, respectively, for some constant $C>0$ independent of $n$. In addition, $a_n \asymp b_n$ means that both $a_n \lesssim b_n$ and $a_n\gtrsim b_n$ hold. For two symmetric matrices $A$ and $B$, we use $A\succcurlyeq B$ to mean that $A-B$ is a positive semi-definite matrix. Let $N(\mu,\Sigma)$ denote the multivariate normal distribution with mean $\mu$ and covariance matrix $\Sigma$.

\subsection{Gibbs posterior for risk minimization}\label{Sec:Gibbs}

Originating in statistical mechanics  and PAC (Probably Approximately Correct)-Bayes literature~\citep{catoni2007pac,guedj2019primer}, the Gibbs posterior~\citep{Alquier_2008,bhattacharya2020gibbs}  arises as the posterior that minimize a certain PAC-Bayesian bound~\citep{guedj2019primer} and is a Bayesian version of empirical risk minimization constructed from a loss function $\ell(x,\theta)$,
 \begin{equation}\label{Gibbs}
     {\pi}_{\rm G}(\theta\,|\,X_{1:n}) = \frac{\exp\big(-n\beta\, \m R_{n}(\theta)\big)\,  \pi(\theta)}{ \int_\Theta \exp\big(-n\beta\, \m R_{n}(\theta)\big)\,  \pi(\theta)\,d\theta}, \quad\mbox{with} \ \ 
     \m R_n(\theta) = n^{-1}\sum_{i=1}^n \ell(X_i;\,\theta),
 \end{equation}
where $\beta$ is the learning rate (inverse temperature) parameter controlling the spread of the distribution. Since the empirical risk function $\m R_n$ provides a good proxy to its population counterpart $\m R$, Bayesian inference via Gibbs posterior aims at minimizing the population risk function $\m R(\theta) = \mb E[\ell(X,\theta)]$ without fully specifying a data generating model.

It is proved in several contexts~\citep{bhattacharya2020gibbs,guedj2019primer,syring2020gibbs} that with certain choice of the learning rate $\beta$ and appropriate conditions on the loss function $\ell$, the Gibbs posterior tends to contract toward the unique minimizer $\theta^\ast=\theta(\m P^\ast)$ of $\m R(\theta)$ over $\Theta$. This ensures the consistency of any reasonable estimator constructed from the Gibbs posterior. The rate of contraction depends on the complexity of parameter space $\Theta$ and is the parametric root-$n$ rate (modulo logarithmic factors) for regular parametric models where $\Theta$ is finite-dimensional. For sparse high-dimensional linear regression, \cite{JMLR:v21:19-152,Martin_2017} show that the Gibbs posterior with suitable $\beta$ achieves the minimax-optimal rate of contraction when a sparsity inducing prior favoring smaller models is employed. 

Regarding uncertainty quantification using credible sets, it is observed in \cite{bissiri2016general,syring2020gibbs,Syring_2018} that the learning rate $\beta$ plays a critical role in calibrating the credible intervals from the Gibbs posterior to be asymptotically valid. Here the asymptotic validity means attaining their frequentist nominal (credible) levels in the limit as $n\to\infty$. 
\cite{Syring_2018}   
 proposes to use a bootstrapping-based algorithm to calibrate the Bayesian credible region of Gibbs posterior by tuning $\beta$. They apply stochastic approximation~\citep{robbins1951} to update $\beta$ until the empirical coverage probability is close enough to the nominal level. 
In another related work, \cite{bhattacharya2020gibbs} shows that the Gibbs posterior is close to a normal distribution centering at the empirical risk minimizer $\hat{\theta}$ with covariance matrix $(n\beta)^{-1} \m H_{\theta^*}^{-1}$, where recall that $\m H_\theta$ denotes the Hessian of $\m R(\theta)$ at $\theta$. Note that this matrix is in general different from $n^{-1}$ times the asymptotic covariance $\m H_{\theta^*}^{-1} \Delta_{\theta^\ast} \m H_{\theta^*}^{-1}$ of $\hat{\theta}$, unless $\Delta_{\theta^\ast} = c \m H_{\theta^*}$ for some constant $c>0$. Here, recall $\Delta_{\theta}= \mb E\big(\nabla_\theta \ell(X,\,\theta)\nabla_\theta \ell(X,\,\theta)^T\big)$. Consequently, unless $\theta$ is one-dimensional, it is impossible to calibrate the covariance structure based on tuning a single parameter $\beta$. Furthermore, different $\beta$'s needed to be tuned in order to calibrate credible intervals corresponding to different components of $\theta$, making the bootstrapping computationally demanding.

\subsection{Bayesian exponentially tilted empirical likelihood}
Conventional Bayesian inference requires the full specification of the likelihood function. However, for complex problems involving complicated dependence structures, it is inevitable to misspecify part of the data generating model, which may lead to inconsistent estimation due to the use of incorrect distributional assumptions. Empirical likelihood methods overcome this issue by producing inference about parameters using the information supplied by moment conditions. They circumvent the need for full knowledge of the likelihood function and are often more robust against model misspecification. \cite{10.2307/20441164} shows that the exponentially tilted empirical likelihood (ETEL), a variant  of the empirical likelihood, shares many desirable properties as the conventional parametric likelihood. In particular, ETEL naturally arises as the nonparametric limit of a Bayesian procedure  for moment condition models with a type of non-informative prior on the space of distributions. For such models, a Bayesian ETEL posterior constructed by combining the ETEL with a prior can be applied to conduct valid statistical inference.  In the following, we briefly review the Bayesian ETEL.

As is common in statistics, we only assume the statistical model $\m P$ to satisfy the moment condition (general estimating) equation $\mb E[g(X,\theta)]=0$ specified by a vector valued moment function $g: \,\m X\times \Theta\to\mb R^d$, where parameter space $\Theta\subset\mb R^d$. In this setup, parameter $\theta$ does not need to fully parametrize the model, and can be certain functional $\theta(\m P)$ of $\m P\in\ms P(\m X)$ such as mean, quantiles and etc.  For a sample $X_{1:n}=\{X_i\}_{i=1}^n$ of size $n$,  the ETEL function $L:\,\m X_{1:n}\times \Theta \to (0,\infty)$  is defined as $L(X_{1:n};\,\theta) = \prod_{i=1}^n p_i(\theta)$, where $\big(p_1(\theta),p_2(\theta),\ldots,p_n(\theta)\big)$ solves the following constrained optimization problem
 \begin{equation}\label{Eqn:ETEL}
   \begin{aligned}
 \max_{(w_1,w_2,\ldots,w_n)} & \sum_{i=1}^n \big[-w_i \log (n w_i)\big]\\
\mbox{subject to} \quad & \sum_{i=1}^n w_i=1,\quad \sum_{i=1}^nw_i g(X_i,\theta) = 0,\\
 & w_1,w_2,\ldots,w_n \geq 0.
 \end{aligned}
\end{equation}
By introducing Lagrange multipliers to the constraints, these probabilities $\{p_i(\theta)\}_{i=1}^n$ can be equivalently expressed as 
 \begin{equation}\label{gen_defetel}
 \begin{aligned}
& p_i(\theta)=\frac{\exp\big([\lambda(\theta)]^T g(X_i,\theta)\big)}{\sum_{i=1}^n \exp\big([\lambda(\theta)]^Tg(X_i,\theta)\big)} \quad\mbox{with}\quad\lambda(\theta)=\underset{\xi \in \mathbb{R}^d}{\arg \min}\Big\{
\sum_{i=1}^n \exp\big(\xi^T  g(X_i,\theta)\big)\Big\}.
 \end{aligned}
 \end{equation}
The unconstrained convex minimization problem~\eqref{gen_defetel} can be solved by a Newton–Raphson procedure.  Here,  $\{p_i(\theta)\}_{i=1}^n$ can be viewed as the probabilities minimizing the KL divergence between the multinomial distribution $(w_1,\cdots, w_n)$, with $w_i$ being assigned to the $i$th observation $X_i$, and the empirical distribution $(n^{-1}, n^{-1},\ldots,n^{-1})$, subject to the constraint that a weighted sample version of the moment condition equation, $\sum_{i=1}^nw_i g(X_i,\theta) = 0$, is satisfied. It is worth mentioning that \cite{https://doi.org/10.1111/insr.12097} and~\cite{10.1214/009053606000001208} provide a unifying perspective by interpreting the EL and the ETEL as minimizing respectively the ``forward” and ``backward'' KL distance between $\big(p_1(\theta),p_2(\theta),\ldots,p_n(\theta)\big)$ and $(n^{-1}, n^{-1},\ldots,n^{-1})$ under each $\theta\in\Theta$. As a consequence, they show that under some regularity conditions, the probabilities $\{p_i(\theta)\}_{i=1}^n$ obtained from the EL and the ETEL are first-order equivalent. Moreover, the point estimators obtained by maximizing the two likelihood functions differ only by a term of order $O_p(n^{-3/2})$.

In the Bayesian framework, ETEL function $L(X_{1:n};\,\theta)$ plays the role of the conventional likelihood function, leading to the Bayesian ETEL posterior density function
\begin{align*}
\pi_{\rm E}(\theta\,|\,X_{1:n})  = \frac{L(X_{1:n};\,\theta)\, \pi(\theta)}{\int_\Theta L(X_{1:n};\,\theta) \,\pi(\theta)\, d\theta}, \quad\forall \theta\in\Theta,
\end{align*}
where recall that $\pi$ denotes the prior density function. On the theoretical side,~\cite{10.1214/009053606000001208} and~\cite{doi:10.1080/01621459.2017.1358172} show that even in the presence of model misspecification (i.e., the equation $\mathbb{E} [g(X,\theta)]=0$ does not admit a solution on $\Theta$), the Bayesian ETEL posterior satisfies the Bernstein–von Mises (BvM) theorem~\citep{10.1214/009053606000001208}. Moreover, when the moment condition model is correctly specified in the sense that  $\mathbb{E}[g(X,\theta)]=0$ admits a unique solution $\theta^*$ over $\Theta$,  the BETEL posterior distribution concentrates on an $n^{-1/2}$-ball centered at $\theta^*$ and is well-approximated by a normal distribution whose data-dependent center is the ETEL maximizer and whose covariance matrix matches the frequentist asymptotic covariance of the center.

 \subsection{Bayesian ETEL for risk minimization}\label{Sec:BETEL}
 
 In this part, we discuss a direct application of the Bayesian ETEL framework to the risk minimization problem and its limitation. In the Section~\ref{Sec:mainmethod}, we will introduce an improved method that overcomes the limitation.
 
\begin{figure}[ht]
\centering  
\includegraphics[trim={0 0.8cm 0 2.3cm}, clip, width=0.7\textwidth]{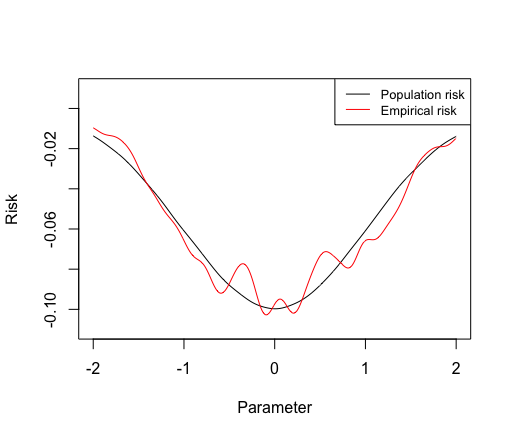}
 \caption{This figure plots the population risk and its empirical counterpart with loss function $\ell(x,\theta)=-\exp\big\{-\frac{(x-\theta)^2}{2\cdot(0.01)^2}\big\}$, where random variable $X\sim N(0,1)$ and $500$ i.i.d. samples of $X$ are used for computing the empirical risk. Although the population risk admits a unique stationary point, the empirical risk has multiple stationary points.}
\label{Fig_Population_emp_risk}
\end{figure}

In the risk minimization problem, if we further assume that loss function $\ell(x,\theta)$ is differentiable with respect to $\theta$ at any point $x\in\m X$ and $\m R(\theta)$ has a unique stationary point, which is its global minimum, then $\theta(\m P)$ can be equivalently defined as the unique solution of the following \emph{first order optimality condition} of minimizing $\m R(\,\cdot\,,\m P)$,
\begin{align*}
\mb E_\m P[\nabla_\theta \ell(X,\theta)] = 0.
\end{align*}
 By supplying the above as the moment condition equation in the Bayesian ETEL with $\nabla_\theta \ell(X,\theta)$ being the moment function, we obtain the following Bayesian ETEL posterior,
 \begin{equation}\label{Eqn:BayesianETEL}
 \begin{aligned}
& \pi_{\rm E}(\theta\,|\,X_{1:n}) = \frac{\pi(\theta)\prod_{i=1}^n p_i(\theta)}{\int_\Theta \pi(\theta) \prod_{i=1}^n p_i(\theta)\,d\theta},\quad \theta\in\Theta,\\
 \mbox{with}\quad & p_i(\theta)=\frac{\exp\big([\lambda(\theta)]^T \nabla_\theta \ell(X_i,\theta)\big)}{\sum_{i=1}^n \exp\big([\lambda(\theta)]^T \nabla_\theta \ell(X_i,\theta)\big)}, \quad i=1,2,\ldots,n\\
\mbox{where} \quad &\lambda(\theta)=\underset{\xi \in \mathbb{R}^d}{\arg \min}\Big\{
\sum_{i=1}^n \exp\big(\xi^T \nabla_\theta \ell(X_i,\theta)\big)\Big\}.
 \end{aligned}
  \end{equation}
 However, this direct application of the Bayesian ETEL suffers from several drawbacks. First, it requires the population level identifiability--- the population risk function $\m R$ has a unique stationary point, which can be difficult to verify and only holds under certain restricted assumptions such as $\m R$ being strongly convex over $\Theta$. Second, even though $\m R$ admits a unique stationary point, it is not guaranteed that the empirical risk function $\m R_n(\cdot) =\mb E_{\m P_n}[\ell(X;\,\theta)] = n^{-1} \sum_{i=1}^n \ell(X_i;\,\theta)$ also admits a unique stationary point (see Figure~\ref{Fig_Population_emp_risk} for an illustration). This may require further restrictive assumptions such as loss function $\ell(x;\,\theta)$ being strongly convex with respect to $\theta$. 
 
Our theoretical result (Theorem~\ref{th1}) in Section~\ref{Sec:TheoryETEL} shows that if the population moment condition equation $\mb E [\nabla_{\theta}\ell(X;\, \theta)]=0$ admits $K$ isolated solutions $\{\tilde\theta_{k}\}_{k=1}^K$ on $\Theta$, then the Bayesian ETEL posterior $\pi_{\rm E}(\theta\,|\,X_{1:n})$ tends to be close to a Gaussian mixture distribution with $K$ components whose means and covariance matrices are $\{\hat\theta_k\}_{k=1}^K$ and $\{ n^{-1} V_k\}_{k=1}^K$, respectively, with $V_k = \m H_{\tilde\theta_k}^{-1}\Delta_{\tilde\theta_k}\m H_{\tilde\theta_k}^{-1}$ taking a sandwiched form. Each mixture component corresponds to one solution $\tilde\theta_{k}$, and one of them is centered at the empirical risk minimizer $\hat \theta$. Moreover, the mixing weight of the $k$th mixture component only depends on $(\pi(\tilde\theta_{k}),V_k)$ for $k=1,\ldots,K$, and does not diminish as sample size $n$ tends to $\infty$. As a consequence, any reasonable estimator, such as the posterior mean, from the Bayesian BETEL posterior $\pi_{\rm E}(\theta\,|\,X_{1:n})$ is not consistent for $\theta^\ast$, let alone statistical inference based on $\pi_{\rm E}(\theta\,|\,X_{1:n})$. On the positive side, the local asymptotic covariance matrix $V_k$ corresponds to $\tilde\theta_k$ matches the asymptotic covariance matrix of the normal center $\hat\theta_k$, meaning that it correctly captures the local random fluctuation. Consequently, if all components other than the one corresponding to the empirical risk minimizer $\hat\theta$ are killed, then the remaining component renders correct uncertainty quantification.
 \section{Bayesian Inference for Risk Minimization}\label{Sec:mainmethod}
 In this section, we propose a new approach of Bayesian inference for solving the risk minimization problem. The proposed method combines merits of the Gibbs posterior and the Bayesian ETEL posterior, leading to consistent estimation and automatically calibrated uncertainty quantification. We also provide its extensions for handling non-smooth loss functions and high-dimensional parameters.
 
 \subsection{Bayesian penalized exponentially tilted empirical likelihood}
 \begin{figure}[t]
\centering  
\subfigure[Risk function]{
 \includegraphics[trim={0.1cm 0.5cm 0.1cm 2.6cm}, clip, width=0.48\textwidth]{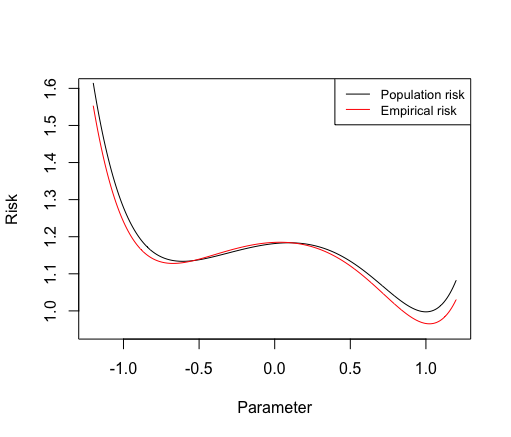}}
\subfigure[Posterior density function]{
 \includegraphics[trim={0.1cm 0.5cm 0.1cm 2.6cm}, clip, width=0.48\textwidth]{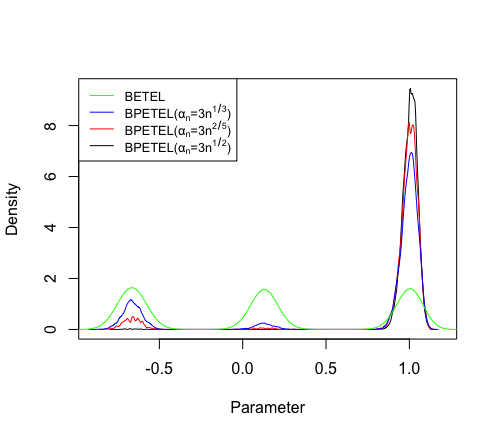}}
 \caption{The figures illustrate the performance of Bayesian ETEL/PETEL when applied to the regression model $Y=f(\theta^* \tilde{X})+e$, where $\theta^*=1$, $f(x)=0.1x^3-0.2x^2-0.2x$, $\tilde X \sim N(0,1)$ and $e\sim N(0,1)$. Figure (a) plots the population risk and its empirical counterpart with loss function $\ell((\tilde x, y),\theta)=(y-f(\theta \tilde{x}))^2$ based $n=500$ i.i.d. samples $\{X_i=(\tilde X_i, Y_i)\}_{i=1}^{500}$. There are three stationary points for both population risk and empirical risk. Figure (b) plots the respective density functions of Bayesian ETEL and PETEL posteriors with $\text{Uniform}(-2,2)$ prior on $\theta$. We can see that the Bayesian ETEL posterior has $3$ equal weighted local modes corresponding to the $3$ stationary points of the empirical risk; while for Bayesian PETEL, the probability mass assigned to the local minimum around $-0.7$ and the local maximum around $0.1$ quickly vanishes as $\alpha_n$ increases. }
\label{Fig_Density_E_PE}
\end{figure}

\begin{figure}[th]
\centering  
 \includegraphics[width=0.7\textwidth]{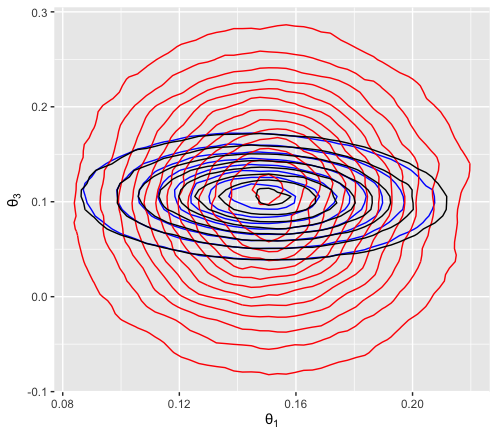}
  \caption{We plot contours of Bayesian PETEL and Calibrated Gibbs  (CG) posteriors when applied to the South African heart disease dataset, described in Section 4.4.2 of~\cite{Hastie2009}. The binary response variable $Y$ is the presence or absence of myocardial infarction (MI) at the time of the survey. Following~\cite{Hastie2009}, we focus here on predictors tobacco,  ldl,  famhist and age. Our parameter $\theta\in\mb R^4$ of interest is the minimizer of population risk from the smoothed hinge loss in the SVM classification~\citep{8637389}, i.e.,  $\frac{1}{2}\lambda \|\theta\|_2^2+\mathbb{E}\frac{1}{2} (\sqrt{U^2+\varepsilon^2}+U)$ where $U=1-Y\theta^TX$, $\lambda=0.5$ and $\varepsilon=0.8$. The blue and red curves are the contours of the joint distribution of $(\theta_1, \theta_3)$ (associated with tobacco and famhist) from the Bayesian PETEL and CG posteriors, respectively. The black curve is the benchmark contour based on bootstrapping samples. As we can see, the CG posterior fails to capture the heterogeneous variance of different $\theta$ components. For example, it overestimates the variance of $\theta_3$.}
\label{Fig_Density_CG_PE}
\end{figure}

From the discussions in Sections~\ref{Sec:Gibbs} and~\ref{Sec:BETEL}, we see that despite the covariance matrix mismatching, the Gibbs posterior has a good concentration property that it places almost all mass on a shrinking neighborhood of $\theta^*$; in contrast, the Bayesian ETEL posterior is susceptible to spurious local minima and is multi-modal. However, the restriction of the BETEL posterior to a local neighborhood around $\theta^*$ carries the correct shape that honestly reflects the uncertainty ---
its local asymptotic covariance matrix $\m H_{\theta^*}^{-1}\Delta_{\theta^*}\m H_{\theta^*}^{-1}$ matches that of its center $\hat{\theta}$ (c.f.~Theorem~\ref{th1}).
This motivate us to propose the following calibrated Gibbs posterior as a Bayesian penalized exponentially tilted empirical likelihood (PETEL), by adding a penalty term $-\alpha_n \mathcal{R}_n(\theta)$ to enforce the concentration of the Bayesian ETEL posterior,
\begin{equation}\label{Bayesian_PETEL}
     \pi_{\rm PE} (\theta\,|\,X_{1:n}) =  \frac{\pi(\theta)\exp\big(-\alpha_n \m R_n(\theta)\big)\prod_{i=1}^n p_i(\theta)}{\int_\Theta \pi(\theta) \exp\big(-\alpha_n \m R_n(\theta)\big) \prod_{i=1}^n p_i(\theta)\,d\theta},\quad \theta\in\Theta,
\end{equation}
where $\{p_i(\theta)\}_{i=1}^n$ are defined in equation~\eqref{Eqn:BayesianETEL}, and $\alpha_n>0$ is a regularization parameter. Here, we add a subscript $n$ in $\alpha_n$ to indicate that it is allowed to be dependent of $n$. We intentionally choose $\alpha_n$ to be $o(n)$ as oppose to $n\beta$ in the Gibbs posterior. As a consequence, the penalty term $\alpha_n \m R_n(\theta)$ has limited impact on the shape of the posterior, since the latter is dominated by the ETEL part $\big\{\prod_{i=1}^n p_i(\theta)\big\}$ which is of order $e^{-O_p(n\|\theta-\hat\theta\|^2)}$. More rigorously, our theoretical analysis in Section~\ref{Sec:Theory} shows that there is a wide range of $\alpha_n$ as $\log n\lesssim \alpha_n\lesssim \sqrt{n\log n}$ to ensure the concentration of the Bayesian PETEL posterior around $\theta^\ast$. Our numerical results in Section~\ref{Sec:simulation} also illustrate the robustness of this procedure to the choice of tuning parameter $\alpha_n$. In contrast, the performance of the Gibbs posterior is quite sensitive to the choice of learning rate $\beta$.
In addition, our theory shows that the Bayesian PETEL posterior is close to a normal distribution with the correct sandwiched covariance matrix; and the coverage probability of the resulting posterior credible region tends to its nominal level in the frequentist sense with the parametric root-$n$ rate (modulo logarithmic factors).

The Bayesian ETEL posterior described in Section~\ref{Sec:BETEL} is a special case of Bayesian PETEL with $\alpha_n=0$. Here, adding a penalty term $\alpha_n \mathcal{R}_n(\theta)$ with appropriate nonzero $\alpha_n$ forces the Bayesian PETEL posterior to empty out mixture components associated with stationary points (local optima and saddle points) of $\m R_n$ that are not the global minimum $\hat\theta$ (see Figure~\ref{Fig_Density_E_PE} for an illustration).  The inclusion of this extra penalty term also comes with computational benefits. For example, suppose we apply Markov Chain Monte Carlo (MCMC) algorithm with local moves to sample from the posteriors. For the Bayesian ETEL, due to the multi-modality, the Markov chain based on local moves may easily get stuck in one mode for a long time. However, for the Bayesian PETEL, the extra penalty term $\alpha_n \mathcal{R}_n(\theta)$ favors points closer to the global minimum $\hat \theta$ of $\m R_n$, and will encourage the Markov chain to move quickly towards $\hat \theta$ in a reasonable amount of steps.

Although the Bayesian PETEL shares a similar component $\exp\big(-\alpha_n \m R_n(\theta)\big)$ as the Gibbs posterior, this component only plays the role of forcing the posterior concentration, and does not contribute to forming its shape. Therefore, the Bayesian PETEL avoids any restrictive assumption, such as the generalized information equality $\Delta_{\theta^\ast} = c \m H_{\theta^*}$ that is required by the Gibbs posterior to exhibit the correct shape for uncertainty quantification, and is suitable for a wider range of problems  (see Figure~\ref{Fig_Density_CG_PE} for a comparison). Furthermore, even if the generalized information equality holds, the Gibbs posterior approach involves the daunting task of selecting the learning rate $\beta$ for calibrating the scale of the covariance. In particular, the performance of Gibbs posterior inference is highly sensitive to the choice of $\beta$~\citep{bhattacharya2020gibbs}. In comparison, our method provably works under a much wider range of $\alpha_n$ values.

 \subsection{Extensions}\label{Sec:extensions}
  In this subsection, we discuss two extensions of our Bayesian PETEL approach.
  
  \vspace{0.5em}
  
  \noindent {\bf Extension to non-smooth loss functions:} When the loss function $\ell(X;\,\theta)$ is not differentiable with respect to $\theta$ at certain pair $(X,\theta)$, we can replace the gradient with any of its subgradient (a subgradient of a function $f:\mb R^d\to \mb R$ at point $x\in\mb R^d$ is a vector $g\in\mb R^d$ such that
$f(y) \geq f(x) + \langle g,\, y-x \rangle + o(\|y-x\|_2)$ as $y\to x$). If $\ell(X;\,\theta)$ is everywhere differentiable with respect to $\theta$, then the gradient $\nabla_\theta \ell(X;\,\theta)$ is the unique subgradient and the method reduces to the Bayesian PETEL with a smooth loss function. Our theory in Section~\ref{Sec:Theory} will cover this case.

\vspace{0.5em}

\noindent {\bf Extension to high-dimensional problems:} In this extension,  our interest is in the high-dimensional setting where dimension $d$ of parameter $\theta$ can be similar or much larger than the sample size $n$. We follow the convention by considering the case where the population risk minimizer $\theta^*$ is $s^*$-sparse with $s^*\ll n$, i.e., the number of non-zero elements in $\theta^*$ is at most $s^*$. Let $s_0$ be a pre-specified upper bound on the sparsity level. For convenience, we consider the following class of sparse priors for achieving consistent estimation, and the method can be straightforwardly carried over to other sparsity inducing priors such as spike and slab priors~\citep{10.1214/009053604000001147} and global-local shrinkage priors~\citep{carvalho2010horseshoe}.
 
 \begin{definition*}[Sparse Prior]\label{sparseprior}
 Prior on $\theta\in\mb R^d$ is induced by: (1) Draw $s$ from a distribution $Q$ on the set $[s_0]$ with probability mass function $q(s)\propto \exp(-\beta_{n,d} \,s)$, for some constant $\beta_{n,d}>0$; (2) Pick uniformly a subset $S$ of cardinality $s$ of $[d]$; (3) Sample $\theta_{S}=\{\theta_j:\,j\in S\}$ from a prior density $\pi_S$ on $\Theta_S$, the $S$-section of $\Theta$, and set $\theta_{S^c} = 0$.
\end{definition*}
 
 \noindent Such sparse priors are employed in many existing works~\citep{JMLR:v21:19-152,Martin_2017,castillo2015,Dellaportas2002} in the Bayesian literature. A correct specification of prior mass $q(s)$ is crucial for controlling the sparsity level of $\theta$, which should decay exponentially fast in $s$~\citep{castillo2015}. In the regression setting, the prior $\pi_S$, for example, can be chosen as Zellner's $g$-prior~\citep{Zellner1986}. 
 
 Let $\{p_i(\theta_S;S)\}_{i=1}^n$ denote the low-dimensional counterpart of the empirical probability functions $\{p_i(\theta)\}_{i=1}^n$ defined in equation~\eqref{Bayesian_PETEL} when $\theta$ is restricted to $\Theta_S\times \{0\}^{d-|S|}$, or
 \begin{equation*}
   \{p_i(\theta_S;S)\}_{i=1}^n=\argmax_{\sum_{i=1}^n w_i=1,\atop \sum_{i=1}^n w_i \nabla_{\theta_S} \ell(X_i;\,\theta)=0} \Big\{\sum_{i=1}^n[-w_i \log (nw_i)]\Big\},\quad \mbox{for}\ \theta = (\theta_S, 0)\in\mb R^d.
\end{equation*}
Now we define the ``model-averaged'' Bayesian PETEL posterior for high-dimensional parameter $\theta$, or equivalently for $(\theta_S, S)$ with $|S|\leq s_0$ and $\theta_S\in \Theta_S$, as
  \begin{equation}\label{Bayesian PETELsp1}
  \begin{aligned}
 &\pi_{\rm PE}(\theta_S,S\,|\,X_{1:n})=\\
 &\frac{\binom{d}{|S|}^{-1}q(|S|)\,\pi_S(\theta_S) \exp\big(-\alpha_{n,d} \m R_n(\theta_S,0)\big)\prod_{i=1}^n p_i(\theta_S;\,S)}{\underset{S\in[d], \,|S|\leq s_0}{\sum}\binom{d}{|S|}^{-1}q(|S|)\int_{\Theta_S}\pi_S(\theta_S) \exp\big(-\alpha_{n,d} \m R_n(\theta_S,0)\big)\prod_{i=1}^n p_i(\theta_S;\,S)\, d\theta_S}.
 \end{aligned}
      \end{equation}
Note that $\theta$ has a one-to-one correspondence with the pair $(\theta_S,S)$ as: 1.~Given $\theta\in \mb R^d$, we have $S=\m S(\theta)$ and $\theta_S = \theta_{\m S(\theta)}$; 2.~Given $(\theta_S,S)$, we have $\theta=(\theta_S,0)\in\mb R^d$.
 
Our theoretical result (Theorem~\ref{th41}) shows that as long as $\log d=o (n)$, there exist ranges of $\alpha_{n,d}$ and $\beta_{n,d}$ to guarantee the concentration of the posterior to the population risk minimizer $\theta^\ast$. In addition, if all nonzero signals in $\theta^\ast$ are suitably large and $\log=o(\sqrt{n})$, then the choice of $\log (d\vee n) \lesssim \alpha_{n,d}\lesssim \sqrt{n\log n}$ and $\log  (d\vee n)\lesssim \beta_{n,d} \lesssim \alpha_{n,d}$ leads to the so-called oracle property: 1.~variable selection consistency, or $\pi_{\rm PE}(S=S^\ast\,|\,X_{1:n}) \approx 1$, where $S^\ast=\m S(\theta^\ast)$ denote the support of $\theta^\ast$; 2.~the condition posterior $\pi_{\rm PE}(\theta_{S^\ast}\,|\,S^\ast,X_{1:n})$ tends to be close to the  normal distribution centering at the constrained minimizer $\hat\theta_{S^\ast}$ of empirical risk $\m R_n$ over $\Theta_{S^\ast}$ with the correct covariance matrix for uncertainty quantification. Consequently, we recommend a default choice of $\alpha_{n,d} =C\sqrt{n}$ and $\beta_{n,d}=C'\log (d\vee n)$ for some suitable constant $C$ and $C'$.

\subsection{Computation}

\begin{algorithm}[ht!]\label{algorithm}
\caption{Metroplis-Hasting algorithm to sample from Bayesian PETEL posterior}
\SetAlgoLined
\SetKwRepeat{Do}{do}{while}%
 \textbf{Input}: Number of iteration $L$, tolerance $\varepsilon$, proposal distribution $p_{prop}(\cdot|\cdot)$, initial state $\theta^0$ and $\lambda(\theta^0)$\;
 \textbf{Data}:$X_1,X_2\cdots,X_n$\;
 \For{$t \leftarrow 0 \,\,to\,\, L-1$}{
   Sample $\tilde{\theta}$ from $p_{prop}(\cdot|\theta^t)$\;
   Generate a uniform random  number $u\in(0,1)$\;
   Define $f(\lambda)\leftarrow \frac{1}{n} \sum_{i=1}^{n} \exp (\lambda^{T} \nabla_{\theta} \ell(x_{i}, \tilde{\theta}))$\;
   $\lambda^0\leftarrow \lambda(\theta^{t})$\;
    $k\leftarrow0$\;
  
   \Repeat{$\|H^{-1}G\|_2\leq \varepsilon$ }{
       $k\leftarrow k+1$\;
       $\gamma=1$\;
       $H\leftarrow\frac{1}{n}\sum_{i=1}^n\exp\left(\nabla_{\theta}\ell(X_i,\tilde{\theta})^T\lambda^{k-1}\right)\nabla_{\theta}\ell(X_i,\tilde{\theta})\nabla_{\theta}\ell(X_i,\tilde{\theta})^T$\;
 $G\leftarrow\frac{1}{n}\sum_{i=1}^n\exp\left(\nabla_{\theta}\ell(X_i,\tilde{\theta})^T\lambda^{k-1}\right)\nabla_{\theta}\ell(X_i,\tilde{\theta})$\;
 \Repeat{$f(\lambda^k)\leq f(\lambda^{k-1})$}
 { $\lambda^k\leftarrow\lambda^{k-1}- \gamma H^{-1}G$\;
$ \gamma=\frac{1}{2}\gamma$;}
       }
   
   $\lambda(\tilde{\theta})\leftarrow \lambda^k$\;
  \eIf{$u\leq \frac{\pi(\tilde{\theta})\exp\left(\sum_{i=1}^n \log \frac{\exp(\lambda(\tilde{\theta})^T \nabla_{\theta} \ell(X_i,\tilde{\theta}))}{\sum_{i=1}^n \exp(\lambda(\tilde{\theta})^T \nabla_{\theta} \ell(X_i,\tilde{\theta}))}-\alpha_n \mathcal{R}_n(\tilde{\theta})\right)p_{prop}(\theta^t|\tilde{\theta})}{\pi(\theta^t)\exp\left(\sum_{i=1}^n \log \frac{\exp(\lambda(\theta^t)^T \nabla_{\theta} \ell(X_i,\theta^t))}{\sum_{i=1}^n \exp(\lambda(\theta^t)^T \nabla_{\theta} \ell(X_i,\theta^t))}-\alpha_n \mathcal{R}_n(\theta^t)\right)p_{prop}(\tilde{\theta}|\theta^t)}$}
  { $\theta^{t+1}\leftarrow\tilde{\theta}$\;
    $\lambda(\theta^{t+1})\leftarrow \lambda(\tilde{\theta})$\;
   }{
    $\theta^{t+1}\leftarrow\theta_t$\;
     $\lambda(\theta^{t+1})\leftarrow \lambda(\theta_t)$\;
  }
 }
 
\end{algorithm}

Since equation~\eqref{Bayesian_PETEL} provides an explicit expression for the Bayesian PETEL posterior up to a normalization constant,  we utilize the Metroplis-Hasting algorithm to draw samples. The major non-trivial part in the algorithm is solving for $\lambda(\theta)$ in the calculation of the ETEL function [c.f.~equation~\eqref{Eqn:BayesianETEL}], which is a convex problem and can be calculated by a modified Newton-Raphson algorithm~\citep{10.1093/biomet/89.1.230}. Algorithm~\ref{algorithm} summarizes the pseudocode for the Metroplis-Hasting steps to sample from Bayesian PETEL posterior, where $\nabla_{\theta} \ell(X,\theta)$ can be replaced by its subgradient if not differentiable. Further details on the computation are provided in Appendix A.

 \section{Theoretical Results and their Consequences}\label{Sec:Theory}
 In this section, we begin with theoretical analysis of the Bayesian ETEL posterior and discuss the consequent limitation. After that, we analyze the proposed Bayesian PETEL posterior with smooth loss, non-smooth loss, and sparse high-dimensional parameters. In Appendix B, we apply these theoretical results to two representative examples, quantile regression and classification using soft-margin SVM.
 
 \subsection{Analysis of Bayesian ETEL posterior}\label{Sec:TheoryETEL}

In this subsection, we study the large sample behavior of the Bayesian ETEL posterior distribution. We first state the following regularity conditions to the loss function, risk function and prior distribution.
 
 \vspace{0.5em}
\noindent \textbf{Assumption A.1}:
The loss function $\ell(X,\theta): \mathcal{X}\times \Theta \to \mathbb{R}$ is thrice differentiable with respect to $\theta$ with bounded mixed partial derivatives up to order three. In addition, the parameter space $\Theta\subset\mb R^d$ is compact.
\\
\noindent
\textbf{Assumption A.2}: (1) The equation $ \nabla_{\theta} \mathcal{R}(\theta)=0$ has $K\geq 1$ isolated solutions $\tilde{\theta}_1,\cdots \tilde{\theta}_K$ on $\Theta$, where for any $1\leq k\leq K$, $\tilde{\theta}_k\in \Theta^{\circ}$; (2) There exist positive constants $(a,b)$ such that for any $k\in[K]$, it holds that $\Delta_{\tilde
{\theta}_k} \succcurlyeq aI_d$ and  $\m H_{\tilde{\theta}_k}^T \m H_{\tilde{\theta}_k} \succcurlyeq b I_d$, where recall that $\m H_{\theta}$ denotes the Hessian matrix of $\mathcal{R}(\theta)$ and $\Delta_{\theta}=\mathbb{E}\left( \nabla_{\theta} \ell(X,\theta) \nabla_{\theta} \ell(X,\theta)^T\right)$.
\vspace{0.5em}
 
\noindent
 \textbf{Assumption A.3}: (1) The prior admits a density function $\pi(\theta)$  with  respect to the Lebesgue measure; (2) There exist positive constants $(c,r,L)$ such that for any $k\in[K]$, it holds that $ \pi(\tilde{\theta}_k)\geq c$ and $\pi(\theta)$ is locally $L$-Lipschitz around $\tilde\theta_k$, or $|\pi(\theta)-\pi(\tilde{\theta}_k)| \leq L \|\theta-\tilde{\theta}_k\|_2$ for all $\theta$ satisfying $\|\theta-\tilde\theta_k\|\leq r$.

The assumptions on the smoothness of the loss function with respect to $\theta$ and the Lipschitz continuity of the prior are common for proving the asymptotic normality of the posterior in parametric models~\citep{Ghosh2003}. Assumption A.2 on the risk function requires the positive definiteness of the sandwich covariance matrix $\m H_{\theta}^{-1}\Delta_{\theta}\m H_{\theta}^{-1}$~\citep{Syring_2018} evaluated at $\{\tilde{\theta}_k\}_{k=1}^K$, so that posterior distributions constrained on neighborhoods of $\tilde{\theta}_k\,(1\leq k\leq K)$ are asymptotically normal. The lower bounds on absolute values of eigenvalues of the Hessian matrix $\m H_{\tilde\theta_k}$ (not necessarily positive semi-definite) requires all saddle points and local optima to be strict, which is a common assumption for analyzing the algorithmic convergence of first-order optimization methods and is satisfied in most applications.

Our first theorem shows that, under these assumptions, the Bayesian ETEL posterior distribution tends to be close to a normal mixture distribution. The center of each mixture component falls into an $n^{-1/2}$-neighborhood centered at one solution of $ \nabla_{\theta} \mathcal{R}(\theta)=0$, either a saddle point or a local optimum. For any $r>0$, we use $B_r(\theta)$ to denote the $\ell_2$-ball with radius $r$ centering at $\theta$. Recall that $V_{\theta}$ denotes the matrix $\m H^{-1}_{\theta}\Delta_{\theta} \m H^{-1}_{\theta}$ for any $\theta\in\Theta$.
 
 \begin{theorem}\label{th1}
Under Assumption A.1, A.2 and A.3, there exists some positive constants $(r,C)$ independent of $n$ such that it holds with probability at least $1-n^{-1}$ that,
 \begin{enumerate}
     \item  For any $1\leq k\leq K$,  the equation $ \nabla_{\theta} \mathcal{R}_n(\theta)=0$ associated with empirical risk $\m R_n$ has a unique solution $\hat{\theta}_k$ in $B_r(\tilde{\theta}_k)$,  where $\sqrt{n}\,(\hat{\theta}_k - \tilde\theta_k) \to N(0, V_{\tilde{\theta}_k})$ in distribution as $n\to\infty$;
\item  $d_{\rm TV}\Big(\pi_{\rm E}(\cdot\,|\,X_{1:n}), \, \sum_{k=1}^K \frac{\pi(\tilde{\theta}_k)\,|V_{\tilde{\theta}_k}|^{1/2}}{\sum_{l=1}^K \pi(\tilde{\theta}_l)\,|V_{\tilde{\theta}_l}|^{1/2} } N\big(\hat{\theta}_k, n^{-1} V_{\tilde{\theta}_k}\big)\Big) \leq C\sqrt{\frac{\log n}{n}}$, where $\pi_{\rm E}(\cdot\,|\,X_{1:n})$ is the Bayesian ETEL posterior defined in~\eqref{Eqn:BayesianETEL}.
 \end{enumerate}
 \end{theorem}
 \noindent According to Theorem~\ref{th1}, each stationary point (saddle point, local minimum or local maximum) of population risk $\m R$ contributes to one component in the normal mixture approximation to the posterior with non-vanishing mixing weight. Moreover, one of these mixture components corresponds to the global minimizer $\theta^\ast$ of $\m R$, which is our estimation target.
As a consequence, the Bayesian ETEL posterior does not concentrate around $\theta^\ast$ unless $\m R$ has a unique stationary point, for example, when $\m R$ is strictly convex over $\Theta$.  A nice property in the theorem is that for each $k\in[K]$, the (rescaled) local covariance matrix $V_{\tilde{\theta}_k}$ matches the asymptotic covariance matrix of the local center $\hat\theta_k$. Therefore, the local shape of the posterior honestly captures the random fluctuation around local center $\hat\theta_k$. 

 \subsection{Analysis of Bayesian PETEL posterior with smooth loss}\label{Sec:BPETELsmooth}
 In this subsection, we establish a Bernstein–von Mises type theorem (asymptotic normality of the posterior) for the Bayesian PETEL posterior when the loss function $\ell(x,\theta)$ is smooth with respect to $\theta$.  We need Assumptions A.1, A.3 and the following.

\vspace{0.5em}
\noindent \textbf{Assumption A.2'}:
(1) The risk function $\mathcal{R}(\theta)$ has a unique global minimizer $\theta^*$ on $\Theta$ and $\theta^* \in \Theta^{\circ}$. (2) There exists a positive constant $a$ such that  $\Delta_{\theta^*} \succcurlyeq aI_d$ and  $\m H_{\theta^*}  \succcurlyeq a I_d$.
 
Assumption A.2' is a counterpart of Assumption A.2 in the previous subsection. However, here we only need matrices $\Delta_{\theta^*}$ and $\m H_{\theta^*}$ at a single point $\theta^\ast$ to be positive definite, which is much weaker. 
 
\begin{theorem}\label{th2}
Under Assumption A.1, A.2' and A.3,  there exist some constants $(C,C_1,C_2)$ independent of $n$, such that if $C_1 \log n \leq \alpha_n \leq C_2 \sqrt{n\log n} $, then it holds with probability at least $1-n^{-1}$ that,
\begin{equation*}
    d_{\rm TV}\Big(\pi_{\rm PE}(\cdot\,|\, X_{1:n}), \, N\big(\hat{\theta}, n^{-1} V_{\theta^*}\big)\Big) \leq  C\sqrt{\frac{\log n}{n}},
\end{equation*}
where recall that $\pi_{\rm PE}(\cdot\,|\, X_{1:n})$ is the Bayesian PETEL posterior distribution defined in equation~\eqref{Bayesian_PETEL} and $\hat{\theta}$ is the empirical risk minimizer on $\Theta$. In addition, we have $\sqrt{n}\,(\hat{\theta} - \theta^\ast) \to N(0, V_{\theta^\ast})$ in distribution as $n\to\infty$.
\end{theorem}
\noindent Theorem~\ref{th2} shows that the Bayesian PETEL posterior distribution of $\sqrt{n}(\theta-\hat{\theta})$ is close to the multivariate normal distribution with center $0$ and covariance matrix $ V_{\theta^*}$ in the total variation metric with rate $O(\sqrt{\frac{\log n}{n}})$. The lower bound requirement of $\alpha_n\geq C_1 \log n$ ensures that the extra penalty from $\m R_n$ is stronger enough to force the concentration of the posterior towards to global minimum $\theta^\ast$ by emptying out other mixture components indicated in Theorem~\ref{th1}.
In contrast, the upper bound requirement of $\alpha_n\leq C_2\sqrt{n\log n}$ guarantees that this extra penalty term will not dominate the ETEL so that it preserves the local shape of the Bayesian ETEL posterior around $\theta^\ast$.

Since the covariance matrix $V_{\theta^*}$ in the normal approximation of $\pi_{\rm PE}(\cdot\,|\, X_{1:n})$ matches the asymptotic covariance  matrix of $\hat{\theta}$, inferential conclusions derived from the Bayesian PETEL distributions are valid in a frequentist sense. The following corollary formalize this statement through characterizing frequentist coverage probabilities of credible regions.  Given a credible level $\alpha\in(0,1)$, let  $q_{\alpha}$ be the $\alpha$-th upper quantile of a $\chi^2$ distribution with $d$ degrees of freedom.  Let $\hat \theta_B$ and $\hat \Sigma_B$ be the mean and covariance matrix of $\theta$ under the Bayesian PETEL posterior distribution. According to Theorem~\ref{th2}, the highest density region of Bayesian PETEL posterior is close to the credible ellipse $\m E_n=\big\{(\theta-\hat\theta_B)^T\hat\Sigma_B^{-1} (\theta-\hat\theta_B) \leq q_{\alpha}\big\}$, and the next  corollary shows that its frequentist coverage is  at most $O((\log n)^{3/2}/\sqrt{n})$ away from $(1-\alpha)$.

\begin{corollary}\label{CI}
Let $\hat\theta_B$ and $\hat\Sigma_B$ be the posterior mean and covariance matrix of the Bayesian PETEL posterior distribution~\eqref{Bayesian_PETEL}. Under the assumptions of Theorem~\ref{th2}, there exists a constant $C_3$ such that
\begin{equation*}
   \Big| \m P^\ast\big(\theta^\ast\in \m E_n\big)-(1-\alpha)\Big| \leq C_3\frac{(\log n)^{3/2}}{\sqrt{n}}.
\end{equation*}
\end{corollary}
\noindent The Bayesian credible region $\m E_n$ in Corollary~\ref{CI} provides a simultaneous inference on the entire parameter vector $\theta$. Similar error bound also applies to the individual credible interval for each coordinate $\theta_j$ in $\theta$ for $j\in[d]$, which is approximately $\big[\hat\theta_{B,j}-z_{\alpha/2}\sqrt{[\hat\Sigma_B]_{jj}},\hat\theta_{B,j}+z_{\alpha/2}\sqrt{[\hat\Sigma_B]_{jj}}\big]$, with $z_{\alpha/2}$ denoting the $\alpha/2$-upper quantile of the standard normal distribution.

 \subsection{Analysis of Bayesian PETEL posterior with non-smooth loss}\label{Sec:BPETELnsmooth}
In practice, non-smooth loss functions are common, for example, in quantile regression and classification via soft-margin SVM~\citep{duda2012pattern}. In this subsection, we address the non-smooth case. 
In this case, it is common that due to the smoothing effect of taking expectation with respect to $X$, the population loss function $\m R(\theta)=\mb E[\ell(X,\theta)]$ remains smooth, which is true in all our considered examples. 
Under such cases, we assume that the moment function $g: \mathcal{X}\times\Theta\rightarrow \mathbb{R}^d$ employed in the ETEL~\eqref{Eqn:ETEL} for forming our Bayesian PETEL~\eqref{Bayesian_PETEL} is any function such that $\mathbb{E} [g(X,\theta)]=\nabla \mathcal{R}(\theta)$, for all $\theta\in\Theta$. For example, this condition can be achieved by choosing $g$ as any subgradient of $\ell(X,\theta)$ with respect to $\theta$ given that subgradients exist everywhere. Let $\Delta_{\theta}=\mathbb{E}\big[g(X,\theta)g(X,\theta)^T\big]$. We make following assumptions on $g$ and $\m R$. Let $\|\cdot\|_{\rm F}$ denote the matrix Frobenius norm.
      
  \vspace{0.5em}
  \noindent    
 \textbf{Assumption B.1}: The parameter space $\Theta$ is compact. The risk function $\mathcal{R}(\theta)$ is bounded and has bounded derivatives up to order three with respect to $\theta$ on $\Theta$. Both $g$ and $\ell$ are bounded over $\Theta$ and $\mathcal{X}$. There exist positive constants $(r, c)$ such that  $\|\Delta_{\theta}-\Delta_{\theta^*}\|_{\rm F} \leq c \|\theta-\theta^*\|_2$ for all $\theta\in B_r(\theta^\ast)$.
 
   \vspace{0.5em}
  \noindent    
 \textbf{Assumption B.2}: Define pseudo-metrics $d_n^g$ and $d_n^{\ell}$ as ${d^g_n}(\theta,\theta')=(n^{-1}\sum_{i=1}^n \|g(X_i,\theta)-g(X_i,\theta')\|_2^2)^{\frac{1}{2}}$ and ${d^{\ell}_n}(\theta,\theta')=(n^{-1}\sum_{i=1}^n (\ell(X_i,\theta)-\ell(X_i,\theta'))^2)^{\frac{1}{2}}$ respectively. There exists some positive constants $(c_0,c_1,\beta)$ such that 
   \begin{enumerate}
     \item[(a)]  The  $\varepsilon$-covering numbers of $\Theta$ with respect to $d_n^g$ and $d_n^{\ell}$ are upper bounded by $(n/\varepsilon)^{c_0}$;
     \item[(b)]  For any  $\theta\in \Theta$, it holds that $\mathbb{E}\big[\|g(X,\theta)-g(X,\theta^*)\|_2^2\big]+\mathbb{E}\big[(\ell(X,\theta)-\ell(X,\theta^*))^2\big]\leq c_1 \|\theta-\theta^*\|^{2\beta}_2$, where $\beta\leq 1$.
     \end{enumerate}

 \noindent Assumptions B.1 and B.2 are similar to the assumptions made in~\cite{10.2307/41000330}. In Assumption B.1, we impose smoothness directly on the risk function $\mathcal{R}(\theta)$ instead of on the loss function $\ell(x,\theta)$. Therefore,  we are able to handle non-smooth loss functions like those involving indicator functions as in quantile regression. Moreover, we only require the Lipschitz continuity of $\Delta_{\theta}=\mathbb{E}(g(X,\theta)g(X,\theta)^T)$ instead of $g(x,\theta)$. The statement in Assumption B.1 and B.2 is a sufficient condition to Assumptions (C4)-(C6) in~\cite{10.2307/41000330} and is easier to verify .

 \begin{theorem}\label{th3}
Under Assumptions A.2', A.3, B.1 and B.2, there exist constants $(C,C_1,C_2)$ independent of $n$, such that if $C_1\log n\leq \alpha_n\leq C_2\sqrt{n\log n}$, then it holds with probability at least $1-n^{-1}$ that,
    \begin{equation*}
    d_{\rm TV}\Big(\pi_{\rm PE}(\cdot\,|\, X_{1:n}), \, N\big(\hat\theta^\diamond, n^{-1} V_{\theta^*}\big)\Big) \leq  C\frac{(\log n)^{1+\beta/2}}{n^{\beta/2}},
\end{equation*}
where $\hat\theta^\diamond = \theta^*-n^{-1}\sum_{i=1}^n \m H_{\theta^*}^{-1}g(X_i,\theta^*)$, $\pi_{\rm PE}(\cdot\,|\, X_{1:n})$ is the Bayesian PETEL posterior distribution with $\nabla_\theta \ell$ replaced by $g$ and $V_{\theta^*}=H^{-1}_{\theta^*}\Delta_{\theta^*}H^{-1}_{\theta^*}$. In addition, we have  $\sqrt{n}\,(\hat{\theta}^\diamond - \theta^\ast) \to N(0, V_{\theta^\ast})$ in distribution as $n\to\infty$.
\end{theorem}
\noindent If the loss function $\ell$ is differentiable with respect to $\theta$ everywhere and $g(x,\theta)$ is chosen to be $\nabla_{\theta} \ell(X,\theta)$, by a standard analysis  of empirical risk minimizer~\citep{RePEc:eee:ecochp:4-36}, $\|\hat\theta^\diamond - \hat\theta\|_2 = O_p(n^{-1})$, and $\hat\theta^\diamond$ in Theorem~\ref{th3} can be replaced with $\hat{\theta}$. Therefore, Theorem~\ref{th2} can be viewed as a special case of Theorem~\ref{th3}.

 \subsection{Analysis of Bayesian PETEL posterior for high-dimensional problem}
 
 We consider the high-dimensional setting as discussed in Section~\ref{Sec:extensions}.  We assume our estimation target $\theta^\ast$, the global minimizer of risk function $\mathcal{R}$ over $\Theta$,  is $s^\ast$-sparse for some $s^\ast\ll n$. The following Theorem~\ref{th41} gives a non-asymptotic analysis to the ``model-averaged'' Bayesian PETEL posterior distribution defined in equation~\ref{Bayesian PETELsp1}. To begin with, we state the following regularity conditions. 
 
 \vspace{0.5em}
 \noindent
 \textbf{Assumption C.1}: There exists an $(n,d)$ independent constant $c$ such that $\Theta$ is contained in $[-c,c]^d$. Moreover, there exist some constants $(c_0,c_1)$ independent of $(n,d)$ such that for any $S\subseteq [d]$ with $|S|\leq s_0$, it holds that $\ell(x,\theta_{S},0)$ is uniformly bounded by $c_0$ and uniformly $c_1$ Lipschitz with respect to $\theta_S$ over $\theta_S \in\Theta_S$ and $x \in \mathcal{X}$.
 
  \vspace{0.5em}
 \noindent
 \textbf{Assumption C.2}:  There exists a positive constant $c_2$ independent of $(n,d)$ such that $\mathcal{R}(\theta)-\mathcal{R}(\theta^*)\geq c_2\|\theta-\theta^*\|^2_2$ holds  for any $\theta\in \Theta$ that is at most $s_0$ sparse.
 
 \vspace{0.5em}
 \noindent
 \textbf{Assumption C.2'}:   There exists  a positive constant $c_3$ independent of $(n,d)$ such that $ \underset{i \in S^*}{\min}\, {\theta^*_i}^2\geq c_3\sqrt{{\log (d\vee n)}/{n}}$. Moreover, there exists a positive constant $c_4$ independent of $(n,d)$ such that $\mathcal{R}(\theta)-\mathcal{R}(\theta^*)\geq c_4\|\theta-\theta^*\|^2_2$ holds for any $\theta\in \Theta$ that is at most $s^\ast$ sparse.
 
 \vspace{0.5em}
 \noindent
 \textbf{Assumption C.3}: There exist some positive constants $(c_5,r,L)$ such that $ \pi_{S^*}(\theta^*_{S^*})\geq c_5$ and $|\pi_{S^*}(\theta_{S^*})-\pi_{S^*}(\theta^*_{S^*})| \leq L \|\theta_{S^*}-\theta^*_{S^*}\|_2$ holds for any $\theta_{S^\ast} \in B_r(\theta^*_{S^*})$, where $S^\ast=\m S(\theta^\ast)$ denotes the support of $\theta^\ast$.
 
Assumption C.2 guarantees the concentration of the ``model-averaged'' Bayesian PETEL posterior to the population risk minimizer $\theta^*$. If all nonzero signals in $\theta^*$ are suitably large as in Assumption C.2', then  Assumption C.2 can be relaxed to the second statement of Assumption C.2'.
%
 \begin{theorem}\label{th41}
Suppose the risk function $\mathcal{R}(\theta): \mathbb{R}^d \to \mathbb{R}$ has a unique global minimizer $\theta^*$ on $\Theta$ that is $s^\ast$ sparse, where  $\theta^*_{S^*}\in \Theta_{S^*}^{\circ}$, $s^\ast \leq s_0$, $d\leq \exp(Cn)$ with an $(n,d)$ independent constant $C$ and Assumption C.1, C.3 holds. Suppose Assumptions A.1 and A.2' hold for the loss function $\ell(x,\theta_{S^*},0)$ and risk function $\m R(\theta_{S^*},0)$ with the parameter space being $\Theta_{S^\ast}$,  then there exist constants $(C_0,C_1,C_2)$ independent of $n$ and $d$ such that if $C_0 \log n \leq \alpha_{n,d}\leq C_1 n$,  then with probability at least $1-n^{-1}$ the ``model-averaged'' Bayesian PETEL posterior in~\eqref{Bayesian PETELsp1} satisfies
 \begin{equation*}
     d_{\rm TV}\bigg(\pi_{\rm PE}(\,\cdot\,|\,S^*,X_{1:n}), N\Big(\hat{\theta}_{S^*},\frac{1}{n} \big({V^{S^*}_{\theta^*_{S^*}}}^{-1}+\frac{\alpha_{n,d}}{n} \m H^{S^*}_{\theta^*_{S^*}}\big)^{-1}\Big)\bigg)\leq C_2 \sqrt{\frac{\log n}{n}} ,
\end{equation*}
where $\mathcal{H}_{\theta_{S^*}^*}^{S^*}$ is the Hessian of $\m R(\theta_{S^*},0)$ at $\theta^*_{S^*}$, $\Delta^{S^*}_{\theta^*_{S^*}}=\mathbb{E}(\nabla_{\theta_{S^*}}\ell (X,\theta^*_{S^*},0)\nabla_{\theta_{S^*}}\ell (X,\theta^*_{S^*},0)^T)$  and $V^{S^*}_{\theta^*_{S^*}}={\mathcal{H}_{\theta^*_{S^*}}^{S^*}}^{-1} \Delta^{S^*}_{\theta_{S^*}^*}{\mathcal{H}_{\theta^*_{S^*}}^{S^*}}^{-1}$. In addition:
\begin{enumerate}
\item  If Assumption C.2 holds, then there exist some positive constants $(C'_0,C'_1,C'_2,C'_3,C'_4)$ independent of $n$ and $d$ such that $C'_1\leq C'_3$ and if $\beta_{n,d}=C'_0\log (d\vee n)$ and $(C'_1 n)\wedge (C'_2\,{\log (d\vee n)}/{\underset{i \in S^*}{\min} \,{\theta^*_i}^2}) \leq \alpha_{n,d}\leq C'_3\, n$,  then it holds with probability at least $1-n^{-1}$  that 
 \begin{equation*}
  \Pi_{\rm PE}\Big(\|\theta-\theta^*\|_2\leq C'_4\sqrt{\frac{\log d\vee \log n}{n}}\,\Big|\, X_{1:n}\Big)\geq 1-\frac{1}{d\vee n}.
  \end{equation*}
  \item  If Assumption C.2' holds with a large enough $c_2$, then there exist  some positive constants $(\bar{C}_0,\bar{C}_1,\bar{C}_2,\bar{C}_3,\bar{C}_4)$ independent of $n$ and $d$ such that   if $\bar{C}_0\,\big((\log (d\vee n))\vee (\alpha_{n,d} \sqrt{\log (d\vee n)/n})\big) \\ \leq \beta_{n,d} \leq \bar{C}_1\, \alpha_{n,d} \,{\underset{i \in S^*}{\min} \,{\theta^*_i}^2}$ and $\bar{C}_2\log (d\vee n)/{\underset{i \in S^*}{\min} \,{\theta^*_i}^2} \leq \alpha_{n,d}\leq \bar{C}_3\, n$,  then it holds with probability at least $1-n^{-1}$  that 
    \begin{equation*}
     \Pi_{\rm PE}(S^*\,|\,X_{1:n})\geq 1-\exp(-\bar{C}_4\,\beta_{n,d}),
 \end{equation*}
  \end{enumerate}
 \end{theorem}

 \noindent Theorem~\ref{th41} shows that when $\log d\leq C\sqrt{n}$ and ${\min}_{i\in S^*}|\theta^*_i|$ is lower bounded by a positive $(n,d)$-independent constant,  if we choose $(\log d\vee \log n)\lesssim\alpha_{n,d}\lesssim \sqrt{n\log n}$ and  $(\log d \vee \log n)\lesssim\beta_{n,d} \lesssim\alpha_{n,d} $, then the Bayesian PETEL posterior of $\theta\in\mb R^d$ converges to a degenerate $s^\ast$-dimensional normal distribution with mean $\hat{\theta}_{S^*}$ and  covariance matrix  $V^{S^*}_{\theta^*_{S^*}}$ with rate $O(\sqrt{\frac{\log n}{n}})$. Since $\sqrt{n}(\hat{\theta}_{S^*}-\theta^*_{S^*}) \to  N(0, V^{S^*}_{\theta^*_{S^*}})$ in distribution as $n\to\infty$, it follows that the highest posterior region derived from the Bayesian PETEL posterior distribution has valid frequentist coverage probability.

 \section{Numerical Studies}
In this section, we  will investigate the performance of the Bayesian PETEL from the frequentist perspective in classification and regression problems, using both synthetic and real datasets.  In addition to Bayesian PETEL, we include three other methods in our comparison. 
\begin{itemize}
\item{CG (calibrated Gibbs posterior)}:  A Bayesian method proposed in~\cite{Syring_2018}, where they estimate the coverage probability by checking if the highest posterior density credible region $\{\theta: \pi_G(\theta|X_{1:n})\geq c_{\alpha}\}$, with $c_{\alpha}$ being chosen such that its posterior coverage  is $1-\alpha$ based on the bootstrapping data, covers the empirical risk minimizer, and apply stochastic approximation to update the learning rate of Gibbs  posterior until the estimated coverage probability is equal to the nominal level.
\item{Bootstrap}:  A frequentist method by bootstrapping the given data and constructing confidence intervals using bootstrapping empirical risk minimizers.
\item{(Misspecfied) ALD}:  A Bayesian method used in quantile regression, where the response distribution is misspecified to be an asymmetric Laplace distribution (ALD)~\citep{10.1214/13-BA817}.
\end{itemize}
 Unless otherwise specified, in the following simulation and real data examples, for Bayesian method, we use Metropolis-Hasting algorithm to generate $3000$ posterior samples, and use their $\frac{\alpha}{2}$ and $1-\frac{\alpha}{2}$ quantiles to construct  $1-\alpha$ Bayes credible intervals independently for each dimension of $\theta$. For Bootstrap,  we resample the data $3000$ times, and construct  $1-\alpha$ confidence intervals using the $\frac{\alpha}{2}$ and $1-\frac{\alpha}{2}$ quantiles of bootstrapping empirical risk minimizers solved by gradient descent. The coverage probabilities (coverage) and average interval lengths (length) are computed based on $1000$ replicates. We use average error  to denote the average of the $\ell_2$ norm of the difference between the resulting point estimates (posterior mean or bootstrapping empirical risk minimizer average) and the population minimizer of the risk function.
  \subsection{Simulation examples}\label{Sec:simulation}
  In our simulation study, we will  use synthetic data to investigate the performance of the Bayesian PETEL in classification, robust regression and quantile regression problem.
 
 \subsubsection{Classification via support vector machine}\label{SVM}
The soft-margin SVM~\citep{duda2012pattern}  minimizes $\frac{1}{2}\lambda \|\theta\|_2^2+\frac{1}{n}\sum_{i=1}^n \max(0, 1-Y_i \theta^T \tilde{X}_i)$  over $\theta \in \mathbb{R}^d$ with given data $\{(\tilde{X}_i, Y_i)\}_{i=1}^n$  and $Y_i=\pm 1$.  The value of $\lambda>0$ controls the $\ell_2$ norm of $\theta$ and the function $\max(0,1-Y\theta^T\tilde{X})$  is called the hinge-loss function. Following~\cite{8637389}, we also consider the smoothed hinge loss $\frac{1}{2}(\sqrt{u^2+\varepsilon^2}+u)$ with $u=1-Y\theta^T\tilde{X}$ and $\varepsilon$ being a small number, so the SVM with smoothed hinge loss minimizes  $\frac{1}{2}\lambda \|\theta\|_2^2+\frac{1}{2n}\sum_{i=1}^n  (\sqrt{u_i^2+\varepsilon^2}+u_i)$ where $u_i=1-Y_i\theta^T\tilde{X}_i$. We generate a synthetic data by creating two centroids  $c_{1}=(0.64,0.45)$ and  $c_{-1}=(-1.18,-0.24)$, then uniformly sampling $Y$ from $\{-1,1\}$ and given $Y=i$, sampling $\tilde{X}\sim N(c_i,I_2)$ with $c_i$ being the respective centroid.  We use the synthetic data to study the performance of Bayesian PETEL posterior for estimation  and inference on  the  global minimizer of  the population level loss function associated  with SVM with hinge loss (SVMH)  problem and smoothed hinge loss (SVMSH)  problem with different $n$ and $\alpha_n$, where $\lambda=0.1$ and $\varepsilon=0.5$. We also include in the comparison two other methods, one is  the classical bootstrapping method (Bootstrap) and the other one is the Calibrated Gibbs posterior (CG).  The Coverage probabilities and average interval lengths with target coverage being $95\%$ are shown in Table~\ref{tablecpilSVMH} and Table~\ref{tablecpilSVMSH}.
\begin{table}[h]
 \caption{\label{tablecpilSVMH}Coverage probabilities (\%) and average interval lengths under SVMH}
    \centering
   \begin{tabular}{cccccccccc}
   \hline
  &&\multicolumn{2}{c}{Bayesian PETEL}&\multicolumn{2}{c}{Bayesian PETEL}&\multicolumn{2}{c}{Bayesian PETEL}&\multicolumn{2}{c}{Bayesian PETEL}\\
  &&\multicolumn{2}{c}{$\alpha_n=0.5n^{\frac{1}{4}}$}&\multicolumn{2}{c}{$\alpha_n=2 n^{\frac{1}{4}}$}&\multicolumn{2}{c}{$\alpha_n=0.5 n^{\frac{1}{3}}$}&\multicolumn{2}{c}{$\alpha_n=2 n^{\frac{1}{3}}$}\\
  \hline
  &&coverage&length&coverage&length&coverage&length&coverage&length\\
  \hline
  \multirow{2}*{$n=500$}&$\theta_1$&96.1&0.171&96.0&0.170&96.6&0.170&95.7&0.169\\
  ~&$\theta_2$&95.5&0.241&95.1&0.237&95.6&0.239&94.8&0.234\\
  
   \multirow{2}*{$n=1000$}&$\theta_1$&95.9&0.121&95.5&0.121&96.0&0.122&95.3&0.120\\
  ~&$\theta_2$&94.1&0.170&93.9&0.169&93.9&0.170&93.9&0.169\\
   \hline
  \hline
  &&\multicolumn{2}{c}{Bayesian PETEL}&\multicolumn{2}{c}{Bayesian PETEL}&\multicolumn{2}{c}{\multirow{2}*{CG}}&\multicolumn{2}{c}{\multirow{2}*{Bootstrap}}\\
&& \multicolumn{2}{c}{$\alpha_n=0.5n^{\frac{1}{2}}$}& \multicolumn{2}{c}{$\alpha_n=2n^{\frac{1}{2}}$}&~&~\\
\hline
 &&coverage&length&coverage&length&coverage&length&coverage&length\\
 \hline
\multirow{2}*{$n=500$}&$\theta_1$&96.0&0.169&95.5&0.163&98.7&0.210&94.2&0.132\\
  ~&$\theta_2$&94.7&0.236&93.5&0.222&92.7&0.215&92.5&0.184\\
  
   \multirow{2}*{$n=1000$}&$\theta_1$&95.6&0.120&95.4&0.117&98.8&0.156&94.5&0.095\\
  ~&$\theta_2$&93.7&0.168&93.2&0.161&93.2&0.159&91.6&0.131\\
  \hline
  \end{tabular}
          \end{table}
 \begin{table}[h]
 \caption{ \label{tablecpilSVMSH}Coverage probabilities (\%) and average interval lengths  under SVMSH}
      \centering
   \begin{tabular}{cccccccccc}
   \hline
  &&\multicolumn{2}{c}{Bayesian PETEL}&\multicolumn{2}{c}{Bayesian PETEL}&\multicolumn{2}{c}{Bayesian PETEL}&\multicolumn{2}{c}{Bayesian PETEL}\\
  &&\multicolumn{2}{c}{$\alpha_n=0.5n^{\frac{1}{4}}$}&\multicolumn{2}{c}{$\alpha_n=2 n^{\frac{1}{4}}$}&\multicolumn{2}{c}{$\alpha_n=0.5 n^{\frac{1}{3}}$}&\multicolumn{2}{c}{$\alpha_n=2 n^{\frac{1}{3}}$}\\
  \hline
  &&coverage&length&coverage&length&coverage&length&coverage&length\\
  \hline
  \multirow{2}*{$n=500$}&$\theta_1$&94.6&0.169&94.8&0.168&94.6&0.169&94.2&0.167\\
  ~&$\theta_2$&95.6&0.243&95.2&0.239&95.3&0.241&94.9&0.237\\
  
   \multirow{2}*{$n=1000$}&$\theta_1$&94.9&0.120&95.1&0.120&95.2&0.120&95.0&0.120\\
  ~&$\theta_2$&94.5&0.173&94.8&0.171&94.7&0.172&94.2&0.170\\
   \hline
  \hline
  &&\multicolumn{2}{c}{Bayesian PETEL}&\multicolumn{2}{c}{Bayesian PETEL}&\multicolumn{2}{c}{\multirow{2}*{CG}}&\multicolumn{2}{c}{\multirow{2}*{Bootstrap}}\\
&& \multicolumn{2}{c}{$\alpha_n=0.5n^{\frac{1}{2}}$}& \multicolumn{2}{c}{$\alpha_n=2n^{\frac{1}{2}}$}&~&~\\
\hline
 &&coverage&length&coverage&length&coverage&length&coverage&length\\
 \hline
\multirow{2}*{$n=500$}&$\theta_1$&94.5&0.168&93.3&0.162&98.5&0.225&95.2&0.171\\
  ~&$\theta_2$&95.1&0.239&94.3&0.222&94.1&0.230&95.4&0.246\\
  
   \multirow{2}*{$n=1000$}&$\theta_1$&95.1&0.120&94.3&0.117&99.0&0.163&94.8&0.121\\
  ~&$\theta_2$&94.5&0.171&93.5&0.164&94.0&0.167&93.8&0.174\\
  \hline
  \end{tabular}
  
     \end{table}
 
 \begin{table}[h]
 
   \caption{  \label{tablecp1}Coverage probabilities (\%) for different target coverages $(90\%, 80\%, 70\%)$ under SVMH and SVMSH }
        \centering
   \begin{tabular}{lcccccccc}
       \hline
   &&&\multicolumn{3}{c}{SVMH}&\multicolumn{3}{c}{SVMSH}\\
 & &&Bayesian&\multirow{2}*{CG}&\multirow{2}*{Bootstrap}&Bayesian&\multirow{2}*{CG}&\multirow{2}*{Bootstrap}\\
 &&&PETEL&~&~&PETEL&~&~\\
   \hline
  \multirow{6}*{$n=500$}& \multirow{2}*{$90\%$}&$\theta_1$&91.7&96.4&88.4&89.0&95.2&88.7\\
&~&$\theta_2$&91.0&86.9&87.1&89.7&87.4&88.8\\
 &\multirow{2}*{$80\%$}&$\theta_1$&81.6&89.6&79.2&80.6&88.7&78.9\\
 &~&$\theta_2$&82.1&76.2&76.7&79.2&76.2&78.6\\
   & \multirow{2}*{$70\%$}&$\theta_1$&71.8&81.7&70.5&68.3&82.0&70.1\\
&  ~&$\theta_2$&71.8&66.7&66.0&69.4&67.3&68.0\\
 \hline
 \hline
  \multirow{6}*{$n=1000$}& \multirow{2}*{$90\%$}&$\theta_1$&91.3&95.5&88.9&89.4&98.2&90.2\\
&~&$\theta_2$&89.8&88.7&87.3&90.2&89.6&89.0\\
  &\multirow{2}*{$80\%$}&$\theta_1$&81.3&89.8&77.5&80.0&92.4&80.9\\
 &~&$\theta_2$&79.0&76.7&73.7&80.8&79.4&78.9\\
   & \multirow{2}*{$70\%$}&$\theta_1$&70.5&81.8&68.1&69.7&84.5&71.7\\
&  ~&$\theta_2$&70.2&66.6&64.8&71.6&67.8&68.4\\
\hline
    \end{tabular}
    \end{table}
    \begin{table}[h]
   \caption{\label{tablecp2}Average testing accuracies under SVMH and SVMSH }
    \begin{tabular}{cccccc}
       \hline
 \multicolumn{3}{c}{SVMH}&\multicolumn{3}{c}{SVMSH}\\
 Bayesian PETEL&CG& Bootstrap&Bayesian PETEL&CG& Bootstrap\\
   \hline
  0.830&0.830&0.832&0.831&0.831&0.832\\
  \hline
     \end{tabular}
       \end{table}
We can see from the tables that, first,  our  method is robust to choices of the penalty parameter $\alpha_n$ in both SVMH and SVMSH, as the coverage is at most $0.018$  away from the target $0.95$  and the change of the interval length is at most $0.021$ when $\alpha_n$ is from $0.5 n^{\frac{1}{4}}$ to $2 n^{\frac{1}{2}}$. Second, the  Calibrated Gibbs posterior (CG) tends to  underestimate the precision of the inference of $\theta_1$ and overestimate the precision of the inference of $\theta_2$, for the reason that  we cannot find a learning rate that simultaneously corrects all entries in the covariance matrix.  Table~\ref{tablecp1} gives coverage probabilities of our method with $\alpha_n=0.5n^{\frac{1}{4}}$,  CG and Bootstrap for  different target coverages, we can see when the sample size $n$ is $500$ and target coverage is $70\%$, the coverage of $\theta_1$ from CG is $11.7\%$ away from $70\%$ in SVMH and $12\%$ away from $70\%$ in SVMSH, while our method is at most $2.1\%$ away from the target. Moreover, coverage probabilities of CG  do not improve when $n$ increases from $500$ to $1000$, while we can see an obvious improvement in our method. Indeed, in the example of SVMSH, the variance vector in the Gaussian limiting distribution of Gibbs posterior with learning rate $\gamma$  and sample size $n$ is approximately $n^{-1}(1.148/\gamma,1.190/\gamma)$ , while the  variance vector in the Gaussian limiting distribution of  the empirical risk minimizer is approximately $n^{-1}(0.953,1.975)$, so no matter how large is the sample size, there does not exist a learning rate that calibrate the credible intervals of $\theta_1$ and $\theta_2$ simultaneously. Moreover, from Table~\ref{tablecp1}, our method performs slightly better than classical bootstrapping method in terms of coverage probabilities, as the deviances of coverage probabilities from the target coverages for our method are in general smaller than that of bootstrapping. It appears that  in Table~\ref{tablecp1},  most coverage probabilities of Bayesian PETEL in the SVMSH column are closer to their nominal values than those in the SVMH column. This phenomenon may attribute to our theoretical results that the Bayesian PETEL with smoothed hinge loss converges to its Gaussian limiting distribution at a faster rate of $O_p(n^{-\frac{1}{2}})$ (c.f. Theorem~\ref{th2}) than the Bayesian PETEL with hinge loss whose rate is $O_p(n^{-\frac{1}{4}})$ (c.f. Corollary 2 in Appendix B.2).   In addition, a larger sample size may be required for improving the performance of all methods for uncertainty quantification when estimating $\theta_2$ (it suffers from noticeable precision overestimation/underestimation across all methods). To study the performance of the resulting point estimators derived from Bayesian PETEL, CG and Bootstrap for correctly classifying the data, we provide in Table~\ref{tablecp2} the average testing accuracies of  the resulting point estimators based on $500$ training samples and $500$ testing samples, where the average testing accuracy means the average of  probabilities that the testing sample is correctly classified  using the corresponding point estimator.  We can see the average testing accuracies are quite similar among the three methods. 

\subsubsection{Robust regression for learning sigmoid unit}\label{Sec:robust_sim}
\begin{table}[h]
\caption{\label{tablerobustcov.1}Coverage probabilities (\%) and average interval lengths under Robust regression}
     \centering
   \begin{tabular}{cccccccc}
    \hline
  &&\multicolumn{2}{c}{Bootstrap}&\multicolumn{2}{c}{Bayesian ETEL}&\multicolumn{2}{c}{Bayesian PETEL}\\
  &&Coverage&Length&Coverage&Length&Coverage&Length\\
  \hline
  \multirow{3}*{Target=$95\%$} & $\theta_1$&98.0&3.67&72.0&1.24&93.4&0.68\\
  &  $\theta_2$&99.5&4.06&77.3&1.13&93.9&1.00\\
& $\theta_3$&96.7&0.70&71.9&0.44&92.7&0.39\\
  \hline
     \hline
     \multirow{3}*{Target=$90\%$} & $\theta_1$&95.8&2.85&67.8&1.04&87.4&0.57\\
  &  $\theta_2$&97.6&3.40&72.5&0.94&88.6&0.84\\
 & $\theta_3$&93.7&0.58&66.6&0.37&87.3&0.33\\
  \hline
      \end{tabular}
            \end{table}
Consider the simple example of learning a sigmoid unit.  Let $S(z)=\frac{\exp(z)}{1+\exp(z)}$  for $z\in\mathbb{R}$. We assume the predictor $\tilde{X} \in\mathbb{R}^2$ follows $N(0,I_2)$ and the response $Y$ is generated by the model $Y=\theta_3^*\cdot S(\theta_1^*\tilde{X}_1+\theta_2^*\tilde{X}_2)+e$, where $\theta^*=(\theta_1^*,\theta_2^*,\theta_3^*)=(1,2,3)$ and the heterogeneous error $e$ follows a Cauchy distribution with location being $0$ and scale being $\frac{\|\tilde{X}\|_2}{\sqrt{6}}$. We consider the Huber loss
\begin{equation*}
\ell(X,\theta)=\left\{\begin{array}{ll}
\frac{1}{2}(Y-\theta_3\cdot S(\theta_1\tilde{X}_1+\theta_2\tilde{X}_2))^{2} & \text { for }|Y-\theta_3\cdot S(\theta_1\tilde{X}_1+\theta_2\tilde{X}_2)| \leq \delta \\
\delta|Y-\theta_3\cdot S(\theta_1\tilde{X}_1+\theta_2\tilde{X}_2)|-\frac{1}{2} \delta^{2} & \text {otherwise}
\end{array}\right.
\end{equation*}
where $X=(\tilde{X},Y)$ and $\delta$ is fixed to be $2$ here.  We sample $n=500$ number of i.i.d samples $\{(\tilde{X}_i, Y_i)\}_{i=1}^{n}$ and  use the synthetic data to study the performance of Bayesian PETEL/ETEL and bootstrapping. For Bayesian PETEL, to achieve fast convergence, we first generate $500$ number of samples from Bayesian PETEL posterior with $\alpha_n=n$  using symmetric random-walk Metropolis algorithm (RMW), where the initial point is randomly selected from $N(2,4 I_3)$, then we use the mean of $400$ to $500$ posterior samples to be the new initial point, and generate $3000$ number of samples from Bayesian PETEL posterior with $\alpha_n$ equal to $2\sqrt{n}$. For Bootstrap, we use gradient descent  to solve the empirical risk minimizer and for Bayesian ETEL, we use RMW  to generate posterior samples, where initial points in gradient descent algorithm and RMW algorithm are randomly selected from $N(2,4 I_3)$ respectively. 
            
The coverage probabilities and average interval lengths are given in Table~\ref{tablerobustcov.1}.  We can see from Table~\ref{tablerobustcov.1} that firstly,  Bayesian PETEL performs notably better than Bootstrap and Bayesian ETEL in terms of coverage probability. Specifically, the Bootstrap tends to underestimate the precision of inferences of $\theta_1$, $\theta_2$ and $\theta_3$ and the average interval lengths are much larger than those of Bayesian PETEL/ETEL. Moreover, the Bayesian ETEL tends to overestimate the precision of inferences of $\theta_1$, $\theta_2$ and $\theta_3$. In addition, the average errors of the resulting point estimators are $ 1.084$ for Bootstrap, $1.825$ for Bayesian ETEL  and $0.3124$ for Bayesian PETEL, we can see that the posterior mean of Bayesian PETEL leads to a much better point estimator of $\theta^*$ than Bayesian ETEL and Bootstrap.  These phenomenons are due to the fact that the risk function is not convex. Indeed, for the  Bootstrap method, the marginal density plots for the first and second dimensions of the bootstrapping empirical risk minimizers solved by gradient descent algorithm are right heavy-tailed, which leads to wider confidence intervals. Specifically,  the gradient vector field of the risk function in region $A=[2.5,4]\times [4,7.5]\times[2.5,2.8]$ is fairly flat (i.e.,~the $\ell_2$ norms of the gradients of the risk function evaluated at points in  set $A$ are all smaller  than $0.1$). For each bootstrapping replicate, if the initial point of the gradient descent algorithm lies in $A$ and the step size is too small for the next iterate to jump over this flat area, the algorithm will converge to some points inside $A$ instead of the true bootstrapping empirical risk minimizer.  For the Bayesian ETEL method, depends on the initial state of the Markov chain, the random walk Metropolis-Hasting algorithm may get stuck in a  local mode of the  Bayesian ETEL posterior that is far away from $\theta^*$, which leads to a large point estimation error for estimating $\theta^*$; while for the Bayesian PETEL method,  the extra penalty term $-\alpha_n\mathcal{R}_n(\theta)$ favors points closer to the empirical risk minimizer. Unlike the gradient descent which may converge to a local minimum or saddle point, the Markov chain has the ability of escaping from any local mode and the generated samples from the Bayesian PETEL after the burn-in period becomes all around $\theta^*$ with marginal densities  for each dimension of $\theta$ being Gaussian-like. Further details are available in Appendix A.5.

%

  \subsubsection{High Dimensional  Quantile Regression}\label{Sec:quantile_sim}
  \begin{table}[h]
     \caption{ \label{tablecpilhdqr.1}Coverage probabilities (\%) and average interval lengths under High Dimensional Quantile Regression}
      \centering
\begin{tabular}{ccccccccc}
   \hline
  &\multicolumn{2}{c}{Bayesian PETEL}&\multicolumn{2}{c}{BIC CG}&\multicolumn{2}{c}{BIC Bootstrapping}&\multicolumn{2}{c}{BIC ALD}\\
  &Coverage&Length&Coverage&Length&Coverage&Length&Coverage&Length\\
  \hline
$\theta_1$ &95.3&0.180&95.3&0.179&96.3&0.182&99.7&0.250\\
$\theta_2$ &94.9&0.138&94.7&0.131&96.0&0.139&98.9&0.184\\
\hline
   \end{tabular}
   \end{table}
\noindent In quantile regression, for fixed $\tau \in (0,1)$, the $\tau^{th}$ quantile of the response $Y\in \mathbb{R}$ given the covariates $\tilde{X}\in \mathbb{R}^d$ is modelled as  
\begin{equation*}
Q_{\tau}(Y|\tilde{X})=\tilde{X}^T\theta^*.
\end{equation*}
Here we consider loss function $\ell(X,\theta)=(Y-\tilde{X}^T\theta)(\tau-\textbf{1}(Y<\tilde{X}^T\theta))$~\citep{Syring_2018} with  $\tau=0.5$. To investigate the performance of our proposed  ``model-averaged'' Bayesian PETEL posterior, we choose $d=1000$ and simulate datasets of $n = 500$ i.i.d observations where each $(\tilde{X}_{i1}, \tilde{X}_{i2})$ is from multivariate Gaussian $N(0, \text{diag}(1,2))$ and $(\tilde{X}_{i3},\cdots,\tilde{X}_{id})$ is  from $N(0, I_{d-2})$. To sample $Y_i=\tilde{X}_i^T\theta^*+e_i$, we use $\theta^*=(2,3,\mathbf{0}_{d-2}^T)^T$ and the heterogeneous error $e_i$ sampled from $N(0,0.5\sqrt{(X^2_{i1}+X^2_{i2})/2})$. To alleviate the curse of dimensionality, we first use stepwise search  to find the model $\tilde{S}$ that maximizes $\exp(-\alpha_{n,d} \mathcal{R}_n(\hat{\theta}_{S},0)-\beta_{n,d}|S|-\log \binom{d}{|S|})$ with  $\alpha_{n,d}=2\sqrt{n}$ and $\beta_{n,d}=1.2\log d$, where $\hat{\theta}_S$ is the constrained empirical risk minimizer on model $S$. We limit the model space to models that have 1-bounded Hamming distances with $\tilde{S}$ and choose the prior to be $\pi(S)\propto \exp(-\beta_{n,d}|S|)\binom{d}{|S|}^{-1}$ and $\pi(\theta_S|S)=N(\mathbf{0}_{|S|}, I_{|S|})$. We  run the  Bayesian PETEL algorithm a thousand times and get that the average  Bayesian PETEL posterior probability of the true model  is $2\times 10^{-3}$ away from $1$. Furthermore,  Table~\ref{tablecpilhdqr.1} gives coverage probabilities and average interval lengths  of  $95\%$ Bayesian PETEL posterior credible intervals of $\theta_1$ and $\theta_2$.  To make comparison, we also consider Calibrated Gibbs posterior,  bootstrapping estimators and misspecified ALD~\citep{10.1214/13-BA817} with the model  selected  by High dimensional BIC~\citep{rigollet2015high} where the penalty parameter  on the number of degrees of freedom is $10\log d$. We can see from Table~\ref{tablecpilhdqr.1} that for the quantile regression problem, our method achieves notably better performance than misspecified ALD,  due to the misspecification of error distribution in the misspecified ALD. Moreover, our method performs similarly with BIC CG and slightly better than  BIC Bootstrap, as coverage probabilities of BIC Bootstrap are  at least $1\%$ away from $95\%$, while those of Bayesian PETEL are at most $0.3\%$ away of $95\%$. In addition, the average errors of the resulting point estimators are $0.1892$ for Bayesian PETEL, $0.1997$ for BIC CG, $0.1950$ for BIC Bootstrap and $0.1898$ for BIC ALD, thus our method achieves the smallest average error among methods considered in this section.


\subsection{Markov chain Monte Carlo convergence and efficiency}\label{Sec:MCMCconvergence}
In this section, we use Gelman–Rubin convergence diagnostic tool~\citep{10.1214/ss/1177011136} to check the convergence of the chains, and use their effective sample sizes and computation times to report the efficiency of the proposed MCMC algorithm. We study the convergence and efficiency of the proposed MCMC algorithm for implementing the proposed Bayesian PETEL posterior for (1) smooth loss function; (2) non-smooth loss function; (3) high-dimensional problems, using examples in Section~\ref{SVM} and Section~\ref{Sec:quantile_sim}. The proposed algorithms are implemented using the R program with a 2.3GHz computer processor.

\subsubsection{Soft-margin SVMs with  hinge loss and smoothed hinge loss}
In this section, we consider the example of soft-margin SVMs with hinge loss (SVMH) and smoothed hinge loss (SVMSH) in Section~\ref{SVM}. We use the Random walk Metropolis-Hasting algorithm with proposal $N(\theta_{\rm old}, \sigma^2 I_d)$ where $\theta_{\rm old}$ is the previous one state in the Markov chain and $\sigma$ is a parameter that is tuned such that the acceptance rate of the Markov chain is close to $0.234$.  For both SVMH and SVMSH, the computation time of a single run with $n=500$, $\alpha_n=2\sqrt{n}$ and $3000$ iterations is $1.24$ min on average. The Gelman–Rubin plots available in Appendix A.5 shows that  the MCMC procedure converges after $1000$ iterations in both SVMSH and SVMH problems. The effective sample sizes of the Markov chain  for each dimension of samples with a total  of $3000$ iterations are on average $(402, 305)$ for SVMSH and $(380,294)$ for SVMH.  We can see the effective sample sizes of the Markov chain for SVMSH are slightly larger than those of SVMH.
  
\subsubsection{High dimensional sparse quantile Regression}
 In this section, we consider the example of high dimensional quantile regression in Section~\ref{Sec:quantile_sim}. We use the independence sampling algorithm with proposal $p_{ prop}(S,\theta_S)=p_{prop}(S)p_{prop}(\theta_S|S)$ being chosen as that described in Appendix A.4. The computation time of  a single run with $n=500$, $d=1000$, $\alpha_{n,d}=2\sqrt{n}$, $\beta_{n,d}=1.2 \log d$ and $3000$ iterations is $1.87$ min on  average. The algorithm generates a sequence of samples $\{(S_i,\theta_{S,i}\}_{i=1}^{3000}$, with $S_i$ being the model and $\theta_{S,i}$ being the parameter corresponds to $S_i$.  We can consider the sequence of $\theta_{S,i}$ that corresponds to the true model $S^*=(1,2)$,  i.e. $\{\theta_{S,i_j}\}_{j=1}^{n'}$ where $n'=\sum_{k=1}^{3000} \textbf{1}(S_k=S^*)$ and $i_j=\{i\,|\, S_i=S^*;\, \sum_{k=1}^{i-1}\textbf{1}(S_k=S^*)=j-1\}$.  
 The number of $n'$ are on average $2997$  and we can learn from the Gelman–Rubin plots for multiple chains of   $\{\theta_{S,i_j}\}_{j=1}^{n'}$  in Appendix A.5  that the MCMC procedure converges after $1000$ iterations. 
 The effective sample sizes for each dimension of $\{\theta_{S,i_j}\}_{j=1}^{n'}$ are on average $(830,785)$ respectively.  The choice of the proposal distribution of the model  $S$ is significant  for  efficiently sampling from Bayesian PETEL  with high dimensional structures. Indeed, if we choose  $p_{ prop}(S)$ to be a uniform distribution among $S\subseteq[d]$, the Markov chain may never generate samples correspond to the true model in any reasonable number of iterations, as the number of candidate models is extremely large. Therefore, we need to adjust the weight of the model in   $\{S\subseteq [d]\}$ to form a reasonable proposal, such that $p_{ prop}(S)$ will only give mass to models that correspond to small constrained minimal empirical risks (i.e., $\mathcal{R}_n(\hat{\theta}_S,0)$  where $\hat{\theta}_S$ is the constrained empirical risk minimizer on model $S$) while in the meantime do not have large complexities. Possible choices of the $p_{prop}(S)$ are described in Appendix A.4.

 \begin{table}[h]
     \caption{\label{tablepb.1}Coverage probabilities (\%) and average interval lengths in the Parking Birmingham Dataset}
         \centering
\begin{tabular}{ccccccccc}
   \hline
  &\multicolumn{2}{c}{Bayesian PETEL}&\multicolumn{2}{c}{CG}&\multicolumn{2}{c}{ALD}&\multicolumn{2}{c}{Bootstrap}\\
   &Coverage&Length &Coverage&Length &Coverage&Length &Coverage&Length\\
  \hline
$\theta_0$   &95.1&0.078&92.1&0.070&89.3&0.065&95.6&0.080\\
$\theta_1$ &95.2&0.092&97.6&0.100&92.5&0.081&95.2&0.093\\
$\theta_2$  &95.6&0.061&97.1&0.065&92.7&0.053&96.1&0.062\\
$\theta_3$ &95.4&0.087&97.7&0.097&92.0&0.077&95.2&0.088\\
\hline
  \end{tabular}
  \end{table}

  \begin{table}[h]
     \caption{ \label{tableod1}Coverage probabilities (\%) and average interval lengths  in the Occupancy Detection Dataset}
       \centering
   \begin{tabular}{ccccccc}
   \hline
  &\multicolumn{2}{c}{Bayesian PETEL}&\multicolumn{2}{c}{CG}&\multicolumn{2}{c}{Bootstrap}\\
  &Coverage&Length &Coverage&Length &Coverage&Length\\
  \hline
$\theta_1\, (Light)$ &93.7&0.0571&93.6&0.0567&95.6&0.0576\\
$\theta_2 \,(CO2)$ &95.2&0.0541&97.2&0.0555&96.0&0.0545\\
$\theta_3\, (Humidity \,Ratio)$ &94.7&0.0640&94.2&0.0622&93.5&0.0644\\
  \hline
     \end{tabular}
  \end{table}

\subsection{Real data analysis}

The good performance of Bayesian PETEL posterior in the simulation examples validates the correctness of our theoretical results in Section 4, that is, the Bayesian PETEL has valid frequentist properties when some regularity conditions are satisfied.  However, it is also a crucial problem of whether our regularity conditions are met in real data applications. To check this, we conduct a real data analysis and study the performance of Bayesian PETEL and its competitors. In the real data analysis, we consider quantile regression with the Parking Birmingham Dataset and classification with the  Occupancy Detection Dataset.  In each example, to show the ``correctness'' of the inference from our method, we sample $n'=2000$ samples with replacement from the original dataset $1000$ times, and in each time, we construct  $95\%$ Bayesian credible intervals from Bayesian PETEL posterior with $\alpha_{n'}=2\sqrt{n'}$ for each dimension of $\theta$ using the resampling dataset and check whether those credible intervals covers each dimension of the empirical risk minimizer $\hat{\theta}$ from the original dataset. Similarly for CG, Bootstrap and ALD.  Moreover, we use average error to denote the average of the $\ell_2$ norm of the difference between resulting point estimates (posterior mean or bootstrapping empirical risk minimizer average) and  $\hat{\theta}$.
  
\subsubsection {Parking Birmingham Dataset}\label{Sec:quantile_realdata} 
We study a dataset comprising Car park occupancy rate from 2016/10/04 to 2016/12/19.  The predictors include time and car park capacity.  The dataset is archived from UCI machine learning repository.  We model the median of the response $Y$ (occupancy rate) given the covariate $T$ (time) and $\tilde{X}$(car park capacity) by the following quantile regression model,
\begin{equation*}
Q_{0.5}(Y|T,\tilde{X})=\theta_0\tilde{X}+\sum_{k=1}^K \theta_k B_k(T),
     \end{equation*}
where $B_k(T)$ denote the $k$th degree of B-spline in $T$.  $K$ is fixed to be $3$ here and the columns of the data matrix are  scaled to be with center $0$ and variance $1$. The coverage probabilities and average interval lengths computed by subsampling are given in Table~\ref{tablepb.1}. To make comparison, we also check coverage probabilities and average interval lengths of CG, ALD  and  Bootstrap with target coverage being $95\%$.  We can see from Table~\ref{tablepb.1} that our method performs  better than CG and  ALD in terms of coverage probabilities, and performs similarly with Bootstrap in terms of coverage probabilities and average interval lengths.  Moreover, the average errors of the resulting point estimators from Bayesian PETEL, CG, ALD and Bootstrap are $0.1368$, $0.1460$, $0.1420$ and $0.1493$ respectively, we can see that Bayesian PETEL has the smallest average error.

\subsubsection{Occupancy Detection Dataset}
 
In this section, we consider the occupancy detection dataset, archived in UCI machine learning repository. The binary response variable $Y$ is  the occupied status of a room which was obtained from time stamped pictures that were taken every minute. We focus here on predictors $\tilde{X}$ including Light, CO2 and Humidity ratio. The goal of this section is to conduct inference to the parameter $\theta$ under the problem of SVM using 
smoothed hinge loss,  where the loss function is  $\frac{1}{2}\lambda \|\theta\|_2^2+\frac{1}{2} (\sqrt{U^2+\varepsilon^2}+U)$ with $U=1-Y\theta^T\tilde{X}$. The tuning parameters $\lambda$ and $\varepsilon$  are chosen to be $0.5$  and $0.1$. The coverage probabilities and average interval lengths computed by subsampling are given in Table~\ref{tableod1}. We also include CG and Bootstrap in comparison.  We can see that our method performs similarly with Bootstrap in terms of coverage probabilities and the average interval lengths of each dimension of $\theta$ in our method are all strictly smaller than those of Bootstrap. Moreover, our method achieves slightly better performance than CG, as the coverage probability of  $\theta_2$ using CG is $2.2\%$  away from  the target while the coverage probability in our method is at most $1.3\%$ away from the target.  In addition,  the averaged errors of the resulting point estimators derived from Bayesian PETEL, CG and Bootstrap are $0.1092$, $0.1059$ and $0.1074$ respectively, thus the average error of Bayesian PETEL is quite similar to that of CG and Bootstrap.

 \section{Discussion}
 In this paper, we propose the Bayesian penalized exponentially tilted empirical likelihood (Bayesian PETEL) posterior, which takes the  exponentially tilted empirical likelihood (ETEL) into a Bayesian framework and uses the empirical risk to exponentially penalize certain ``loss'' of parameter $\theta$ on the training data. Our model is free from the underlying distribution and is theoretically justified in the sense that it can be approximated by a normal distribution centered at the empirical risk minimizer, and its covariance matrix  matches the frequentist asymptotic covariance matrix of its mean vector. As a consequence, the posterior credible regions derived from Bayesian PETEL posteriors  have approximately correct  frequentist coverage.  The theory we provided can adapt to the case that the loss function is non-smooth, which includes quantile regression and soft-margin SVM as two representative examples. 
Our method naturally extends to the sparse high dimensional model: we show that the proposed ``model-averaged'' Bayesian PETEL posterior converges to a normal distribution under the true model, and the accompanied Bayesian credible region has valid frequentist coverage. Compared with methods based on Gibbs posterior, our method does not require the generalized information equality and is thus insusceptible to the model misspecification biases. Furthermore,  we show in the simulation study  that the corresponding posterior inference  from our method is notably more accurate than the calibrated Gibbs posterior and performs comparably to the bootstrapping.  Although the current paper focused on the exponentially tilted empirical likelihood, using the empirical likelihood or some other variants may work as well, which will be left as a future direction.

 \bibliography{references}{}
\bibliographystyle{abbrvnat} 
\newpage

\appendix
\begin{center}
{\bf\Large Appendix}
\end{center}


\section{Computational Details}\label{App:computation}

In this section, we will discuss computational aspects of sampling from the Bayesian PETEL posterior distribution.
\subsection{Algorithm Overview}
Since the Bayesian PETEL provides an explicit expression for the posterior up to a normalisation constant,  we can utilize the Metroplis-Hasting algorithm to draw posterior samples. In each step, we propose a new parameter $\tilde{\theta}$ from the proposal $p_{prop}(\theta|\theta_{old})$, where  $\theta_{old}$ is the parameter value
from the previous step. A uniform random number $u\in (0,1)$ is drawn, if 
\begin{equation*}
    u<\frac{\pi(\tilde{\theta})\exp(\log L(X^n;\tilde{\theta})-\alpha_n \mathcal{R}_n(\tilde{\theta}))p_{prop}(\theta_{old}|\tilde{\theta})}{\pi(\theta_{old})\exp\left(\log L(X^n;\theta_{old})-\alpha_n \mathcal{R}_n(\theta_{old})\right)p_{prop}(\tilde{\theta}|\theta_{old})},
\end{equation*}
then we accept the proposed $\tilde{\theta}$, otherwise, we retain $\theta_{old}$ in the chain. \\
\quad\\
One major difficulty in the sampling from Bayesian PETEL posterior distribution is the computation of the log ETEL, as it involves solving the Lagrange multiplier $\lambda(\theta)={\arg \min }_{\lambda}\, n^{-1} \sum_{i=1}^{n} \exp (\lambda^{T} \nabla_{\theta} L(x_{i}, \theta))$.  Since  solving $\lambda(\theta)$ is a convex problem, it can be calculated by modified Newton-Raphson algorithm~\citep{10.1093/biomet/89.1.230}.  Algorithm 1 summarizes the pseudocode for the Metroplis-Hasting steps to sample from Bayesian PETEL posterior and $\nabla_{\theta} \ell(X,\theta)$ is replaced by its subgradient when the loss function is not differentiable at  $\theta$.

\begin{algorithm}[ht!]\label{algorithm}
\caption{Metroplis-Hasting algorithm to sample from Bayesian PETEL posterior}
\SetAlgoLined
\SetKwRepeat{Do}{do}{while}%
 \textbf{Input}: Number of iteration $L$, tolerance $\varepsilon$, proposal distribution $p_{prop}(\cdot|\cdot)$, initial state $\theta^0$ and $\lambda(\theta^0)$\;
 \textbf{Data}:$X_1,X_2\cdots,X_n$\;
 \For{$t \leftarrow 0 \,\,to\,\, L-1$}{
   Sample $\tilde{\theta}$ from $p_{prop}(\cdot|\theta^t)$\;
   Generate a uniform random  number $u\in(0,1)$\;
   Define $f(\lambda)\leftarrow \frac{1}{n} \sum_{i=1}^{n} \exp (\lambda^{T} \nabla_{\theta} \ell(x_{i}, \tilde{\theta}))$\;
   $\lambda^0\leftarrow \lambda(\theta^{t})$\;
    $k\leftarrow0$\;
  
   \Repeat{$\|H^{-1}G\|_2\leq \varepsilon$ }{
       $k\leftarrow k+1$\;
       $\gamma=1$\;
       $H\leftarrow\frac{1}{n}\sum_{i=1}^n\exp\left(\nabla_{\theta}\ell(X_i,\tilde{\theta})^T\lambda^{k-1}\right)\nabla_{\theta}\ell(X_i,\tilde{\theta})\nabla_{\theta}\ell(X_i,\tilde{\theta})^T$\;
 $G\leftarrow\frac{1}{n}\sum_{i=1}^n\exp\left(\nabla_{\theta}\ell(X_i,\tilde{\theta})^T\lambda^{k-1}\right)\nabla_{\theta}\ell(X_i,\tilde{\theta})$\;
 \Repeat{$f(\lambda^k)\leq f(\lambda^{k-1})$}
 { $\lambda^k\leftarrow\lambda^{k-1}- \gamma H^{-1}G$\;
$ \gamma=\frac{1}{2}\gamma$;}
       }
   
   $\lambda(\tilde{\theta})\leftarrow \lambda^k$\;
  \eIf{$u\leq \frac{\pi(\tilde{\theta})\exp\left(\sum_{i=1}^n \log \frac{\exp(\lambda(\tilde{\theta})^T \nabla_{\theta} \ell(X_i,\tilde{\theta}))}{\sum_{i=1}^n \exp(\lambda(\tilde{\theta})^T \nabla_{\theta} \ell(X_i,\tilde{\theta}))}-\alpha_n \mathcal{R}_n(\tilde{\theta})\right)p_{prop}(\theta^t|\tilde{\theta})}{\pi(\theta^t)\exp\left(\sum_{i=1}^n \log \frac{\exp(\lambda(\theta^t)^T \nabla_{\theta} \ell(X_i,\theta^t))}{\sum_{i=1}^n \exp(\lambda(\theta^t)^T \nabla_{\theta} \ell(X_i,\theta^t))}-\alpha_n \mathcal{R}_n(\theta^t)\right)p_{prop}(\tilde{\theta}|\theta^t)}$}
  { $\theta^{t+1}\leftarrow\tilde{\theta}$\;
    $\lambda(\theta^{t+1})\leftarrow \lambda(\tilde{\theta})$\;
   }{
    $\theta^{t+1}\leftarrow\theta_t$\;
     $\lambda(\theta^{t+1})\leftarrow \lambda(\theta_t)$\;
  }
 }
 
\end{algorithm}

\subsection{ Choice of $\alpha_n$}
According to Theorem 2, the penalty parameter $\alpha_n$ should be in the range $\log n\lesssim \alpha_n \lesssim \sqrt{n\log n}$. In practice, we could choose $\alpha_n=Cn^{-c}$ with $0<c \leq 0.5$ and $C$ being a positive constant  (e.g., $\alpha_n=0.5\sqrt{n}$). We show in the simulation study in Section 5 that the performance of Bayesian PETEL is robust to the choice of $C$ and $c$. In the case of small dataset, by the result in Theorem 1,   the BETEL posterior distribution, which is equivalent to the Bayesian PETEL posterior with $\alpha_n=0$, is asymptotically mixture of Gaussian, with centers being  solutions of $ \nabla_{\theta} \mathcal{R}_n(\theta)=0$, so the posterior mean of BETEL will mismatch that of the empirical risk minimizer when  $ \nabla_{\theta} \mathcal{R}_n(\theta)=0$ has multiple solutions on $\Theta$. So intuitively, with a small value of $\alpha_{n}$, the Bayesian PETEL posterior of $\theta$ may have several modes, while a large value of $\alpha_{n}$ may lead to the invalidity of inference due to the mismatching of covariance matrix.  According to this fact, we could tune the penalty parameter by starting with a small number (e.g. $\log n$)  and increase it with step size $s_n$ until the posterior mean matches that of the empirical risk minimizer, in  which  the empirical risk minimizer can be  solved by subgradient/gradient descent or estimated by  the posterior mean of the Gibbs posterior given by  $\pi_{\rm G}(\theta\,|\,X^n)\propto \pi(\theta)\exp(-n \mathcal{R}_n(\theta))$. Since the target range  $\log n\lesssim \alpha_n \lesssim \sqrt{n\log n}$  is wide, it's safe to choose a large step size (e.g., $0.5\sqrt{n}$),  and a reasonable $\alpha_n$ could be found in few steps.

\subsection{Choice of the proposal distribution}\label{A.3}
It has long been recognized that the choice of the proposal distribution is crucial to the rapid convergence of the Metropolis-Hastings algorithm. The most common case involves a symmetric random-walk Metropolis algorithm (RMW) in which the proposal is given by $\theta=\theta_{old}+e$, where the increment $e$ is follow some fixed symmetric distribution (e.g.  $N(0, \sigma^2 \Sigma)$ with $\Sigma$ being a positive definite  $d\times d$ matrix). In this case, the crucial issue is to how to properly scale the proposal (e.g., how to choose $\sigma$) for avoiding extreme cases that the chain moves too slowly or the proposal is usually be rejected. A simple way to avoid the extremes is to monitor the acceptance rate of the algorithm. In our case, by Theorem 2, the Bayesian PETEL posterior distribution can be well approximated by $N(\hat{\theta}, \frac{1}{n}\m H_{\theta^*}\Delta_{\theta^*}^{-1}\m H_{\theta^*})$, so a reasonable choice of $\sigma$ would be $\sigma\asymp n^{-\frac{1}{2}}$ and we could start with $\sigma=c n^{-\frac{1}{2}}$ with positive constant $c$ and adjust $c$ until the acceptance rate is close to $0.234$~\citep{10.1214/aoap/1034625254}.  Apart from guaranteeing the quick convergence of Metropolis-Hastings algorithm, choosing $\sigma\asymp n^{-\frac{1}{2}}$ can guarantee the rapid convergence of the Newton-Raphson algorithm for computing $\lambda(\tilde{\theta})$ if we choose the initial value of $\lambda$ in the Newton-Raphson algorithm at time $t$ to be the $\lambda$ value computed in the last step (i.e., $\lambda(\theta^{t-1})$), and only one step update will give a estimate that at most $n^{-1}$ away from $\lambda(\tilde{\theta})$.

 \subsection{ Sampling from ``model-averaged'' Bayesian PETEL with sparse prior}
 A  Metropolis–Hastings procedure can be used to sample from the ``model-averaged'' posterior under model uncertainty~\citep{Dellaportas2002}. Given the current value of a proposal $(S,\theta_S)$, a proposal $(S',\theta'_{S'})$ is generated from some proposal distribution $p_{prop}(\cdot|(S,\theta_S))$,  the proposal is accepted as the next observation of the chain with the conventional Metropolis–Hastings acceptance probability 
\begin{equation*}
    \alpha=\frac{q(|S'|)\binom{d}{|S'|}^{-1}\pi_{S'}(\theta'_{S'})\exp(-\alpha_{n,d} \mathcal{R}_n(S',\theta'_{S'}))\prod_{i=1}^n p_i(\theta'_{S'};\, S')p_{prop}((S,\theta_S)|(S',\theta'_{S'}))}{q(|S|)\binom{d}{|S|}^{-1}\pi_{S}(\theta_{S})\exp(-\alpha_{n,d} \mathcal{R}_n(S,\theta_{S}))\prod_{i=1}^n p_i(\theta_{S};\, S)p_{prop}((S',\theta'_{S'})|(S,\theta_S))}.
\end{equation*}
  In practice, the proposal is constructed  as a proposal for model $S$, followed by a proposal for model parameters $\theta_S$, i.e., $p_{prop}((S',\theta'_{S'})|(S,\theta_S))=p_{prop}(S'|(S,\theta_S))p_{prop}(\theta'_{S'}|S',S,\theta_S)$~\citep{Dellaportas2002}.  The independence sampler~\citep{tierney1994} is a special case of this approach and is straightforward to implement.   By independence sampler, we mean the Markov chains with proposal that is not allowed to depend on the previous states,  i.e., $p_{prop}((S',\theta'_{S'})|(S,\theta_S))=p_{prop}(S')p_{prop}(\theta'_{S'}|S')$. The independence sampler is closely related to the corresponding important sampling process and works best if the proposal $p_{prop}$ is a reasonable approximation to the target posterior distribution~\citep{Dellaportas2002}. Therefore, we should choose $p_{ prop}(S)$ such it will only give mass to models that correspond to small constrained minimal empirical risks (i.e., $\mathcal{R}_n(\hat{\theta}_S,0)$  where $\hat{\theta}_S$ is the constrained empirical risk minimizer on model $S$) while in the meantime do not have large complexities. One possible choice for the proposal of model $S$ could be $p_{prop}(S'|(S,\theta_S))=p_{prop}(S') \propto   \exp(-\alpha_{n,d} \mathcal{R}_n(\hat{\theta}_{S'},0)-\beta_{n,d}|S'|- \log\binom{d}{|S'|})$.  To alleviate the curse of dimensionality, one can first use (stochastic) local search algorithms~\citep{bottolo2010} to find the model $\tilde{S}$ that maximizes $\exp(-\alpha_{n,d} \mathcal{R}_n(\hat{\theta}_{S},0)-\beta_{n,d}|S|- \log\binom{d}{|S|})$ and limit the model space to models that have bounded Hamming distances with $\tilde{S}$. Moreover,   Based on Theorem 4, the target posterior distribution could be approximated by a Gaussian distribution on the true model with mean $\hat{\theta}_{S^*}$ and covariance matrix $\frac{1}{n} (\m H_{\theta^*_{S^*}}^{S^*})^{-1} \Delta_{\theta^*_{S^*}}^{S^*}(\m H_{\theta^*_{S^*}}^{S^*})^{-1} $, so when the loss function is twice-differentiable w.r.t to $\theta$,  we could use  the empirical counterpart $(\hat{H}_{\theta_{S^*}}^{S^*}, \hat{\Delta}_{\theta_{S^*}}^{S^*})$ of $(\m H_{\theta_{S^*}}^{S^*},\Delta_{\theta_{S^*}}^{S^*})$ evaluated at $\hat{\theta}_{S^*}$ in place of $\m H_{\theta^*_{S^*}}^{S^*}$ and $\Delta_{\theta^*_{S^*}}^{S^*}$ . Thus the proposal distribution of $\theta_S$ given $S$ could be chosen as $N(\hat{\theta}_S, n^{-1}(\hat{H}_{\hat{\theta}_S}^S)^{-1}\hat{\Delta}_{\hat{\theta}_S}^S(\hat{H}_{\hat{\theta}_S}^S)^{-1})$.   For the non-smooth loss function, the strategy for estimating the covariance matrix of the  Gaussian limiting distribution of  Bayesian PETEL posterior with smooth loss function may not apply, as the empirical risk function in general dos not admit a Hessian matrix.  While if we can find a twice differentiable function $\ell_{\epsilon}$ such that $\lim _{\epsilon \rightarrow 0} \lim _{n \rightarrow \infty}\frac{1}{n}\sum_{i=1}^n {\rm Hess}_{\theta}\ell_{\epsilon}(X_i,\theta)={\rm Hess}_{\theta}\mathbb{E} \ell(\theta)$, then we can estimate the Hessian of the risk function by the Hessian of  $\frac{1}{n}\sum_{i=1}^n \ell_{\epsilon}(X_i,\theta)$ where $\epsilon$ can decrease with $n$ at suitable rate.  For example, for the hinge loss $\ell((\tilde{X},Y),\theta)=\max(0,1-Y \theta^T \tilde{X})$, by  equation (1.7) of~\cite{8637389}, we can choose $\ell_{\epsilon}((\tilde{X},Y),\theta)=\frac{1}{2}(u+\sqrt{\epsilon^2+u^2})$ where $u=1-Y\theta^T\tilde{X}$. Similarly for the loss function in quantile regression $\ell((\tilde{X},Y),\theta)=\tau(Y-\tilde{X}^T\theta)+\max(0,\tilde{X}^T\theta-Y)$, we can choose  $\ell_{\epsilon}((\tilde{X},Y),\theta)=-\tau u+\frac{1}{2}(u+\sqrt{\epsilon^2+u^2})$ where $u=\tilde{X}^T\theta-Y$.

\subsection{Additional Plots}
\begin{figure}[H]
\centering  
\subfigure[SVM under hinge loss (SVMH).]{
  \includegraphics[trim={0 0 0 2cm}, clip, width=0.45\textwidth,height=0.45\textwidth]{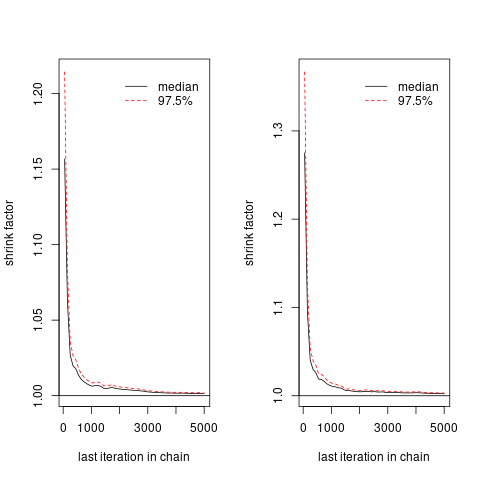} }
\subfigure[SVM under smoothed hinge loss (SVMSH).]{
 \includegraphics[trim={0 0 0 2cm}, clip, width=0.45\textwidth,,height=0.45\textwidth]{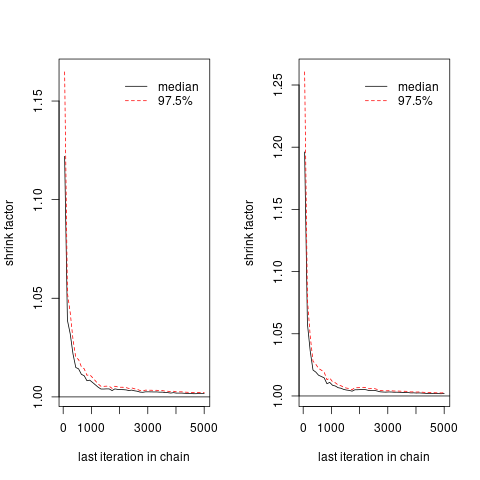}}
  \subfigure[High dimensional quantile regression.]{
 \includegraphics[trim={0 0 0 2cm}, clip, width=0.45\textwidth,,height=0.45\textwidth]{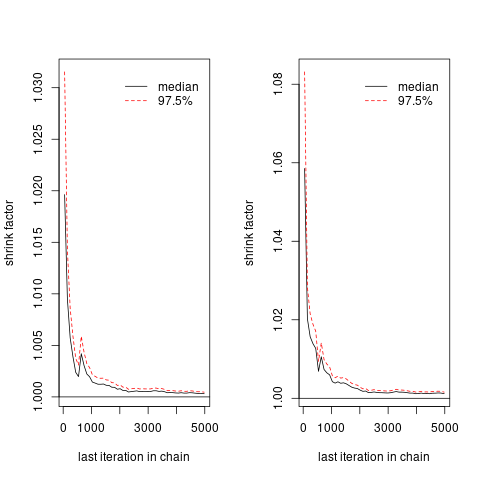}}
 \caption{Gelman-Rubin diagnostic plots for each dimension of parameters (left:  the first dimension; right: the second dimension) after $5000$ iterations }
\label{GRplot}
\end{figure}

 \paragraph{Diagnostic Plots on MCMC Convergence}
 We provide the  Gelman–Rubin plots  mentioned in Section 5.2.  The Gelman–Rubin diagnostic evaluates MCMC convergence by analyzing the difference between multiple Markov chains. The convergence is assessed by checking whether the $50\%$ and $97.5\%$ quantiles of the sampling distribution of the Markov chain for the shrink factor (the estimated potential scale reduction) is close to $1$.
Plots showing the evolution of Gelman and Rubin’s shrink factor as the number of iterations increases for the SVMH, SVMSH and high dimensional quantile regression described in Section 5.2 are presented  in Figure~\ref{GRplot}.

\begin{figure}[H]
\centering  
\subfigure[Marginal density plots of bootstrapping empirical risk minimizers solved by gradient descent algorithm.]{
  \includegraphics[width=0.6\textwidth,height=0.4\textwidth]{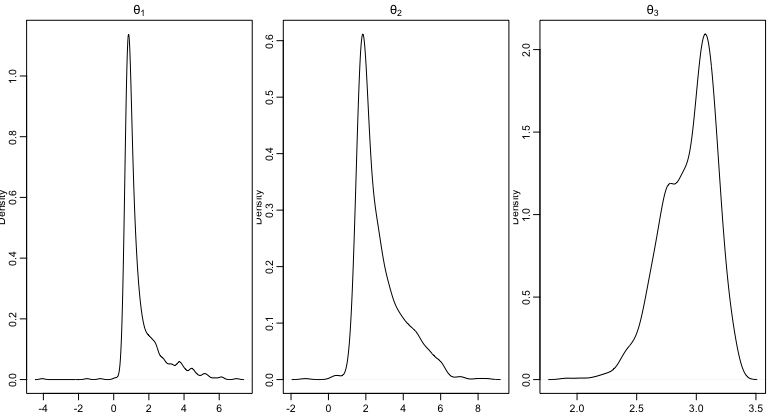} }
\subfigure[Marginal density plots of posterior samples from Bayesian ETEL using  RMW algorithm.]{
 \includegraphics[width=0.6\textwidth,,height=0.4\textwidth]{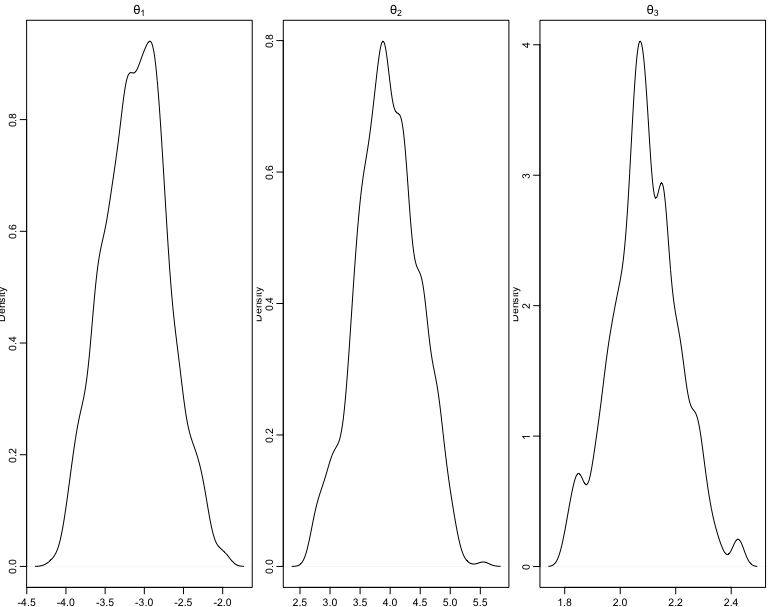}}
  \label{densityplot1}
   \subfigure[Marginal density plots of posterior samples from Bayesian PETEL using  RMW algorithm with $\alpha_n=2\sqrt{n}$.]{
 \includegraphics[width=0.6\textwidth,,height=0.4\textwidth]{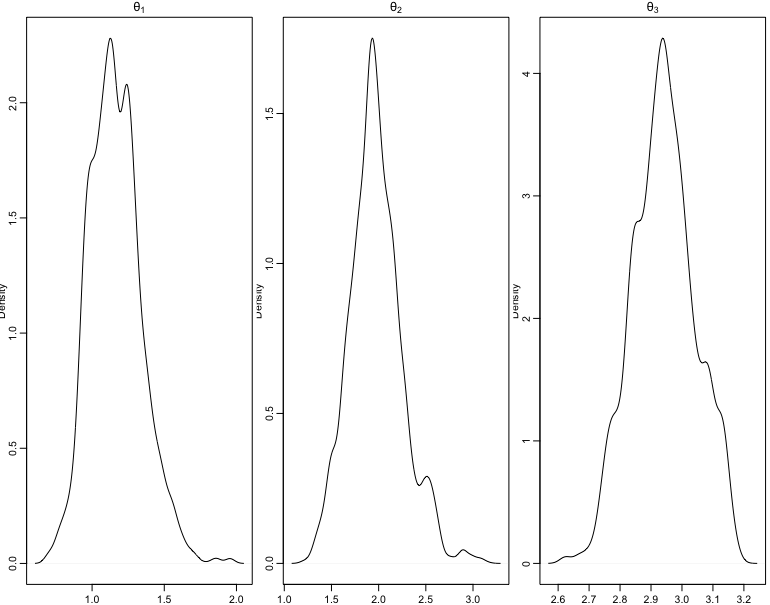}}
 \caption{Marginal density plots for each dimension of parameters (Bootstrap, Bayesian ETEL and Bayesian PETEL). }
\label{densityplot2}
\end{figure}

\paragraph{Plots of Robust regression for learning sigmoid unit}
We provide in Figure~\ref{densityplot2} the marginal density plots of each dimension of bootstrapping empirical risk minimizers  and posterior samples from Bayesian ETEL and Bayesian PETEL for a single run. We can see that for the  Bootstrap, the marginal density plots for the first and second dimensions of the bootstrapping empirical risk minimizers solved by gradient descent algorithm are right heavy-tailed; for the Bayesian ETEL, when the initial state  is near $(-3,4,2)$, then the samples generated in the Markov chain may all included in a small neighborhood of  $(-3,4,2)$; while for the Bayesian PETEL, when we choose $\alpha_n=n$ and the initial state to be  $(-3,4,2)$, the Markov chain  converges to a neighborhood of $\theta^*=(1,2,3)$ in less than 200 number of iterations. Then when we use the posterior mean of the previous Markov chain to be the new initial state and change $\alpha_n$ to $2\sqrt{n}$,  the samples generated in the new Markov chain are all around $\theta^*$ and the marginal density plots are all Gaussian-like.


\section{Applications of our Theoretical Results}
In this section, we will apply our theoretical results to two representative examples: quantile regression and soft-margin support vector machine.
\subsection{Example: Quantile Regression}
In quantile regression, for fixed $\tau \in (0,1)$, the $\tau^{th}$ quantile of the response $Y\in \mathbb{R}$ given the covariates $\tilde{X}\in \mathbb{R}^d$ is modelled as  
\begin{equation*}
Q_{\tau}(Y|\tilde{X})=\tilde{X}^T\theta^*.
\end{equation*}
The main difficulty in putting the Bayesian framework to work for quantile regression is  that no parametric likelihood is given in the model on  $Q_{\tau}(Y|\tilde{X})$, which is essential to the validity of Bayesian inference. Several authors have attempted to use a misspecified asymmetric Laplace likelihood  as a working likelihood in the Bayesian quantile regression framework~\citep{YU2001437, 10.1214/13-BA817},  which corresponds to a Gibbs posterior using the empirical risk function $\mathcal{R}_n(\theta)=\frac{1}{n}\sum_{i=1}^n (Y_i-\tilde{X}_i^T\theta)(\tau-\textbf{1}(Y_i<\tilde{X}_i^T\theta))$ given data $\{X_i=(\tilde{X}_i, Y_i)\}_{i=1}^n$~\citep{Syring_2018}.  However,  the inference derived from Gibbs posterior may be invalid as the generalized information equality~\citep{CHERNOZHUKOV2003293} may not hold.  Our method overcomes this issue by taking log ETEL into the Gibbs posterior framework, so the validity of inference is not affected by whether the generalized information equality holds or not. We consider $g(X,\theta)=(\textbf{1}(Y<\tilde{X}^T\theta)-\tau)\tilde{X}$ with  moment conditions $\mathbb{E} g(X,\theta)=0$  and  the loss function  $\ell(X,\theta)=(Y-\tilde{X}^T\theta)(\tau-\textbf{1}(Y<\tilde{X}^T\theta))$ in the Bayesian PETEL posterior distribution,  where $X=(\tilde{X}, Y)$. In this case, Assumption B.1 and B.2 could be simplified to the following Assumption B.q. 

\vspace{0.5em}
\noindent
\textbf{Assumption B.q:}  (1) The support of data $\tilde{X}$ and $Y$, denoted by $\tilde{\mathcal{X}}$ and $\mathcal{Y}$ respectively, are bounded. (2) The conditional density of $Y$ given $\tilde{X}$: $p(t|\tilde{X})$ is bounded and has bounded derivatives with respect to $t$ over $t\in \mathbb{R}$ and $\tilde{X}\in \tilde{\mathcal{X}}$.  

 \begin{corollary}\label{coquantile}
Consider loss function $\ell(X,\theta)$ and $g(X,\theta)$ defined in the above Quantile regression example, under Assumption B.q, Assumption A.2' and Assumption A.3, there exists some constants $(c,c_1,c_2)$ independent of $n$, such that when $c_1\log n\leq \alpha_n\leq c_2\sqrt{n\log n}$, it holds with probability at least $1-\frac{1}{n}$ that,
    \begin{equation*}
    TV\big(\pi_{\rm PE}(\cdot\,|\, X^n), \, N(\theta^*-n^{-1}\m H_{\theta^*}^{-1}\sum_{i=1}^n g(X_i,\theta^*), n^{-1} V_{\theta^*})\big) \leq  c\frac{(\log n)^{\frac{5}{4}}}{n^{\frac{1}{4}}},
\end{equation*}
where $\pi_{\rm PE}(\cdot\,|\, X^n)$ is the Bayesian PETEL posterior distribution and $V_{\theta^*}=\m H^{-1}_{\theta^*}\Delta_{\theta^*}\m H^{-1}_{\theta^*}$.
\end{corollary}
\noindent
The convergence rate in Corollary~\ref{coquantile} is slower than the case of smooth loss functions,  for the reason that $g(X,\theta)$ involves indicators and Assumption B.2 holds with $\beta=\frac{1}{2}$. Asymptotically, Corollary~\ref{coquantile} justifies the validity of the inference derived from the Bayesian PETEL approach in a frequentist sense, which is a property that is not shared by all working likelihoods. Indeed, when the $\tau$th quantile of $Y$ given $\tilde{X}$ is equal to $\tilde{X}^T\theta^*$, by simple computation, we can get $\m H_{\theta^*}=\mathbb{E} (p(\tilde{X}^T \theta^*|\tilde{X}) \tilde{X}\tilde{X}^T)$ and $\Delta_{\theta^*}=\mathbb{E} ((\tau-\textbf{1}(Y<\tilde{X}^T\theta^*))^2 \tilde{X}\tilde{X}^T)=(\tau-\tau^2)\mathbb{E} \tilde{X}\tilde{X}^T$.  So the generalized information equality holds when $p(\tilde{X}^T \theta^*|\tilde{X})$ is constant for all $\tilde{X}\in \tilde{\mathcal{X}}$ (e.g., homoscedastic error models), while there is no guarantee of the validity of the inference derived from  misspecified asymmetric Laplace likelihood or calibrated Gibbs posterior when the homoscedasticity assumption is invalid. Conversely, our method  is insusceptible to  model misspecification biases.

 \subsection{Example: Soft-Margin Support Vector Machine}
 The soft-margin SVM~\citep{duda2012pattern}  minimizes $\mathcal{R}_n(\theta)=\frac{1}{2}\lambda \|\theta\|_2^2+\frac{1}{n}\sum_{i=1}^n \max(0, 1-Y_i \theta^T \tilde{X}_i)$  over $\theta \in \mathbb{R}^d$ given data $\{X_i=(\tilde{X}_i, Y_i)\}_{i=1}^n$  and $Y_i=\pm 1$.  The value of $\lambda>0$ controls the $\ell_2$ norm of $\theta$ and the function $\max(0,1-y\theta^T\tilde{x})$  is called the hinge-loss function. ~\cite{10.1214/11-BA601, Syring_2018} proposed to taking a pseudo-likelihood $\exp(-\alpha n\mathcal{R}_n(\theta))$ with a learning rate $\alpha$ into a Bayesian framework, while there is no reason that posterior credible regions derived from it will be calibrated even though the learning rate is tuned to be the optimal, as the generalized information equality is generally not guaranteed. In our method,  we consider $g(X,\theta)=\lambda\theta-Y\textbf{1}(Y\theta^T\tilde{X}\leq 1)\tilde{X}$ with moment conditions $\mathbb{E} g(X,\theta)=0$ and  the loss function $\ell(X,\theta)=\frac{1}{2}\lambda\|\theta\|_2^2+\max(0,1-Y\theta^T\tilde{X})$ in the Bayesian PETEL posterior distribution, where $X=(\tilde{X}, Y)$.  In this case,  Assumption B.1 and B.2 could be simplified to the following Assumption B.s. 
 
 \vspace{0.5em}
\noindent
 \textbf{Assumption B.s:} (1) The support of covariant $\tilde{X}$, denoted by $\tilde{\mathcal{X}}$, is bounded. (2) There exist some positive constants $(c_1,c_2)$ such that the parameter space $\Theta\subseteq \{\theta\in \mathbb{R}^d, c_1\leq |\theta_i| \leq c_2 \,(1\leq i\leq d)\}$. (3) Let $p^j_{1}(\tilde{X}_j)$ and $p^j_{-1}(\tilde{X}_j)$ denotes the conditional density of $\tilde{X}_j$ given $\tilde{X}_{-j}$ and $Y=\pm 1$ respectively, where $\tilde{X}_j$ denotes the $j$th dimension of $\tilde{X}$ and $\tilde{X}_{-j}$ denotes the element of $\tilde{X}$ except for $\tilde{X}_j$. Let $ \tilde{\mathcal{X}}_{-j}$ denote the support of $\tilde{X}_{-j}$,  $\{p^j_1 (\tilde{X}_j)\}_{j=1}^d$ and $\{p^j_{-1} (\tilde{X}_j)\}_{j=1}^d$ are bounded and  have bounded first order derivatives with respect to $\tilde{X}_j$ over $\tilde{X}_j\in \mathbb{R}$ and $\tilde{X}_{-j} \in \tilde{\mathcal{X}}_{-j}$.
 
 \begin{corollary}\label{cosvm}
Under Assumption B.s, Assumption A.2' and Assumption A.3, there exists some constants $(c,c_1,c_2)$ independent of $n$, such that when $c_1\log n\leq \alpha_n\leq c_2\sqrt{n\log n}$, it holds with probability at least $1-\frac{1}{n}$ that,
    \begin{equation*}
    TV\big(\pi_{\rm PE}(\cdot\,|\, X^n), \, N(\theta^*-n^{-1}\m H_{\theta^*}^{-1}\sum_{i=1}^n g(X_i,\theta^*), n^{-1} V_{\theta^*})\big) \leq  c\frac{(\log n)^{\frac{5}{4}}}{n^{\frac{1}{4}}},
\end{equation*}
where $\pi_{\rm PE}(\cdot\,|\, X^n)$ is the Bayesian PETEL posterior distribution and $V_{\theta^*}=\m H^{-1}_{\theta^*}\Delta_{\theta^*}\m H^{-1}_{\theta^*}$.
\end{corollary}
\noindent
Corollary~\ref{cosvm} gives theoretical guarantee to the inference from our method for soft-margin SVM and in general, the calibrated Gibbs posterior would
 not work for this example. Indeed,  it can be shown that $\Delta_{\theta^*}=\mathbb{E} ((\lambda\theta^*-Y\textbf{1}_{Y{\theta^*}^TX\leq 1}\tilde{X})(\lambda\theta^*-Y\textbf{1}_{Y{\theta^*}^TX\leq 1}\tilde{X})^T)$, while the diagonal elements of $\m H_{\theta^*}$  are  $\lambda+\mathbb{E} ((1-Y\sum_{k\neq i}^d \theta_k \tilde{X}_k)^2p_{Y}^i((Y-\sum_{k\neq i}^d \theta_k \tilde{X}_k)/\theta_i)/|\theta_i|^3)$ with $1\leq i\leq d$ respectively. So generally,  adjusting the learning rate of Gibbs posterior couldn't exactly correct for the covariance matrix mismatching. 

 \section{Proof of Main results}
\subsection{Proof of Theorem 1}
 Let $\nabla_{\theta} \ell(X,\theta)=g(x,\theta)$ and $L(X^n;\theta)=\prod_{i=1}^n p_i(\theta)$. We begin the proof of Theorem 1 with the following lemmas.

 \begin{lemma}\label{l1}
Under Assumption A.1 and A.2 , for any $\tilde{\theta} \in \Theta$ such that $\nabla_{\theta} R(\tilde{\theta})=0$, if the prior $\pi({\theta})$ has support $B_r(\tilde{\theta})$ and there exist some positive constants $(c,L)$ such that $\pi(\tilde{\theta})\geq c$ and for any $\theta \in B_r(\tilde{\theta})$, it holds that $|\pi(\theta)-\pi(\tilde{\theta})|\leq L \|\theta-\tilde{\theta}\|_2$, $\m H_\theta^T \m H_\theta\succcurlyeq c I_d$ and $\m H_{\theta}^T \m H_{\tilde{\theta}}^{-1} \succcurlyeq c I_d$. Then there exist some constants $(c_1, c_2,c_3)$ independent of n, so that it holds with probability at least $1-\frac{c_3}{n^2}$ that,
\begin{enumerate}
    \item   $\nabla \mathcal{R}_n(\theta)=0$ has unique solution $\hat{\theta}$ on $B_r(\tilde{\theta})$ and  $\|\hat{\theta}-\tilde{\theta}\|\leq c_1\sqrt{\frac{\log n}{n}}$.
    \item
 $ \mathlarger \int {\left|\pi(\hat{\theta}+\frac{h}{\sqrt{n}})\exp\left(\log \frac{L(X^n;\,\hat{\theta}+h/\sqrt{n})}{L(X^n;\,\hat{\theta})}\right)-\pi(\hat{\theta})\exp\left(-\frac{h^TV^{-1}_{\tilde{\theta}}h}{2}\right)\right|}dh\leq c_2\sqrt{ \frac{\log n}{n}}$.
 \end{enumerate}
\end{lemma}
 
\begin{lemma}\label{l1.2}
Under Assumption A.1, A.2 and A.3, let $A=\Sigma_{k=1}^K B_r(\tilde{\theta}_k)$, if there exists a positive constant $c$ such that for any $\theta \in A^c$, $\|\nabla_{\theta} \mathcal{R}(\theta)\|_2\geq c$, then there exist constants $(c_1,c_2)$ so that it holds with probability at least $1-\frac{c_2}{n^2}$ that $\Pi_{\rm E}(\theta \in A^c|\, X^n) \leq \exp(-c_1 n^{\frac{1}{3}})$.
 \end{lemma}
 
 By Assumption A.1 and A.2, there exists a small enough  positive constant $r$ such that 
 for any $1\leq k \leq K$ and  $\theta \in B_r(\tilde{\theta}_k)$, it holds that $\m H_\theta^T \m H_\theta\succcurlyeq \frac{b}{2} I_d$ and $\m H_{\theta}^T \m H_{\tilde{\theta}_k}^{-1} \succcurlyeq \frac{1}{2} I_d$. Also, for any $1\leq k,k'\leq K$ and $k\neq k'$, $B_r(\tilde{\theta}_k)\cap B_r(\tilde{\theta}_{k'})=\emptyset$. Moreover, by the assumption that the equation $\nabla \mathcal{R}(\theta)=0$ has exact $K$ number of isolated solutions, there exists a positive constant $c$ such that for any $\theta \in \{ \Sigma_{k=1}^K B_r(\tilde{\theta}_k)\}^c$ , $\|\nabla_{\theta} \mathcal{R}(\theta)\|_2\geq c$.  Denote the posterior distribution of $\theta$ constrained on $B_r(\tilde{\theta}_k)$ by $\pi_k(\theta|X^n)$, then 
 \begin{equation*}
     \pi_k(\theta|X^n)=\frac{\pi_k(\theta)L(X^n;\,\theta)}{\int_{B_r(\tilde{\theta}_k)}\pi(\theta)L(X^n;\,\theta)d\theta},
 \end{equation*}
 where $\pi_k(\theta)=\pi(\theta)\textbf{1}_{B_r(\tilde{\theta}_k)}$. Let $\hat{\theta}_k$ be the solution of $\nabla_{\theta} \mathcal{R}_n(\theta)=0$ on  $B_r(\tilde{\theta})$,  then by Lemma~\ref{l1}, it holds with probability at least $1-\frac{c_3K}{n^2}$ that for any $1\leq k\leq K$, 
 \begin{equation*}
\left|\int \left( \pi_k(\hat{\theta}_k+\frac{h}{\sqrt{n}})\exp\left(\log \frac{L(X^n;\,\hat{\theta}_k+h/\sqrt{n})}{L(X^n;\,\hat{\theta}_k)}\right)-\pi(\hat{\theta}_k)\exp\left(-\frac{h^TV^{-1}_{\tilde{\theta}_k}h}{2}\right)\right)dh\right|\lesssim \sqrt{ \frac{\log n}{n}}.
    \end{equation*}
    \noindent
So combined with Lemma~\ref{l1} and the shift and scale invariance of the total variation distance, we have 
    \begin{equation}\label{f14}
    d_{\rm TV}(\pi_k(\theta|X^n), N(\hat{\theta}_k, \frac{1}{n} V_{\tilde{\theta}_k})) \leq  c_2 \sqrt{ \frac{\log n}{n}}.
    \end{equation}
    \noindent
 We then compute $ \Pi_{\rm E}(\theta \in B_r(\tilde{\theta}_k)|X^n)$.
 \begin{equation*}
     \begin{aligned}
           \Pi_{\rm E}(\theta \in B_r(\tilde{\theta}_k)|X^n)=\frac{n^{\frac{d}{2}}\int_{\theta \in B_r(\tilde{\theta}_k)} \pi(\theta) \exp\left(\log\frac{L(X^n;\theta)}{\left(\frac{1}{n}\right)^n}\right)d\theta}{n^{\frac{d}{2}}\int_{\Sigma_{k=1}^K B_r(\tilde{\theta}_k)} \pi(\theta) \exp\left(\log\frac{L(X^n;\theta)}{\left(\frac{1}{n}\right)^n}\right)d\theta}\cdot  \Pi_{\rm E}(\theta \in \Sigma_{k=1}^K B_r(\tilde{\theta}_k)|\, X^n).
     \end{aligned}
     \end{equation*}
         \noindent
Then by Lemma~\ref{l1}, it holds  with probability at least $1-\frac{c_3K}{n^2}$ that for any  $1\leq k\leq K$, $\|\tilde{\theta}_k-\hat{\theta}_k\|_2\leq c_1 \sqrt{\frac{\log n}{n}}$. Then there exist some constant $(c_4, c_5)$ such that 
\begin{equation*}
     \begin{aligned}
    & \int_{\|h\|_2\leq c_4 \sqrt{n}} \pi_k(\hat{\theta}_k+\frac{h}{\sqrt{n}})\exp\left(\log \frac{L(X^n;\,\hat{\theta}_k+h/\sqrt{n})}{L(X^n;\,\hat{\theta}_k)}\right) dh\\
     &\leq (\sqrt{n})^d  \int_{\theta \in B_r(\tilde{\theta}_k)} \pi(\theta) \exp\left(\log\frac{L(X^n;\theta)}{\left(\frac{1}{n}\right)^n}\right)d\theta
    \\
    &\leq \int_{\|h\|_2\leq c_5 \sqrt{n}} \pi_k(\hat{\theta}_k+\frac{h}{\sqrt{n}})\exp\left(\log \frac{L(X^n;\,\hat{\theta}_k+h/\sqrt{n})}{L(X^n;\,\hat{\theta}_k)}\right) dh.
\end{aligned}
     \end{equation*}
 Then, by Lemma~\ref{l1} and Lemma~\ref{l1.2}, it holds with probability at least $1-\frac{c_3 K}{n^2}$ that for any $1\leq k\leq K$, 
 \begin{equation}\label{f1}
 \begin{aligned}
       &\left|(\sqrt{n})^d  \int_{\theta \in B_r(\tilde{\theta}_k)} \pi(\theta) \exp\left(\log\frac{L(X^n;\theta)}{\left(\frac{1}{n}\right)^n}\right)d\theta-\pi(\tilde{\theta}_k)(2\pi)^{\frac{d}{2}}|V_{\tilde{\theta}_k}|^{\frac{1}{2}}\right| \lesssim \sqrt{\frac{\log n}{n}},\\
         &\left|(\sqrt{n})^d  \int_{\theta \in \sum_{k=1}^K B_r(\tilde{\theta}_k)} \pi(\theta) \exp\left(\log\frac{L(X^n;\theta)}{\left(\frac{1}{n}\right)^n}\right)d\theta-\sum_{k=1}^K \pi(\tilde{\theta}_k)(2\pi)^{\frac{d}{2}}|V_{\tilde{\theta}_k}|^{\frac{1}{2}}\right| \lesssim  \sqrt{\frac{\log n}{n}},\\
        &\left| \Pi_{\rm E}(\theta \in B_r(\tilde{\theta}_k)|X^n)-  \frac{\pi(\tilde{\theta}_k)|V_{\tilde{\theta}_k}|^{\frac{1}{2}}}{\sum_{k=1}^K \pi(\tilde{\theta}_k)|V_{\tilde{\theta}_k}|^{\frac{1}{2}}}\right|\lesssim  \sqrt{\frac{\log n}{n}}.\\
    \end{aligned}
  \end{equation}
 Let $A= \sum_{k=1}^K B_r(\tilde{\theta}_k)$, for any measurable set $A'\subseteq \mathbb{R}^d$, $A'$ can written as
  \begin{equation*}
      A'=A'\cap A +A'\cap A^c.\\
  \end{equation*}
 So,
 \begin{equation*}
      \Pi_{\rm E}(\theta\in A'|X^n) = \Pi_{\rm E}(\theta\in A'\cap A^c|X^n)+\sum_{k=1}^K   \Pi_{\rm E}(\theta \in A' \cap B_r(\tilde{\theta}_k)|X^n)
 \end{equation*}
 
 \begin{equation*}
     0\leq  \Pi_{\rm E}(\theta\in A'\cap A^c|X^n)\leq  \Pi_{\rm E}(\theta \in A^c|X^n)\leq \exp(-c_1 n^{\frac{1}{3}}).
 \end{equation*}
          \noindent
 Then by equation~\eqref{f14} and equation~(\ref{f1}), there exist positive constants $(c_0,c_1,c_2)$ such that it holds with probability at least $ 1-\frac{c_0}{n^2}$ that for any $1\leq k\leq K$ and $A' \subseteq \mathbb{R}^d$, 
 
\begin{equation*}
    \begin{aligned}
        &\left| \Pi_{\rm E}(\theta\in  A' \cap B_r(\tilde{\theta}_k)|X^n)- \tilde{\pi}_k  \Pi_{N(\hat{\theta}_k, \frac{1}{n} V_{\tilde{\theta}_k})}(\theta\in  A' \cap B_r(\tilde{\theta}_k))\right|\lesssim \sqrt{\frac{\log n}{n}}\\
        &\left|\tilde{\pi}_k  \Pi_{N(\hat{\theta}_k, \frac{1}{n} V_{\tilde{\theta}_k})}(\theta\in  A' \cap B_r(\tilde{\theta}_k)) - \Pi_{\sum_{k=1}^K \tilde{\pi}_k N(\hat{\theta}_k, \frac{1}{n} V_{\tilde{\theta}_k})}(\theta\in  A' \cap B_r(\tilde{\theta}_k))\right| \lesssim \exp(-c_1 n)\\
         &\Pi_{\sum_{k=1}^K \tilde{\pi}_k N(\hat{\theta}_k, \frac{1}{n} V_{\tilde{\theta}_k})}(\theta\in  A^c)\leq \exp(-c_2 n)\\
       &\tilde{\pi}_k=\frac{\pi(\tilde{\theta}_k)|V_{\tilde{\theta}_k}|^{\frac{1}{2}}}{\sum_{k=1}^K \pi(\tilde{\theta}_k)|V_{\tilde{\theta}_k}|^{\frac{1}{2}}} 
    \end{aligned}
\end{equation*}
Take supreme over $A'$, we can get  with probability at least $1-\frac{1}{n}$,
\begin{equation*}
    TV\left(\pi_{\rm E}(\theta|X^n), \sum_{k=1}^K \frac{\pi(\tilde{\theta}_k)|V_{\tilde{\theta}_k}|^{\frac{1}{2}}}{\sum_{i=1}^K \pi(\tilde{\theta}_i)|V_{\tilde{\theta}_i}|^{\frac{1}{2}}}  N(\hat{\theta}_k, \frac{1}{n} V_{\tilde{\theta}_k})\right)\lesssim  \sqrt{\frac{\log n}{n}}.
\end{equation*}
Moreover, by Lemma~\ref{l1} and $0=\nabla \mathcal{R}_n(\tilde{\theta}_k)+\mathcal{H}^n_{\tilde{\theta}_k}(\hat{\theta}_k-\tilde{\theta}_k)+O(\|\tilde{\theta}_k-\hat{\theta}_k\|_2^3)$, we can get that $\|\sqrt{n}(\hat{\theta}_k-\tilde{\theta}_k)+\mathcal{H}_{\tilde{\theta}_k}^{-1} \frac{1}{\sqrt{n}}\sum_{i=1}^n g(X_i, \tilde{\theta}_k)\|_2=o_p(1)$, then the statement that $\sqrt{n}(\hat{\theta}_k-\tilde{\theta}_k)$ converges to $N(0,V_{\tilde{\theta}_k})$ in distribution is followed from standard Central limit theorem and Slutsky's  theorem.

 \subsubsection{Proof of Lemma~\ref{l1}}
Let $\mathcal{H}^n_{\theta}$ be the Hessian matrix of $\mathcal{R}_n(\theta)$.   By Assumption A.1, we can get  for any $\theta,\theta' \in \Theta$ and $1\leq j,k \leq d$,
\begin{equation*}
\begin{aligned}
 &\sqrt{\frac{1}{n}\sum_{i=1}^n \left(\frac{\partial \ell(X_i, \theta)}{\partial \theta_j}-\frac{\partial \ell(X_i, \theta')}{\partial \theta'_j}\right)^2}\lesssim \|\theta-\theta'\|_2\\
 & \sqrt{\frac{1}{n}\sum_{i=1}^n \left(\frac{\partial^2 \ell(X_i, \theta)}{\partial \theta_j\partial \theta_k}-\frac{\partial^2 \ell(X_i, \theta')}{\partial \theta'_j\partial\theta'_k}\right)^2}\lesssim \|\theta-\theta'\|_2.
 \end{aligned}
\end{equation*}
Since $\Theta$ is compact, w.l.o.g, we can assume $\Theta=B_1(\bold{0}_d)$.
Then by standard symmetrization (see for example,  8.3.24 of~\citep{vershynin_2018}) and Dudley's inequality (see for example, 8.1.3 of~\citep{vershynin_2018}), we can get for any $1\leq j,k\leq d$,
\begin{equation*}
    \begin{aligned}
  &\mathbb{E}\, \underset{\theta \in \Theta}{\sup} \left|\frac{1}{n}\sum_{i=1}^n\frac{\partial \ell(X_i, \theta)}{\partial \theta_j}
    -\mathbb{E} \frac{\partial \ell(X, \theta)}{\partial \theta_j}\right| 
    +\mathbb{E}\, \underset{\theta \in \Theta}{\sup}\left|\frac{1}{n}\sum_{i=1}^n \frac{\partial^2 \ell(X_i, \theta)}{\partial \theta_j\partial \theta_k}-\mathbb{E} \frac{\partial^2 \ell(X, \theta)}{\partial \theta_j\partial \theta_k}\right|\\
   &  \lesssim \frac{1}{\sqrt{n}}\int \sqrt{\log \mathcal{N}(\Theta,\|\cdot\|_2,\varepsilon)}d\varepsilon,
 \end{aligned}
\end{equation*}
where $\mathcal{N}(\Theta,\|\cdot\|_2,\varepsilon)$ denotes the $\varepsilon$-covering number of $\Theta$ with respect to $\ell_2$ norm, which is upper bounded by $\left(\frac{3}{\varepsilon}\right)^d$~\citep{vershynin_2018}. Then using Bernstein inequality~\citep{wainwright_2019}, there exists a constant $c_1$, such that it holds with probability at least $1-\frac{1}{n^2}$ that
\begin{equation}\label{f12}
\underset{\theta\in \Theta}{\sup}(\|\nabla_{\theta} \mathcal{R}(\theta)-\nabla_{\theta} \mathcal{R}_n(\theta)\|_2 +\| \m H_\theta-\mathcal{H}^n_{\theta}\|_{\rm F})\leq c_1 \sqrt{\frac{\log n}{n}}
\end{equation}
 Then we have for any $\theta \in B_r(\tilde{\theta})$,  $\mathcal{H}^n_{\theta} \mathcal{H}^n_\theta\succcurlyeq \frac{b}{4} I_d\, (b>0)$ and $\|\nabla \mathcal{R}_n(\tilde{\theta})\|\leq c_1 \sqrt{\frac{\log n}{n}}$.  Let $\theta^0=\tilde{\theta}$, and for $k=1,2,\cdots$, we recurring define $\theta^k=\theta^{k-1} -(\mathcal{H}^n_{\theta^{k-1}})^{-1}\nabla_{\theta} \mathcal{R}_n(\theta^{k-1})$. Then with probability at least $1-\frac{1}{n^2}$, for any $k\geq 1$, it holds that $\|\theta^k-\theta^{k-1}\|_2\lesssim (\frac{\log n}{n})^{2^{k-2}}$ and $\|\nabla_{\theta} \mathcal{R}_n(\theta^k)\|_2\lesssim  (\frac{\log n}{n})^{2^{k-1}}$, so we can define $\hat{\theta}=\underset{k\rightarrow +\infty}{\lim} \theta^{k}$ and we have $\nabla \mathcal{R}_n(\hat{\theta})=0$ and $\hat{\theta} \in B_r(\tilde{\theta})$.
We  now prove the uniqueness of the solution of $\nabla \mathcal{R}_n(\theta)=0$ on $B_r(\tilde{\theta})$ in the following lemma.
\begin{lemma}\label{l2}
Under Assumption A.1 and A.2 , for any $\tilde{\theta} \in \Theta$ such that $\nabla_{\theta} R(\tilde{\theta})=0$,  if there exist some positive constants $(r,c)$ such that for any $\theta \in B_r(\tilde{\theta})$, it holds that $\m H_\theta^T \m H_\theta\succcurlyeq c I_d$ and $\m H_{\theta}^T \m H_{\tilde{\theta}}^{-1} \succcurlyeq c I_d$. There exists a positive constant $c_0$, such that 
 \begin{enumerate}
     \item For any $\theta \in B_r(\tilde{\theta})$,  $\|\nabla_{\theta} \mathcal{R}(\theta)\|_2 \geq c_0 \|\theta-\tilde{\theta}\|_2$. 
      \item It holds with probability at least $1-\frac{1}{n^2}$ that  for any $\theta \in B_r(\tilde{\theta})$, $\|\nabla_{\theta} \mathcal{R}_n(\theta)\|_2 \geq c_0 \|\theta-\hat{\theta}\|_2$.
  \end{enumerate}
 \end{lemma}
 \noindent
So, by Lemma~\ref{l2}, we could get the conclusion of the first statement.\\
\quad\\
For the second statement, since
 \begin{equation*}
 \begin{aligned}
   \pi_{\rm E} (\sqrt{n}(\theta-\hat{\theta})|\,X^n) =\frac{\pi(\hat{\theta}+\frac{h}{\sqrt{n}})\exp(\log \frac{L(X^n;\,\hat{\theta}+h/\sqrt{n})}{L(X^n;\,\hat{\theta})})}{\int\pi(\hat{\theta}+\frac{h}{\sqrt{n}})\exp(\log \frac{L(X^n;\,\hat{\theta}+h/\sqrt{n})}{L(X^n;\,\hat{\theta})})dh},
    \end{aligned}
 \end{equation*}
 we then bound $\mathlarger{\int}\left|\pi(\hat{\theta}+\frac{h}{\sqrt{n}})\exp(\log \frac{L(X^n;\,\hat{\theta}+h/\sqrt{n})}{L(X^n;\,\hat{\theta})})-\pi(\hat{\theta})\exp(-\frac{h^TV_{\tilde{\theta}}^{-1}h}{2})\right|dh$.\\
 \quad\\
 Define the following set of $h$,
\begin{equation*}
 \begin{aligned}
 &A_1=\left\{\|h\|_2\leq \delta_1\sqrt{\log n}\right\},\\
 &A_2=\left\{\delta_1\sqrt{\log n}\leq \|h\|_2\leq \delta_2(\log n)^{1.5}\right\},\\
 &A_3=\left\{\|h\|_2\geq \delta_2 (\log n)^{1.5}\right\}.\\
   \end{aligned}
 \end{equation*}
 \\
 \textbf{Step 1}:  Consider $A_3$. let $\theta'=\hat{\theta}+\frac{h}{\sqrt{n}}$, where $\theta' \in B_r(\tilde{\theta})$, then by Lemma~\ref{l2}, with probability at least $1-\frac{1}{n^2}$, $\|\theta'-\tilde{\theta}\|\geq  \frac{\delta_2}{2} \frac{(\log n)^{1.5}}{\sqrt{n}}$.
 \begin{equation*}
 \begin{aligned}
&\log \frac{L(X^n;\,\hat{\theta}+h/\sqrt{n})}{L(X^n;\,\hat{\theta})}=\sum_{i=1}^n \log p_i(\theta')-n(\log \frac{1}{n})\\
&\sum_{i=1}^n p_i(\theta')g(X_i,\theta')=0.
 \end{aligned}
 \end{equation*}
So, 
\begin{equation*}
    \sum_{i=1}^n  \left(p_i(\theta')-\frac{1}{n}\right)g(X_i,\theta')=\nabla \mathcal{R}_n(\theta').
\end{equation*}
By Lemma~\ref{l2}, there exists a positive constant $c$ such that it holds with probability at least $1-\frac{1}{n^2}$ that, $\|\nabla_{\theta} \mathcal{R}(\theta')\|_2\geq  \frac{c\delta_2}{2} \frac{(\log n)^{1.5}}{\sqrt{n}}$ and  $\|\nabla_{\theta} \mathcal{R}_n(\theta')\|_2\geq  \frac{c\delta_2}{4} \frac{(\log n)^{1.5}}{\sqrt{n}}$. So, 
\begin{equation*}
 \begin{aligned}
&\sum_{i=1}^n \left(p_i(\theta')-\frac{1}{n}\right)^2 \sum_{i=1}^n \|g(X_i,\theta)\|_2^2 \geq  \frac{c^2\delta^2_2}{16} \frac{(\log n)^{3}}{n}\\
&\sum_{i=1}^n \left(p_i(\theta')-\frac{1}{n}\right)^2 \geq c_2\delta_2^2 \frac{(\log n)^{3}}{n^2}.
 \end{aligned}
 \end{equation*}
 Define $q(p_1, \cdots, p_{n-1})=\sum_{i=1}^{n-1} \log p_i +\log (1-\sum_{i=1}^{n-1} p_i)$. The Hessian matrix of function $q$ at point  $(p_1, \cdots, p_{n-1})$ is 
 \begin{equation*}
     \m H_q|_{(p_1, \cdots, p_{n-1})}= \text{Diag}(-\frac{1}{p_1^2},\cdots,-\frac{1}{p_{n-1}^2})-\frac{1}{(1-\sum_{i=1}^{n-1}p_i)^2}\textbf{1}_{(n-1)\times(n-1)},
 \end{equation*}
 where $\textbf{1}_{(n-1)\times(n-1)}$ denotes the $(n-1)\times(n-1)$ matrix with all entries  being $1$. Let $p=(p_1, \cdots, p_{n})$ and $p_{-n}=(p_1, \cdots, p_{n-1})$. If  $\|p\|_{\infty} \geq 2d \frac{\log n}{n}$, then 
 \begin{equation*}
 \sum_{i=1}^n \log p_i \leq \log \frac{2d\log n}{n} +(n-1) \log \frac{1-2d\frac{\log n}{n}}{n-1}.
 \end{equation*}
So,
 \begin{equation}\label{f2}
 \begin{aligned}
  -n\log n -\sum_{i=1}^n \log p_i  &\geq -\log(2d\log n) -(n-1) \log\left( (1-2d \frac{\log n}{n}) \frac{n}{n-1}\right)\\
  &\geq d \log n.
  \end{aligned}
 \end{equation}
 \quad\\
 If $\|p\|_{\infty} \leq 2d \frac{\log n}{n}$, then when $\delta_2$ is large enough, we have $\sum_{i=1}^{n-1} (p_i-\frac{1}{n})^2\geq \frac{8d^3(\log n)^3}{n^2}$, so by mean value theorem,  
 \begin{equation}\label{f3}
 \begin{aligned}
&q(\frac{1}{n}, \cdots ,\frac{1}{n})-q(p_{-n})\\
&=-\frac{1}{2}(p_{-n}-\frac{1}{n}\textbf{1}_{(n-1)})^T \m H_q|_{(cp_{-n}+(1-c)\frac{1}{n}\textbf{1}_{(n-1)})}(p_{-n}-\frac{1}{n}\textbf{1}_{(n-1)})\\
&\geq d\log n.
  \end{aligned}
 \end{equation}
 So,
 \begin{equation*}
 \begin{aligned}
  &\int_{A_3}\left|\pi(\hat{\theta}+\frac{h}{\sqrt{n}})\exp(\log \frac{L(X^n;\,\hat{\theta}+h/\sqrt{n})}{L(X^n;\,\hat{\theta})})-\pi(\hat{\theta})\exp(-\frac{h^TV_{\tilde{\theta}}^{-1}h}{2})\right|dh\\
  & \leq \exp(-d\log n)(\sqrt{n})^d+\int_{A_3} \pi(\hat{\theta})\exp(-\frac{h^TV_{\tilde{\theta}}^{-1}h}{2})dh\\
  &\leq \frac{1}{\sqrt{n}}.
 \end{aligned}
 \end{equation*}
  
  \quad\\
 \textbf{Step 2}: Consider $A_1$ and $A_2$. let $\theta=\hat{\theta}+\frac{h}{\sqrt{n}}$, we have with probability at least $1-\frac{1}{n^2}$, $\|\theta-\tilde{\theta}\|_2\leq \frac{2\delta_2(\log n)^{1.5}}{\sqrt{n}}$.
 \begin{lemma}\label{l3}
If (1) $\|g(x,\theta)\|_2$ is uniformly bounded over $x\in \mathcal{X}$ and $\theta\in \Theta$; (2) each element of $\Delta_{\theta}$ and $\mathbb{E} g(X,\theta)$ are uniformly Lipschitz over a neighborhood of $\tilde{\theta}$ and $\Delta_{\tilde{\theta}}\succcurlyeq a I_{d}$ with a positive constant $a$; (4) $\mathbb{E} g(X,\tilde{\theta})=0$. Then there exist some positive constants $(\delta_0, c_2, c_3)$, such that 
 \begin{enumerate}
     \item For any $\lambda \in \mathbb{S}^{d-1}$ and $\theta \in   B_{\delta_0} (\tilde{\theta})$, $\m P^*(\lambda^T g(X,\theta)\geq c_2)\geq c_3$, where recall that $\m P^*$ denotes the  underlying data distribution.
     \item If in addition each element of $g(X,\theta)$ is uniformly Lipschitz with respect to $\theta$ over  a neighborhood of $\tilde{\theta}$ and $X\in \mathcal{X}$, then it holds with probability at least  $1-\frac{1}{n^2}$ that for any $\lambda \in \mathbb{S}^{d-1}$ and $\theta \in   B_{\delta_0} (\tilde{\theta})$,  it satisfies that $\frac{1}{n}\sum_{i=1}^n \textbf{1}_{ \lambda^T g(X_i,\theta)\geq \frac{c_2}{2}}\geq \frac{c_3}{2}$.
  \end{enumerate}
 \end{lemma}
 \noindent
 Since $\lambda(\theta)=\underset{\lambda\in \mathbb{R}^d} {\arg\min} \frac{1}{n}\sum_{i=1}^n \exp(\lambda^T g(X_i,\theta))$. By Lemma~\ref{l3},  $\lambda(\theta)$ exists and 
 \begin{equation*}
 \frac{1}{n}\sum_{i=1}^n \exp(\lambda(\theta)^T g(X_i,\theta))g(X_i, \theta)=0\textbf{1}_{d}.
 \end{equation*}
 Let $\tilde{\lambda}(\theta)=\frac{\lambda(\theta)}{\|\lambda(\theta)\|_2}$, then
     \begin{equation*}
 \frac{1}{n}  \sum_{i=1}^n \exp\left(\|\lambda(\theta)\|_2\tilde{\lambda}(\theta)^T g(X_i,\theta)\right)\tilde{\lambda}(\theta)^T g(X_i, \theta)=0.
      \end{equation*}
  So, by Lemma~\ref{l3}, it holds with  probability larger than $1-\frac{1}{n^2}$ that for any $\theta \in\left\{\theta\,\big|\, \|\theta-\tilde{\theta}\|_2\leq \frac{2\delta_2(\log n)^{1.5}}{\sqrt{n}}\right\}$, 
  \begin{equation}\label{boundlambda}
  \begin{aligned}
   &\frac{1}{n}  \underset{i\in[n]\atop \tilde{\lambda}(\theta)^T g(X_i,\theta) \geq \frac{c_2}{2} }{\sum}  \exp\left(\|\lambda(\theta)\|_2\tilde{\lambda}(\theta)^T g(X_i,\theta)\right)\tilde{\lambda}(\theta)^T g(X_i, \theta)\\
   &\leq \frac{1}{n}  \underset{i\in[n]\atop \tilde{\lambda}(\theta)^T g(X_i,\theta) \geq 0 }{\sum}  \exp\left(\|\lambda(\theta)\|_2\tilde{\lambda}(\theta)^T g(X_i,\theta)\right)\tilde{\lambda}(\theta)^T g(X_i, \theta)\\
   &=-\frac{1}{n}  \underset{i\in[n]\atop \tilde{\lambda}(\theta)^T g(X_i,\theta) \leq   0}{\sum}  \exp\left(\|\lambda(\theta)\|_2\tilde{\lambda}(\theta)^T g(X_i,\theta)\right)\tilde{\lambda}(\theta)^T g(X_i, \theta)\\
   &\leq \sqrt{\frac{1}{n} \sum_{i=1}^n \|g(X_i, \theta)\|_2^2}\\
   &\leq 2\sqrt{tr(\Delta_{\tilde{\theta}})}.
  \end{aligned}
   \end{equation}
 So we can get 
 \begin{equation*}
  \begin{aligned}
 & 2\sqrt{tr(\Delta_{\tilde{\theta}})}\geq \frac{c_2c_3}{4}\exp(\frac{c_2}{2}\|\lambda(\theta)\|_2)\\
& \|\lambda(\theta)\|_2\leq \frac{2\log\frac{8\sqrt{tr(\Delta_{\tilde{\theta}})}}{c_2c_3}}{c_2}=\lambda_0
  \end{aligned}
   \end{equation*}
   \begin{lemma}\label{l4}
   Under Assumption A.1 and A.2, for any $\tilde{\theta} \in \Theta$ such that $\nabla_{\theta} R(\tilde{\theta})=0$, there exist  constants $(c_5,c_6)$, such that it holds with probability larger than $1-\frac{c_6}{n^2}$ that for any $\theta \in \left\{\theta\,\big|\, \|\theta-\tilde{\theta}\|_2\leq \frac{2\delta_2(\log n)^{1.5}}{\sqrt{n}}\right\}$, it holds that  $\underset{1\leq i\leq d}{\max}\|\frac{\partial\lambda(\theta)}{\partial  \theta_i}\|_2 \leq c_5$ and $\underset{1\leq i\leq d\atop 1\leq j\leq d}{\max}\|\frac{\partial^2\lambda(\theta)}{\partial  \theta_i\partial \theta_j}\|_2 \leq c_5$.
   \end{lemma}
   \quad\\
   Since $\log L(X^n;\theta)=-n\log n+\sum_{i=1}^n \log \frac{\exp(\lambda(\theta)^T g(X_i,\theta))}{\frac{1}{n}\sum_{i=1}^n \exp(\lambda(\theta)^T g(X_i,\theta))}$. Let $l(X,\theta)=\log \frac{\exp(\lambda(\theta)^T g(X,\theta))}{\frac{1}{n}\sum_{i=1}^n \exp(\lambda(\theta)^T g(X_i,\theta))}$, then 
   \begin{equation*}
       \log \left(\frac{L(X^n;\hat{\theta}+\frac{h}{\sqrt{n}})}{L(X^n;\hat{\theta})}\right)=\sum_{i=1}^n l(X_i, \hat{\theta}+\frac{h}{\sqrt{n}})-\sum_{i=1}^n l(X_i, \hat{\theta}).
   \end{equation*}
   \begin{equation*}
   \|h\|\leq \delta_2(\log n)^{1.5}.
   \end{equation*}
   \quad\\
   Since $\frac{1}{n}\sum_{i=1}^n g(X_i, \hat{\theta})=0$ and $\lambda(\hat{\theta})=0$, let $l^{(1)}(x,\theta)$ and $l^{(2)}(x,\theta)$ denote the gradient and Hessian matrix of $l(X,\theta)$ with respect to $\theta$,  we have $\sum_{i=1}^n l^{(1)} (X_i, \hat{\theta})=0$. 
  Let $\theta_t=\hat{\theta}+\frac{th}{\sqrt{n}}$ with some $t\in[0,1]$. Let $\tau_i(\theta)=\exp(\lambda(\theta)^T g(X_i, \theta))$ and $\tau_n (\theta)=\frac{1}{n} \sum_{i=1}^n \tau_i(\theta)$.  Then, we have 
        \begin{equation*}
       \begin{aligned}
     &\frac{1}{n}\sum_{i=1}^n l^{(2)} (X_i, \theta_t)\\
      &=\frac{1}{n}\sum_{i=1}^n \Big(1-\frac{\tau_i(\theta_t)}{\tau_n(\theta_t)}\Big)\Big(\sum_{j=1}^d \lambda_j(\theta_t) g^{(2)}_j(X_i, \theta_t)+\lambda^{(1)}(\theta_t)^T g^{(1)}(X_i, \theta_t)\\
        &+g^{(1)}(X_i, \theta_t)^T\lambda^{(1)}(\theta_t)+\sum_{j=1}^dg_j(X_i, \theta_t)\lambda_j^{(2)}(\theta_t)\Big)\\
        &-\frac{1}{n}\sum_{i=1}^n \frac{\tau_i(\theta_t)}{\tau_n(\theta_t)}\left(\lambda(\theta_t)^T g^{(1)} (X_i,\theta_t)+g(X_i, \theta_t)^T \lambda^{(1)} (\theta_t)\right)^T\left(\lambda(\theta_t)^T g^{(1)} (X_i,\theta_t)+g(X_i, \theta_t)^T \lambda^{(1)} (\theta_t)\right)\\
        &+\frac{1}{n}\sum_{i=1}^n \frac{\tau_i(\theta_t)}{(\tau_n(\theta_t))^2} 
       \left( \frac{1}{n}\sum_{j=1}^n \tau_j(\theta_t) \left(\lambda(\theta_t)^T g^{(1)} (x_j,\theta_t)+g(x_j, \theta_t)^T \lambda^{(1)} (\theta_t)\right)^T\right)\\
      &\cdot \left(\lambda(\theta_t)^T g^{(1)} (X_i,\theta_t)+g(X_i, \theta_t)^T \lambda^{(1)} (\theta_t)\right).\\
 \end{aligned}
   \end{equation*}
  By Lemma~\ref{l4} and the facts that $\lambda(\hat{\theta})=0$ and $\frac{1}{n}\sum_{i=1}^n g(X_i, \hat{\theta})=0$, we have for any $1\leq i \leq n$,
   \begin{equation*}
       \begin{aligned}
       & \|\lambda(\theta_t)\|_2 \lesssim \frac{\|h\|_2}{\sqrt{n}}\\
        &|\tau_i(\theta_t)-1| \lesssim \frac{\|h\|_2}{\sqrt{n}}\\
        &|\tau_n(\theta_t)-1|\lesssim \frac{\|h\|_2}{\sqrt{n}}\\
        &\|\frac{1}{n} \sum_{j=1}^n g(x_j, \theta_t) -0\|_2\lesssim \frac{\|h\|_2}{\sqrt{n}}.\\
       \end{aligned}
   \end{equation*}
   \quad\\
   So, we can get 
   \begin{equation*}
       \begin{aligned}
       &\frac{1}{n}\sum_{i=1}^n l^{(2)} (X_i, \theta_t)=O(\frac{\|h\|_2}{\sqrt{n}})-\frac{1}{n}\sum_{i=1}^n \frac{\tau_i(\theta_t)}{\tau_n(\theta_t)} \left(\lambda^{(1)} (\theta_t)^T g(X_i, \theta_t)g(X_i, \theta_t)^T \lambda^{(1)} (\theta_t)\right)\\
       \end{aligned}
   \end{equation*}
   Since $\lambda^{(1)}(\hat{\theta})=-\left(\frac{1}{n}\sum_{i=1}^n  g(X_i,\hat{\theta}) g(X_i, \hat{\theta})^T\right)^{-1}\left(\frac{1}{n} \sum_{i=1}^n g^{(1)} (X_i, \hat{\theta})\right)$, then by Bernstein inequality~\citep{wainwright_2019} and the first statement of Lemma~\ref{l1},  there exists a constant $c_0$ such that it holds with probability at least $ 1-\frac{c_0}{n^2}$ that,
   \begin{equation*}
       \|\lambda^{(1)}(\hat{\theta})-(-\Delta_{\tilde{\theta}}^{-1}\m H_{\tilde{\theta}})\|_{\rm F}\lesssim \sqrt{\frac{\log n}{n}}.
   \end{equation*}
   Then by Lemma~\ref{l4} and the mean value theorem,   there exists a constant $c_0$ such that it holds with probability at least $ 1-\frac{c_0}{n^2}$ that for any $t \in [0,1]$ and $\|h\|_2\leq \delta_2 (\log n)^{1.5}$,
   \begin{equation}\label{f13}
     | \sum_{i=1}^n l(X_i, \hat{\theta}+\frac{h}{\sqrt{n}})- \sum_{i=1}^n l(X_i, \hat{\theta})+\frac{1}{2}h^T\m H_{\tilde{\theta}}\Delta_{\tilde{\theta}}^{-1}\m H_{\tilde{\theta}}h| \lesssim \frac{\|h\|^3_2+\|h\|_2^2\sqrt{\log n}}{\sqrt{n}}.
   \end{equation}
    So for set $A_2$,
     \begin{equation*}
       \begin{aligned}
    &\int_{A_2}\left|\pi(\hat{\theta}+\frac{h}{\sqrt{n}})\exp\left(\log \frac{L(X^n;\,\hat{\theta}+h/\sqrt{n})}{L(X^n;\,\hat{\theta})}\right)-\pi(\hat{\theta})\exp\left(-\frac{h^TV^{-1}_{\tilde{\theta}}h}{2}\right)\right|dh\\
    &\leq \int_{A_2} \pi(\hat{\theta}+\frac{h}{\sqrt{n}}) \exp\left( \sum_{i=1}^n l(X_i, \hat{\theta}+\frac{h}{\sqrt{n}})-\sum_{i=1}^n l(X_i, \hat{\theta})\right) dh + \int_{A_2}  \pi(\hat{\theta})\exp\left(-\frac{h^TV^{-1}_{\tilde{\theta}}h}{2}\right) dh.
      \end{aligned}
   \end{equation*}
   
   \quad\\
   When $\delta_1$ is large enough, we have with probability at least $ 1-\frac{c_0}{n^2}$ that, 
   \begin{equation*}
       \begin{aligned}
       &\int_{A_2}  \pi(\hat{\theta})\exp\left(-\frac{h^TV^{-1}_{\tilde{\theta}}h}{2}\right) dh\leq \frac{1}{n},\\
       & \int_{A_2} \pi(\hat{\theta}+\frac{h}{\sqrt{n}}) \exp\left( \sum_{i=1}^n l(X_i, \hat{\theta}+\frac{h}{\sqrt{n}})-\sum_{i=1}^n l(X_i, \hat{\theta})\right) dh \\
        &\leq \int_{A_2} \pi(\hat{\theta}+\frac{h}{\sqrt{n}}) \exp\left(-\frac{h^TV^{-1}_{\tilde{\theta}}h}{2}+c \frac{(\log n)^{4.5}}{\sqrt{n}}\right)dh\\
        &\leq \frac{1}{n}
       \end{aligned}
   \end{equation*}
   \noindent
For set $A_1$, we have with probability at least $ 1-\frac{c_0}{n^2}$ that,
   \begin{equation*}
       \begin{aligned}
       &\int_{A_1}\left|\pi(\hat{\theta}+\frac{h}{\sqrt{n}})\exp\left(\log \frac{L(X^n;\,\hat{\theta}+h/\sqrt{n})}{L(X^n;\,\hat{\theta})}\right)-\pi(\hat{\theta})\exp\left(-\frac{h^TV^{-1}_{\tilde{\theta}}h}{2}\right)\right|dh\\
       &\leq \int_{A_1}\pi(\hat{\theta}+\frac{h}{\sqrt{n}})\left|\exp\left( \sum_{i=1}^n l(X_i, \hat{\theta}+\frac{h}{\sqrt{n}})-\sum_{i=1}^n l(X_i, \hat{\theta})\right)-\exp\left(-\frac{h^TV^{-1}_{\tilde{\theta}}h}{2}\right)\right|dh\\
       &+\int_{A_1} \left|\pi(\hat{\theta})-\pi(\hat{\theta}+\frac{h}{\sqrt{n}})\right| \exp\left(-\frac{h^TV^{-1}_{\tilde{\theta}}h}{2}\right)dh \\
       & \lesssim  \int_{A_1}\pi(\hat{\theta}+\frac{h}{\sqrt{n}})\exp\left(-\frac{h^TV^{-1}_{\tilde{\theta}}h}{2}\right)\frac{\|h\|^3_2+\|h\|_2^2\sqrt{\log n}}{\sqrt{n}} dh+\sqrt{\frac{\log n}{n}}\\
       &\lesssim \sqrt{\frac{\log n}{n}}.
       \end{aligned}
   \end{equation*}
\noindent
So, it holds with probability at least $ 1-\frac{c_0}{n^2}$ that 
\begin{equation*}
        \int \left|\pi(\hat{\theta}+\frac{h}{\sqrt{n}})\exp\left(\log \frac{L(X^n;\,\hat{\theta}+h/\sqrt{n})}{L(X^n;\,\hat{\theta})}\right)-\pi(\hat{\theta})\exp\left(-\frac{h^TV^{-1}_{\tilde{\theta}}h}{2}\right)\right|dh\lesssim \sqrt{\frac{\log n}{n}}.
  \end{equation*}

\subsection{Proof of Theorem 2}
 We use the notation $L(X^n; \theta)$ to denote $\prod_{i=1}^n p_i(\theta)$. The statement that $\sqrt{n}(\hat{\theta}-\theta^*)$ converge to $N(0,V_{\theta^*})$ in distribution is followed from Theorem 7.1 of~\cite{RePEc:eee:ecochp:4-36}.  Moreover, we have the following lemma.
 \begin{lemma}\label{lemmath2}
 Under Assumption A.1, A.2' and A.3, there exists a constant $C_1$ such that for any constant $C_2$,  there exists a constant $C$ such that if $C_1\log n\leq \alpha_n\leq C_2 n$, then it holds with probability larger than $1-\frac{1}{n}$ that, 
 \begin{equation*}
\begin{aligned}
  d_{\rm TV}\bigg(\pi_{\rm PE}(\sqrt{n}(\theta-\hat{\theta})|\,X^n), N\Big(0,\big(V^{-1}_{\theta^*}+\frac{\alpha_n}{n}\m H_{\theta^*}\big)^{-1}\Big)\bigg)\leq C \sqrt{\frac{\log n}{n}}.
 \end{aligned}
 \end{equation*}
 \end{lemma} 
 \noindent So, when $C_1\log n\leq\alpha_n\leq C_2\sqrt{n\log n}$, it holds that $d_{\rm TV}(\pi_{\rm PE}(\sqrt{n}(\theta-\hat{\theta})|\,X^n), N(0,V_{\theta^*}))\lesssim \sqrt{\frac{\log n}{n}}$. The desired conclusion is then followed from the shift and scale invariance of the total variation distance.
 \subsubsection{Proof of Lemma~\ref{lemmath2}}
 \begin{equation*}
   \pi_{\rm PE} (\sqrt{n}(\theta-\hat{\theta})|\,X^n) =\frac{\pi(\hat{\theta}+\frac{h}{\sqrt{n}})\exp\left(\log \frac{L(X^n;\,\hat{\theta}+h/\sqrt{n})}{L(X^n;\,\hat{\theta})}-\alpha_n\left(\mathcal{R}_n(\hat{\theta}+h/\sqrt{n})-\mathcal{R}_n(\hat{\theta})\right)\right)}{\int\pi(\hat{\theta}+\frac{h}{\sqrt{n}})\exp\left(\log \frac{L(X^n;\,\hat{\theta}+h/\sqrt{n})}{L(X^n;\,\hat{\theta})}-\alpha_n\left(\mathcal{R}_n(\hat{\theta}+h/\sqrt{n})-\mathcal{R}_n(\hat{\theta})\right)\right)dh},
\end{equation*}
\noindent
we then bound
\begin{equation*}
 \begin{aligned}
& \mathlarger{\int}\Big|\pi(\hat{\theta}+\frac{h}{\sqrt{n}})\exp\left(\log \frac{L(X^n;\,\hat{\theta}+h/\sqrt{n})}{L(X^n;\,\hat{\theta})}-\alpha_n\left(\mathcal{R}_n(\hat{\theta}+h/\sqrt{n})-\mathcal{R}_n(\hat{\theta})\right)\right)\\
& -\pi(\hat{\theta})\exp\big(-\frac{h^T(V_{\theta^*}^{-1}+\frac{\alpha_n}{n}\m H_{\theta^*}\big)h}{2})\Big|dh.
   \end{aligned}
 \end{equation*}
\noindent
 Define the following set of $h$,
\begin{equation*}
 \begin{aligned}
 &A_1=\left\{\|h\|_2\leq \delta_1\sqrt{\log n}\right\},\\
 &A_2=\left\{\delta_1\sqrt{\log n}\leq \|h\|_2\leq \delta_2(\log n)^{1.5}\right\},\\
 &A_3=\left\{\|h\|_2\geq \delta_2 (\log n)^{1.5}\right\}.\\
   \end{aligned}
 \end{equation*}
 \noindent
First for $A_3$, when $ \delta_2 (\log n)^{1.5} \leq \|h\|_2\leq \delta_3 \sqrt{n}$, let $\theta=\hat{\theta}+\frac{h}{\sqrt{n}}$, then $\|\theta-\hat{\theta}\|_2\leq \delta_3$.
Also, by the positive definiteness of $\m H_{\theta^*}$ and Lemma~\ref{l1}, it holds with probability at least $1-\frac{c_0}{n^2}$ that, 
\begin{equation}\label{f6}
\|\theta^*-\hat{\theta}\|_2\lesssim \sqrt{\frac{\log n}{n}}.
\end{equation}
\noindent
So we can choose $\delta_3$ to be small enough such that there exists a positive constant $c$ so that for any $\theta \in B_{2\delta_3}(\theta^*)$,
\begin{equation*}
    \m H_{\theta} \succcurlyeq cI_d.
\end{equation*}
Also, by the fact that $\|h\|_2 \geq \delta_2 (\log n)^{1.5}$ and equation~\eqref{f2},~\eqref{f3}, we can get that when $\delta_2$ is large enough,  there exists a constant $c_0$ such that it holds with probability larger than $1-\frac{c_0}{n^2}$ that for any  $\|h\|_2 \geq \delta_2 (\log n)^{1.5}$, it satisfies that
\begin{equation}\label{eqnco0}
  \log \frac{L(X^n;\,\hat{\theta}+h/\sqrt{n})}{L(X^n;\,\hat{\theta})} \geq 2 d\log n.
\end{equation}
When $\|h\|_2\geq \delta_3\sqrt{n}$, by the assumption that $\theta^*$ is the unique minimizer of $\mathcal{R}(\theta)$, there exists a positive constant $c$ such that it holds with probability at least $1-\frac{1}{n^2}$ that for any $\|h\|_2\geq \delta_3\sqrt{n}$, it satisfies that
\begin{equation*}
   R\left(\hat{\theta}+\frac{h}{\sqrt{n}}\right)-\mathcal{R}(\theta^*)\geq c.
\end{equation*}
\noindent
Similar as equation~\eqref{f12}, by Dudley's inequality and Bernstein inequality, it holds with probability at least $1-\frac{1}{n^2}$ that,
\begin{equation*}
    \underset{\theta\in \Theta}{\sup}|\mathcal{R}(\theta)-\mathcal{R}_n(\theta)|\lesssim \sqrt{\frac{\log n}{n}}.
\end{equation*}
\noindent
Then combined with equation~\eqref{f6}, it holds with probability at least $ 1-\frac{1}{n^2}$ that for any $\theta \in \Theta$ such that $\|\theta-\hat{\theta}\|_2\geq \delta_3$,
\begin{equation}\label{eqnco1}
\begin{aligned}
    &\exp\left(\log \frac{L(X^n;\, \theta)}{L(X^n;\,\hat{\theta})}-\alpha_n\left(\mathcal{R}_n(\theta)-\mathcal{R}_n(\hat{\theta})\right)\right)\\
    &\leq \exp\left(-\alpha_n\left(\mathcal{R}_n(\theta)-\mathcal{R}_n(\hat{\theta})\right)\right)\\
    &\leq \exp(-\alpha_n \frac{c}{2}).
    \end{aligned}
\end{equation}
\noindent
So if we choose $\alpha_n\geq \frac{3}{c} d\log n$,  there exists a constant $c_0$ such that it holds with probability at least $ 1-\frac{c_0}{n^2}$ that,
\begin{equation*}
\begin{aligned}
 &\int_{A_3} \bigg|\pi(\hat{\theta}+\frac{h}{\sqrt{n}})\exp\Big(\log \frac{L(X^n;\,\hat{\theta}+h/\sqrt{n})}{L(X^n;\,\hat{\theta})}-\alpha_n\big(\mathcal{R}_n(\hat{\theta}+h/\sqrt{n})-\mathcal{R}_n(\hat{\theta})\big)\Big)\\
& -\pi(\hat{\theta})\exp(-\frac{h^TV_{\theta^*}^{-1}h}{2})\bigg|dh \lesssim \frac{1}{n}
     \end{aligned}
\end{equation*}
\noindent
For set $A_1$ and $A_2$, use the same strategy of the proof of Lemma~\ref{l1},  there exists a constant $c_0$ such that it holds with probability at least $ 1-\frac{c_0}{n^2}$ that  for any $\|h\|_2\leq \delta_2 (\log n)^{1.5}$,
\begin{equation}\label{eqnco2}
\left|\log \frac{L(X^n;\,\hat{\theta}+h/\sqrt{n})}{L(X^n;\,\hat{\theta})}-\frac{h^TV^{-1}_{\theta^*}h}{2}\right|\lesssim \frac{\|h\|_2^2(\|h\|_2+\sqrt{\log n})}{\sqrt{n}}.
\end{equation}
\noindent
Also,  since $\nabla_{\theta} \mathcal{R}_n(\hat{\theta})=0$ and $\alpha_n\leq C_2 n$, 
\begin{equation*}
    \begin{aligned}
        \mathcal{R}_n(\hat{\theta}+\frac{h}{\sqrt{n}})-\mathcal{R}_n(\hat{\theta})= \frac{1}{2n^2} h^T \sum_{i=1}^n \ell^{(2)} (X_i, \hat{\theta}+\frac{ch}{\sqrt{n}}) h,
    \end{aligned}
\end{equation*}
\noindent
it holds with  probability larger than $ 1-\frac{c_0}{n^2}$ that 
\begin{equation}\label{eqnco3}
   \left| \alpha_n(\mathcal{R}_n(\hat{\theta}+\frac{h}{\sqrt{n}})-\mathcal{R}_n(\hat{\theta}))-\frac{\alpha_n}{n}\frac{h^T \m H_{\theta^*}h}{2}\right|\lesssim \frac{\|h\|_2^2(\|h\|_2+\sqrt{\log n})}{\sqrt{n}}.
    \end{equation}
Then if $\alpha_n\leq C_2 n$ and $\delta_1$ is large enough, we have 
  \begin{equation*}
    \begin{aligned}
         &\int_{A_1} \bigg|\pi(\hat{\theta}+\frac{h}{\sqrt{n}})\exp\Big(\log \frac{L(X^n;\,\hat{\theta}+h/\sqrt{n})}{L(X^n;\,\hat{\theta})}-\alpha_n\big(\mathcal{R}_n(\hat{\theta}+h/\sqrt{n}) -\mathcal{R}_n(\hat{\theta})\big)\Big)\\
         &-\pi(\hat{\theta})\exp(-\frac{h^T(V^{-1}_{\theta^*}+\frac{\alpha_n}{n}\m H_{\theta^*})h}{2})\bigg|dh \\
         &\leq \int_{A_1}\pi(\hat{\theta}+\frac{h}{\sqrt{n}})\bigg|\exp\Big(\log \frac{L(X^n;\,\hat{\theta}+h/\sqrt{n})}{L(X^n;\,\hat{\theta})}-\alpha_n\big(\mathcal{R}_n(\hat{\theta}+h/\sqrt{n})-\mathcal{R}_n(\hat{\theta})\big)\Big)\\
         &- \exp(-\frac{h^T (V^{-1}_{\theta^*}+\frac{\alpha_n}{n}\m H_{\theta^*})h}{2})\bigg|dh +\int_{A_1}\left|\pi(\hat{\theta}+\frac{h}{\sqrt{n}})-\pi(\hat{\theta})\right| \exp(-\frac{h^T(V^{-1}_{\theta^*}+\frac{\alpha_n}{n}\m H_{\theta^*})h}{2})dh\\
        &\lesssim\int_{A_1}\pi(\hat{\theta}+\frac{h}{\sqrt{n}})\exp(-\frac{h^T (V^{-1}_{\theta^*}+\frac{\alpha_n}{n}\m H_{\theta^*})h}{2})  \frac{\|h\|_2^2(\|h\|_2+\sqrt{\log n})}{\sqrt{n}} dh + \sqrt{\frac{\log n}{n}}\\
        &\lesssim \sqrt{\frac{\log n}{n}}
    \end{aligned}
\end{equation*}
  \begin{equation*}
    \begin{aligned}
           &\int_{A_2} \bigg|\pi(\hat{\theta}+\frac{h}{\sqrt{n}})\exp\Big(\log \frac{L(X^n;\,\hat{\theta}+h/\sqrt{n})}{L(X^n;\,\hat{\theta})}-\alpha_n\big(\mathcal{R}_n(\hat{\theta}+h/\sqrt{n}) -\mathcal{R}_n(\hat{\theta})\big)\Big)\\
         &-\pi(\hat{\theta})\exp(-\frac{h^T(V^{-1}_{\theta^*}+\frac{\alpha_n}{n}\m H_{\theta^*})h}{2})\bigg|dh \\
          &\leq \int_{A_2} \pi(\hat{\theta}+\frac{h}{\sqrt{n}})\exp\Big(\log \frac{L(X^n;\,\hat{\theta}+h/\sqrt{n})}{L(X^n;\,\hat{\theta})}-\alpha_n\big(\mathcal{R}_n(\hat{\theta}+h/\sqrt{n})-\mathcal{R}_n(\hat{\theta})\big)\Big)dh\\
          &+\int_{A_2}\pi(\hat{\theta})\exp(-\frac{h^T(V^{-1}_{\theta^*}+\frac{\alpha_n}{n}\m H_{\theta^*})h}{2})dh\\
          &\leq \int_{A_2} \pi(\hat{\theta}+\frac{h}{\sqrt{n}})\exp\left(-\frac{h^T(V^{-1}_{\theta^*}+\frac{\alpha_n}{n}\m H_{\theta^*})h}{2}+c\frac{(\log n)^{4.5}}{\sqrt{n}}\right) dh+\frac{1}{n}\lesssim \frac{1}{n}
 \end{aligned}
\end{equation*}
 \noindent
So, it holds with probability at least $1-\frac{1}{n}$ that, 
\begin{equation*}
\begin{aligned}
 &\int\bigg|\pi(\hat{\theta}+\frac{h}{\sqrt{n}})\exp\left(\log \frac{L(X^n;\,\hat{\theta}+h/\sqrt{n})}{L(X^n;\,\hat{\theta})}-\alpha_n\left(\mathcal{R}_n(\hat{\theta}+h/\sqrt{n})-\mathcal{R}_n(\hat{\theta})\right)\right)\\
& -\pi(\hat{\theta})\exp(-\frac{h^T (V^{-1}_{\theta^*}+\frac{\alpha_n}{n}\m H_{\theta^*})h}{2})\bigg|dh \lesssim  \sqrt{\frac{\log n}{n}}, \\
  \end{aligned}
 \end{equation*}
 \begin{equation*}
\begin{aligned}
 & \bigg|\int \pi(\hat{\theta}+\frac{h}{\sqrt{n}})\exp\left(\log \frac{L(X^n;\,\hat{\theta}+h/\sqrt{n})}{L(X^n;\,\hat{\theta})}-\alpha_n\left(\mathcal{R}_n(\hat{\theta}+h/\sqrt{n})-\mathcal{R}_n(\hat{\theta})\right)\right)\\
& -\pi(\hat{\theta})\exp(-\frac{h^T (V^{-1}_{\theta^*}+\frac{\alpha_n}{n}\m H_{\theta^*})h}{2}) dh\bigg| \lesssim \sqrt{\frac{\log n}{n}}, \\
 \end{aligned}
 \end{equation*}
\begin{equation*}
\begin{aligned}
  d_{\rm TV}\bigg(\pi_{\rm PE}(\sqrt{n}(\theta-\hat{\theta})|\,X^n), N\Big(0,\big(V^{-1}_{\theta^*}+\frac{\alpha_n}{n}\m H_{\theta^*}\big)^{-1}\Big)\bigg)\lesssim \sqrt{\frac{\log n}{n}}.
 \end{aligned}
 \end{equation*}

\subsection{Proof of Corollary 1}
Define sets
\begin{equation*}
    \begin{aligned}
     &A_1=\left\{X^n\Big|\, \|\hat{\theta}-\theta^*\|_2\leq c_1 \sqrt{\frac{\log n}{n}}\right\},\\
       \end{aligned}
\end{equation*}
     \begin{equation*}
    \begin{aligned}
     &A_2=\left\{X^n\Big|\, d_{\rm TV}( \pi_{\rm PE}(\cdot\,|\,X^n), N(\hat{\theta},\frac{1}{n}V_{\theta^*}))\leq c_2 \sqrt{\frac{\log n}{n}}\right\},\\
       \end{aligned}
\end{equation*}
     \begin{equation*}
    \begin{aligned}
     &A_3=\left\{X^n\Big|\, \| \hat{\theta}_B-\hat{\theta}\|\leq c_3 \frac{\sqrt{\log n}}{n}, \, \|n\hat{\Sigma}_B -V_{\theta^*}\|_{\rm F} \leq c_3\sqrt{\frac{\log n}{n}}\right\}\\
       \end{aligned}
\end{equation*}
     \begin{equation*}
    \begin{aligned}
     &A_4=\left\{X^n\Bigg|\, \left\|\frac{1}{n}\sum_{i=1}^n \ell^{(2)} (X_i, \theta^*)- \m H_{\theta^*}\right\|_{\rm F} \leq c_4 \sqrt{\frac{\log n}{n}}\right\}.
    \end{aligned}
\end{equation*}

 Let $A=A_1\cap A_2\cap A_3\cap A_4$. Then by equation~\eqref{eqnco0},~\eqref{eqnco1},~\eqref{eqnco2},~\eqref{eqnco3} and Bernstein inequality, when $C_1\log n\leq\alpha_n\leq C_2 \sqrt{n\log n}$ and $(c_1,c_2,c_3,c_4)$ is  large enough, we have $\m P^*(A)\geq 1-\frac{1}{n}$. Then
 \begin{equation*}
 \begin{aligned}
  &P\left(\left\{(\theta^*- \hat{\theta}_B)^T \hat{\Sigma}_B^{-1}(\theta^*- \hat{\theta}_B) \leq q_{\alpha}\right\}\cap A\right)\\
  &\leq P\left( (\theta^*- \hat{\theta}_B)^T \hat{\Sigma}_B^{-1}(\theta^*- \hat{\theta}_B) \leq q_{\alpha}\right)\\
  &\leq  P\left(\left\{(\theta^*- \hat{\theta}_B)^T \hat{\Sigma}_B^{-1}(\theta^*- \hat{\theta}_B) \leq q_{\alpha}\right\}\cap A\right)+\frac{1}{n}\\
 \end{aligned}
 \end{equation*}

 So there exists a positive constant $c$ such that 
 
\begin{equation*}
    \begin{aligned}
    & P\left(\left\{(\theta^*-\hat{\theta})^T n V^{-1}_{\theta^*} (\theta^*-\hat{\theta})\leq q_{\alpha}-c\frac{(\log n)^{\frac{3}{2}}}{\sqrt{n}}\right\}\cap A\right)\\
    &\leq P\left(\left\{(\theta^*- \hat{\theta}_B)^T \hat{\Sigma}_B^{-1}(\theta^*- \hat{\theta}_B)\leq q_{\alpha}\right\}\cap A\right)\\
    &\leq P\left(\left\{(\theta^*-\hat{\theta})^T n V_{\theta^*}^{-1} (\theta^*-\hat{\theta})\leq q_{\alpha}+c\frac{(\log n)^{\frac{3}{2}}}{\sqrt{n}}\right\}\cap A\right)\\
    \end{aligned}
\end{equation*}

Since under set $A$, $\|\hat{\theta}-\theta^*\|_2\lesssim \sqrt{\frac{\log n}{n}}$, $\hat{\theta}$ is an interior point of $\Theta$ and $\nabla \mathcal{R}_n(\hat{\theta})=0$. So we have 
\begin{equation*}
\begin{aligned}
&0=\frac{1}{n}\sum_{i=1}^n g(X_i, \hat{\theta})= \frac{1}{n}\sum_{i=1}^n g(X_i, \theta^*)+\frac{1}{n}\sum_{i=1}^n g^{(1)} (X_i, \theta^*)(\hat{\theta}-\theta^*)+\frac{1}{2n}\sum_{i=1}^n g^{(2)} (X_i, \theta')(\hat{\theta}-\theta^*)^{\otimes 2},\\
&\sqrt{n}(\theta^*-\hat{\theta}) =\left(\frac{1}{n} \sum_{i=1}^n g^{(1)} (X_i, \theta^*)\right)^{-1} \left(\frac{1}{\sqrt{n}} \sum_{i=1}^n g(X_i, \theta^*) +\frac{1}{2\sqrt{n}}\sum_{i=1}^n g^{(2)}(X_i, \theta') (\hat{\theta}-\theta^*)^{\otimes 2}\right),
    \end{aligned}
\end{equation*}
where $g(x,\theta)=\nabla \ell(X,\theta)$. Since under set $A$,  $\left\|\frac{1}{n}\sum_{i=1}^n g^{(1)} (X_i, \theta^*)- \m H_{\theta^*}\right\|_{\rm F} \leq c_4 \sqrt{\frac{\log n}{n}}$,  there exists a constant $c_5$ such that 

\begin{equation*}
    \begin{aligned}
   & P\left(\left\{(\theta^*-\hat{\theta})^T n V^{-1}_{\theta^*} (\theta^*-\hat{\theta})\leq q_{\alpha}+c\frac{(\log n)^{\frac{3}{2}}}{\sqrt{n}}\right\}\cap A\right)\\
    &\leq P\left(\left\{ \left(\frac{1}{\sqrt{n}} \sum_{i=1}^n V_{\theta^*}^{-\frac{1}{2}} \m H_{\theta^*}^{-1}g(X_i, \theta^*)\right)^T\left(\frac{1}{\sqrt{n}} \sum_{i=1}^n V_{\theta^*}^{-\frac{1}{2}} \m H_{\theta^*}^{-1}g(X_i, \theta^*)\right)\leq q_{\alpha}+c_5 \frac{(\log n)^{\frac{3}{2}}}{\sqrt{n}}  \right\}\cap A\right)\\
    &\leq P\left(\left(\frac{1}{\sqrt{n}} \sum_{i=1}^n V_{\theta^*}^{-\frac{1}{2}} \m H_{\theta^*}^{-1}g(X_i, \theta^*)\right)^T\left(\frac{1}{\sqrt{n}} \sum_{i=1}^n V_{\theta^*}^{-\frac{1}{2}} \m H_{\theta^*}^{-1}g(X_i, \theta^*)\right)\leq q_{\alpha}+c_5 \frac{(\log n)^{\frac{3}{2}}}{\sqrt{n}}  \right)
    \end{aligned}
\end{equation*}
Since $\mathbb{E} g(X, \theta^*)=\nabla \mathcal{R}(\theta^*)=0$, $\text{Cov}(g(X, \theta^*))=\Delta_{\theta^*}$ and $V_{\theta^*}^{-1}=\m H_{\theta^*}\Delta_{\theta^*}^{-1}\m H_{\theta^*}$, we have 
\begin{equation*}
    \begin{aligned}
       &E(V_{\theta^*}^{-\frac{1}{2}} \m H_{\theta^*}^{-1}g(X_i, \theta^*))=0\\
     & \text{Cov}( V_{\theta^*}^{-\frac{1}{2}} \m H_{\theta^*}^{-1}g(X_i, \theta^*))=I_d.\\
    \end{aligned}
    \end{equation*}
    Then by Berry-Esseen theorem~\citep{Rai__2019}, there exists a constant $c_5$ such that
    \begin{equation*}
        \begin{aligned}
        &P\left((\theta^*- \hat{\theta}_B)^T \hat{\Sigma}_B^{-1}(\theta^*- \hat{\theta}_B) \leq q_{\alpha}\right)\\
           &\leq P\left(\left(\frac{1}{\sqrt{n}} \sum_{i=1}^n V_{\theta^*}^{-\frac{1}{2}} \m H_{\theta^*}^{-1}g(X_i, \theta^*)\right)^T\left(\frac{1}{\sqrt{n}} \sum_{i=1}^n V_{\theta^*}^{-\frac{1}{2}} \m H_{\theta^*}^{-1}g(X_i, \theta^*)\right)\leq q_{\alpha}+c_4 \frac{(\log n)^{\frac{3}{2}}}{\sqrt{n}}  \right)+\frac{1}{n} \\
           &\leq P\left(\chi^2_d\leq q_{\alpha}\right)+c_5 \frac{(\log n)^{\frac{3}{2}}}{\sqrt{n}}\\
           &=1-\alpha+c_5\frac{(\log n)^{\frac{3}{2}}}{\sqrt{n}}\\
        \end{aligned}
    \end{equation*}
Similarly, there exists a constant $c_6$ such that 
\begin{equation*}
        \begin{aligned}
        &P\left((\theta^*- \hat{\theta}_B)^T \hat{\Sigma}_B^{-1}(\theta^*- \hat{\theta}_B) \leq q_{\alpha}\right)\\
            &\geq 1-\alpha-c_6 \frac{(\log n)^{\frac{3}{2}}}{\sqrt{n}}.\\
        \end{aligned}
    \end{equation*}
We then get the desired conclusion.

\subsection{Proof of Theorem 3}
We first state an assumption that is similar to the Assumptions (C4)-(C6) in~\cite{10.2307/41000330}.\\
\noindent
\textbf{Assumption B.2'}: There exist constant $c$ and $0<\beta\leq 1$ such that it holds with probability at least $1-n^{-2}$ that 
    \begin{enumerate}
     \item[(a)] $\underset{\theta\in \Theta} {\sup}\, \big\|n^{-1}\sum_{i=1}^n g(X_i,\theta)g(X_i,\theta)^T-\mathbb{E} \big[g(X,\theta)g(X,\theta)^T\big]\big\|_{\rm F} \leq c\sqrt{\frac{\log n}{n}}$;
    \item[(b)]  $\underset{\theta\in \Theta} {\sup} \big\| n^{-1}\sum_{i=1}^n g(X_i,\theta)-n^{-1}\sum_{i=1}^n g(X_i,\theta^*)-\mathbb{E} [g(X,\theta)]+\mathbb{E} [g(X,\theta^*)]\|_2 \leq c\,\Big(\sqrt{\frac{\log n}{n}} \, \|\theta-\theta^*\|^\beta_2+\frac{\log n}{n}\Big)$;
      \item[(c)]  $\underset{\theta\in \Theta} {\sup} \big|n^{-1}\sum_{i=1}^n \ell(X_i,\theta)-n^{-1}\sum_{i=1}^n \ell(X_i,\theta^*)-\mathbb{E} [\ell(X,\theta)]+\mathbb{E} [\ell(X,\theta^*)]\big| \leq c\,\Big(\sqrt{\frac{\log n}{n}} \, \|\theta-\theta^*\|^\beta_2+\frac{\log n}{n}\Big)$.
    \end{enumerate}
\noindent
We then state a lemma to prove that the statement in Assumption B.2 and Assumption B.1 is a sufficient condition to the statement in Assumption B.2'.
     \begin{lemma}\label{lemmaC.3}
 Define 
  \begin{equation*}
        \begin{aligned}
&d_n^g(\theta,\theta')=\sqrt{\frac{1}{n}\sum_{i=1}^n \|g(X_i,\theta)-g(X_i,\theta')\|_2^2},\\
&d_n^{\ell}(\theta,\theta')=\sqrt{\frac{1}{n}\sum_{i=1}^n (\ell(X_i,\theta)-\ell(X_i,\theta'))^2}.
 \end{aligned}
    \end{equation*}
 If  (1) ${\sup}_{x\in \mathcal{X}, \theta\in \Theta}(\|g(x,\theta)\|_2+|\ell(X,\theta)|)\leq C$. (2) The $\varepsilon$-covering numbers with respect to distance $d_n^g$ and $d_n^{\ell}$ of $\Theta$, denoted by $\mathcal{N}(\Theta,d_n^g,\varepsilon)$ and  $\mathcal{N}(\Theta,d_n^{\ell},\varepsilon)$ respectively, are bounded by $(\frac{n}{\varepsilon})^c$ with a constant $c$. (3) $\sqrt{\mathbb{E}\|g(X,\theta)-g(X,\theta^*)\|_2^2}+ \sqrt{\mathbb{E}(\ell(X,\theta)-\ell(X,\theta^*))^2} \leq c_1\|\theta-\theta^*\|_2^{\beta}$, then Assumption B.2' holds.
 \end{lemma}
 \quad\\
  Let $\hat\theta^\diamond =\theta^*-\frac{1}{n}\m H_{\theta^*}^{-1}\sum_{i=1}^n g(X_i,\theta^*)$. We then bound
  \begin{small}
 \begin{equation*}
  \begin{aligned}
  \int\bigg|\pi(\hat\theta^\diamond +\frac{h}{\sqrt{n}})\exp\Big(\log \frac{L(X^n;\,\hat\theta^\diamond +h/\sqrt{n})}{(\frac{1}{n})^n}-\alpha_n\big(\mathcal{R}_n(\hat\theta^\diamond +h/\sqrt{n})-\mathcal{R}_n(\theta^* )\big)\Big)-\pi(\hat\theta^\diamond)\exp(-\frac{h^TV_{\theta^*}h}{2})\bigg|dh.
     \end{aligned}
  \end{equation*}
 \end{small}
 Define the following set of $h$,
\begin{equation*}
 \begin{aligned}
 &A_1=\left\{\|h\|_2\leq \delta_1\sqrt{\log n}\right\},\\
 &A_2=\left\{\delta_1\sqrt{\log n}\leq \|h\|_2\leq \delta_2(\log n)^{1.5}\right\},\\
 &A_3=\left\{\|h\|_2\geq \delta_2 (\log n)^{1.5}\right\}.\\
   \end{aligned}
 \end{equation*}  
  To begin with, we state the following lemmas.
  
 \begin{lemma}\label{le5.1}
 Suppose Assumption B.1, B.2 and A.2' holds, then there exist some positive constants $r$ and $C$, such that it holds with probability at least $1-\frac{1}{n^2}$ that, 
  \begin{equation*}
  \underset{\theta\in B_r(\theta^*)}{\sup}\|\lambda(\theta)\|_2 \leq C.
  \end{equation*}
  \end{lemma}
  \begin{lemma}\label{le5.2}
   Suppose Assumption B.1, B.2 and A.2' holds. Define $\tilde{\lambda}(\theta)=-\Delta_{\theta^*}^{-1}\big(\frac{1}{n}\sum_{i=1}^n g(X_i,\theta^*)+\m H_{\theta^*}(\theta-\theta^*)\big)=-\Delta_{\theta^*}^{-1} \m H_{\theta^*}(\theta-\hat\theta^\diamond )$. There exist positive constants $r_0$ , $c_0$ and $c$ such that it holds with probability at least $1-\frac{c_0}{n^2}$ that, 
 \begin{equation*}
 \underset{\theta \in B_{r_0}(\theta^*)}{\sup} \|\lambda(\theta)-\tilde{\lambda}(\theta)\|_2\leq c(\|\theta-\theta^*\|_2^2+\sqrt{\frac{\log n}{n}}\|\theta-\theta^*\|^{\beta}_2+\frac{\log n}{n}).
 \end{equation*}
  \end{lemma}
    \noindent
 Let $\mathcal{A}_1$ be the event  $\{\|\frac{1}{n} \sum_{i=1}^n g(X_i,\theta^*)-\mathbb{E} g(X,\theta^*)\|_2\leq c\sqrt{\frac{\log n}{n}}\}\cap \{|\frac{1}{n}\sum_{i=1}^n L(X_i,\theta^*)-\mathbb{E} L(X,\theta^*)| \leq c\sqrt{\frac{\log n}{n}}\}$, then by Assumption B.1, there exists a large enough $c$ such that $\m P^*(\mathcal{A}_1)\geq 1-\frac{1}{n^2}$. Let $\mathcal{A}_2$ be the event that statements in (a), (b), (c) of Assumption B.2' hold, then by  Lemma~\ref{lemmaC.3},  $\m P^*(\mathcal{A}_2)\geq 1-\frac{1}{n^2}$. Unless otherwise specified, the following analysis is under event $\mathcal{A}_1\cap \mathcal{A}_2$.\\
\quad\\  
\textbf{Step 1}: Consider set $A_3=\left\{\|h\|_2\geq \delta_2 (\log n)^{1.5}\right\}$. We first consider the case that $\delta_2 (\log n)^{1.5}\leq \|h\|_2 \leq \delta_3\sqrt{n}$ and let $\theta=\hat\theta^\diamond +\frac{h}{\sqrt{n}}$, then by  Assumption B.2' and $\mathbb{E} g(X,\theta^*)=0$, we can get  $\|\hat\theta^\diamond -\theta^*\|\lesssim \sqrt{\frac{\log n}{n}}$ and $\underset{\theta\in \Theta}{\inf}\mathcal{R}_n(\theta) -\mathcal{R}_n(\theta^*)\gtrsim -\sqrt{\frac{\log n}{n}}$. Then by the thirce differentiability of $\mathcal{R}(\theta)$ and $\m H_{\theta^*}\succcurlyeq aI_d$ with a positive constant $a$, when $\delta_3$ is small enough, it holds that $\m H_{\theta}\succcurlyeq \frac{a}{2}I_d$. So, by $\alpha_n \lesssim \sqrt{n\log n}$ and (b) of Assumption B.2', same as Step 1 of  the proof of Lemma~\ref{l1}, we can get when $\delta_2$ is large enough,  for any $h\in \mathbb{R}^d$ such that  $\delta_2 (\log n)^{1.5}\leq \|h\|_2 \leq \delta_3\sqrt{n}$, it holds that 
  \begin{equation*}
  \exp\left(\log \frac{L(X^n;\,\hat\theta^\diamond +h/\sqrt{n})}{(\frac{1}{n})^n}-\alpha_n\left(\mathcal{R}_n(\hat\theta^\diamond +h/\sqrt{n})-\mathcal{R}_n(\theta^* )\right)\right)\leq \exp(-2d \log n).
  \end{equation*}
 For the case that  $\|h\|_2 \geq \delta_3\sqrt{n}$, by Assumption A.2', B.1 and B.2' we can get that there exists a positive constant $c$ such that
 \begin{equation*}
 \underset{\theta\in \Theta}{\sup}|\mathcal{R}_n(\theta)-\mathcal{R}_n(\theta^* )-\mathcal{R}(\theta)+R(\theta^* )|\lesssim \sqrt{\frac{\log n}{n}},
 \end{equation*}

 \begin{equation*}
 \underset{\theta\in \Theta\atop \|\theta-\theta^*\|_2\geq \frac{\delta_3}{2}}{\inf} \mathcal{R}(\theta)- \mathcal{R}(\theta^*)\geq c>0.
 \end{equation*}
 So, when $\alpha_n \geq \frac{4d}{c}\log n$, for any $h$ such that $\|h\|_2 \geq \delta_3\sqrt{n}$, it holds that 
  \begin{equation*}
  \exp\left(\log \frac{L(X^n;\,\hat\theta^\diamond +h/\sqrt{n})}{(\frac{1}{n})^n}-\alpha_n\left(\mathcal{R}_n(\hat\theta^\diamond +h/\sqrt{n})-\mathcal{R}_n(\theta^* )\right)\right)\leq \exp(-2d \log n).
  \end{equation*}
 So we can get 
  \begin{equation*}
  \begin{aligned}
 & \int_{A_3}\bigg|\pi(\hat\theta^\diamond +\frac{h}{\sqrt{n}})\exp\Big(\log \frac{L(X^n;\,\hat\theta^\diamond +h/\sqrt{n})}{(\frac{1}{n})^n}-\alpha_n\big(\mathcal{R}_n(\hat\theta^\diamond +h/\sqrt{n})-\mathcal{R}_n(\theta^* )\big)\Big)\\
 & -\pi(\hat\theta^\diamond )\exp(-\frac{h^TV_{\theta^*}h}{2})\bigg|dh \leq \frac{1}{\sqrt{n}}.
   \end{aligned}
  \end{equation*}
 
\quad\\
\textbf{Step 2}: Consider set $A_1$ and $A_2$, when $\|h\|_2\leq \delta_2 (\log n)^{1.5}$, let $\theta=\hat\theta^\diamond +\frac{h}{\sqrt{n}}$, then we have $\|\theta-\theta^*\|_2\leq \frac{\|h\|_2}{\sqrt{n}}+c\sqrt{\frac{\log n}{n}} $. By Lemma~\ref{le5.2},  we can get $\|\lambda(\theta)\|\lesssim \sqrt{\frac{\log n}{n}}+\frac{\|h\|_2}{\sqrt{n}}$. Moreover,
\begin{equation*}
 \log \frac{L(X^n;\theta)}{(\frac{1}{n})^n}=\sum_{i=1}^n \lambda(\theta)^T g(X_i,\theta)- n \log \left(\frac{1}{n}\sum_{i=1}^n \exp(\lambda(\theta)^T g(X_i,\theta))\right).
 \end{equation*}
 Since
 \begin{equation*}
 \exp(\lambda(\theta)^T g(X_i,\theta))=1+\lambda(\theta)^T g(X_i,\theta)+\frac{1}{2} (\lambda(\theta)^T g(X_i,\theta))^2+O\left(\|h\|_2^3n^{-\frac{3}{2}}+(\frac{\log n}{n})^{\frac{3}{2}}\right).
 \end{equation*}
We have
\begin{equation*}
\begin{aligned}
 \log \left(\frac{1}{n}\sum_{i=1}^n \exp(\lambda(\theta)^T g(X_i,\theta))\right)&=\frac{1}{n}\sum_{i=1}^n \lambda(\theta)^T g(X_i,\theta)+\frac{1}{2n}\sum_{i=1}^n ( \lambda(\theta)^T g(X_i,\theta))^2\\
& - \frac{1}{2}\left(\frac{1}{n} \sum_{i=1}^n \lambda(\theta)^T g(X_i,\theta)\right)^2+O\left(\|h\|_2^3n^{-\frac{3}{2}}+(\frac{\log n}{n})^{\frac{3}{2}}\right).
 \end{aligned}
 \end{equation*}
 So
\begin{equation*}
\begin{aligned}
\log \frac{L(X^n;\theta)}{(\frac{1}{n})^n}=-\frac{1}{2}\sum_{i=1}^n ( \lambda(\theta)^T g(X_i,\theta))^2+\frac{n}{2}\left(\frac{1}{n} \sum_{i=1}^n \lambda(\theta)^T g(X_i,\theta)\right)^2+O\left(\|h\|_2^3n^{-\frac{1}{2}}+\frac{(\log n)^{\frac{3}{2}}}{\sqrt{n}}\right).
 \end{aligned}
\end{equation*}
For the first term, by Assumption B.1 and B.2', we have 
\begin{equation*}
\begin{aligned}
 &\left|\frac{1}{n}\sum_{i=1}^n ( \lambda(\theta)^T g(X_i,\theta^*))^2-\frac{1}{n}\sum_{i=1}^n ( \lambda(\theta)^T g(X_i,\theta))^2-\mathbb{E} (\lambda(\theta)^T g(X,\theta^*))^2+\mathbb{E} (\lambda(\theta)^T g(X,\theta))^2\right|\\
&=\bigg|\lambda(\theta)^T\Big(\frac{1}{n}\sum_{i=1}^n g(X_i,\theta^*)g(X_i,\theta^*)^T-\frac{1}{n}\sum_{i=1}^n g(X_i,\theta)g(X_i,\theta)^T-\mathbb{E} g(X,\theta^*)g(X,\theta^*)^T\\
&+\mathbb{E} g(X,\theta)g(X,\theta)^T\Big)\lambda(\theta)\bigg|\lesssim \sqrt{\frac{\log n}{n}} \frac{\|h\|_2^2}{n}+(\frac{\log n}{n})^{\frac{3}{2}}.
 \end{aligned}
\end{equation*}
\noindent
Also,
\begin{equation*}
\begin{aligned}
&|\mathbb{E} (\lambda(\theta)^T g(X,\theta^*))^2-\mathbb{E} (\lambda(\theta)^T g(X,\theta))^2|\\
&=|\lambda(\theta)^T(\Delta_{\theta}-\Delta_{\theta^*})\lambda(\theta)|\\
&\lesssim \|h\|_2^3 n^{-\frac{3}{2}}+ (\frac{\log n}{n})^{ \frac{3}{2}}.
 \end{aligned}
\end{equation*}
\begin{equation*}
\begin{aligned}
&\frac{1}{n}\sum_{i=1}^n ( \lambda(\theta)^T g(X_i,\theta^*))^2\\
&=\lambda(\theta)^T \frac{1}{n} \sum_{i=1}^n g(X_i,\theta^*) g(X_i,\theta^*)^T \lambda(\theta)\\
&=\lambda(\theta)^T \Delta_{\theta^*} \lambda(\theta)+O\left(\sqrt{\frac{\log n}{n}}\frac{\|h\|_2^2}{n}+(\frac{\log n}{n})^{\frac{3}{2}}\right)
\end{aligned}
\end{equation*}
So we can get 
\begin{equation*}
-\frac{1}{2}\sum_{i=1}^n ( \lambda(\theta)^T g(X_i,\theta))^2=-\frac{n}{2} \lambda(\theta)^T \Delta_{\theta^*} \lambda(\theta)+O\left(\frac{\|h\|_2^3}{\sqrt{n}}+\frac{(\log n)^{\frac{3}{2}}}{\sqrt{n}}\right)
\end{equation*}
\noindent
For the second term, Since 
\begin{equation*}
\begin{aligned}
&\|\frac{1}{n}\sum_{i=1}^n g(X_i,\theta)-\mathbb{E} g(X,\theta)\|_2\lesssim \sqrt{\frac{\log n}{n}}\\
&\|\mathbb{E} g(X,\theta)\|_2\lesssim \frac{\|h\|_2}{\sqrt{n}}+ \sqrt{\frac{\log n}{n}},\\
\end{aligned}
\end{equation*}
we have 
\begin{equation*}
\frac{n}{2}\left(\frac{1}{n} \sum_{i=1}^n \lambda(\theta)^T g(X_i,\theta)\right)^2\lesssim \frac{\|h\|_2^4}{n}+\frac{(\log n)^2}{n}
\end{equation*}
\noindent
So we can get 
\begin{equation*}
\log \frac{L(X^n;\theta)}{(\frac{1}{n})^n}=-\frac{n}{2}\lambda(\theta)^T \Delta_{\theta^*}\lambda(\theta)+O\left(\|h\|_2^3n^{-\frac{1}{2}}+\frac{(\log n)^{\frac{3}{2}}}{\sqrt{n}}\right).
\end{equation*}
\noindent
Also by Lemma~\ref{le5.2}, we have 
\begin{equation*}
\|\lambda(\theta)-(-\Delta_{\theta^*}^{-1}\m H_{\theta^*}(\theta-\hat\theta^\diamond ))\|_2 \lesssim    \|\theta-\theta^*\|_2^2+\sqrt{\frac{\log n}{n}}\|\theta-\theta^*\|^{\beta}_2+\frac{\log n}{n}. 
\end{equation*}
So we can get 
\begin{equation*}
\log \frac{L(X^n;\hat\theta^\diamond +h/\sqrt{n})}{(\frac{1}{n})^n}=-\frac{1}{2}h^T V^{-1}_{\theta^*}h+O\left(\frac{\|h\|_2^{2+\beta}+{\log n}^{1+\frac{\beta}{2}}}{n^{\frac{\beta}{2}}}\right).
\end{equation*}
Moreover, by Assumption B.1 and B.2', 
\begin{equation*}
\begin{aligned}
&|\mathcal{R}_n(\hat\theta^\diamond +h/\sqrt{n})-\mathcal{R}_n(\theta^* )-\m R(\hat\theta^\diamond +h/\sqrt{n})+\m R(\theta^* )|\lesssim \left(\frac{\log n}{n}\right)^{\frac{1+\beta}{2}}+\sqrt{\frac{\log n}{n}}\left(\frac{\|h\|_2}{\sqrt{n}}\right)^{\beta}\\
&|\m R(\hat\theta^\diamond +h/\sqrt{n})-\m R( \theta^* )|\lesssim \frac{\log n}{n}+\frac{\|h\|_2^2}{n}.
\end{aligned}
\end{equation*}
So we have 
\begin{equation*}
\mathcal{R}_n(\hat\theta^\diamond +h/\sqrt{n})-\mathcal{R}_n(\theta^* )\gtrsim -\left(\left(\frac{\log n}{n}\right)^{\frac{1+\beta}{2}}+\sqrt{\frac{\log n}{n}}\left(\frac{\|h\|_2}{\sqrt{n}}\right)^{\beta}+\frac{\|h\|_2^2}{n}\right).
\end{equation*}
\noindent
Then by $\alpha_n\lesssim \sqrt{n\log n}$, similar as the proof of Theorem 2, we can get 
\begin{equation*}
\begin{aligned}
 & \int_{A_1\cup A_2}\bigg|\pi(\hat\theta^\diamond +\frac{h}{\sqrt{n}})\exp\Big(\log \frac{L(X^n;\,\hat\theta^\diamond +h/\sqrt{n})}{(\frac{1}{n})^n}-\alpha_n\big(\mathcal{R}_n(\hat\theta^\diamond +h/\sqrt{n})-\mathcal{R}_n(\theta^* )\big)\Big)\\
& -\pi(\hat\theta^\diamond )\exp(-\frac{h^TV_{\theta^*}h}{2})\bigg|dh\lesssim \frac{(\log n)^{1+\frac{\beta}{2}}}{n^{\frac{\beta}{2}}}.
  \end{aligned}
  \end{equation*}
So by Assumption A.2' and B.1, we could get the first statement. The second statement  that $\sqrt{n}(\hat{\theta}^{\diamond}-\theta^*)$ converges to $N(0, V_{\theta^*})$ in distribution is followed from standard central limit theorem.

 \subsection{Proof of Theorem 4}
 The first statement of Theorem 4 is a direct result from Lemma~\ref{lemmath2}.  W.l.o.g, we can assume $d>n$, otherwise we could replace $\log d$ with $\log n$ in the following analysis.  By the definition of the ``model-averaged'' Bayesian PETEL in Section 3.2, we have
 \begin{equation} 
   \begin{aligned}
 &\Pi_{\rm PE}(S\,|\,X^n)=\\
 &\frac{\binom{d}{|S|}^{-1}q(|S|)\,\int_{\Theta_S}\pi_S(\theta_S) \exp\big(-\alpha_{n,d} \m (R_n(\theta_S,0)-R_n(\theta^*))\big)\prod_{i=1}^n p_i(\theta_S;\,S)/(\frac{1}{n})^n\, d\theta_S}{\underset{S\in[d], \,|S|\leq s_0}{\sum}\binom{d}{|S|}^{-1}q(|S|)\int_{\Theta_S}\pi_S(\theta_S) \exp\big(-\alpha_{n,d} \m (R_n(\theta_S,0)-R_n(\theta^*))\big)\prod_{i=1}^n p_i(\theta_S;\,S)/(\frac{1}{n})^n\, d\theta_S}.
 \end{aligned}
      \end{equation} 
\begin{equation} 
   \begin{aligned}
 &\Pi_{\rm PE}(\|\theta-\theta^*\|_2\geq \delta\,|\,X^n)=\\
 &\frac{\underset{S\in[d], \,|S|\leq s_0}{\sum}\binom{d}{|S|}^{-1}q(|S|)\,\int_{\|(\theta_S,0)-\theta^*\|_2\geq \delta}\pi_S(\theta_S) \exp\big(-\alpha_{n,d} \m (R_n(\theta_S,0)-R_n(\theta^*))\big)\prod_{i=1}^n p_i(\theta_S;\,S)/(\frac{1}{n})^n\,d\theta_S}{\underset{S\in[d], \,|S|\leq s_0}{\sum}\binom{d}{|S|}^{-1}q(|S|)\int_{\Theta_S}\pi_S(\theta_S) \exp\big(-\alpha_{n,d} \m (R_n(\theta_S,0)-R_n(\theta^*))\big)\prod_{i=1}^n p_i(\theta_S;\,S)/(\frac{1}{n})^n\, d\theta_S}.
 \end{aligned}
      \end{equation}
Let $L(X^n; \theta_S, S)=\prod_{i=1}^n p_i(\theta_S; \, S)$

\noindent
\textbf{Step 1:} Lower bound the denominator.\\
 \noindent
 By $\alpha_{n,d}\leq C_3 n$,  similar as the analysis of Lemma~\ref{l1} and Theorem 2,  it holds with probability larger than $1-\frac{1}{n^2}$ that, 
\begin{equation}\label{lboudenominator}
    \begin{aligned}
   &  \underset{S\subseteq\{1,\cdots,d\}\atop |S|\leq s_0}{\sum}q(|S|)\binom{d}{|S|}^{-1}\int \pi_S(\theta_S) \exp\left(\log \frac{L(X^n; \theta_S, S)}{\left(\frac{1}{n}\right)^n}-\alpha_{n,d} (\mathcal{R}_n(\theta_S,0)-\mathcal{R}_n(\theta^*))\right) d\theta_S\\
   &\geq 
 q(s^*)\binom{d}{s^*}^{-1}\int_{\|\theta_{S^*}-\hat{\theta}_{S^*}\|_2\leq\frac{1}{n}} \pi_{S^*}(\theta_{S^*}) \exp\left(\log \frac{L(X^n; \theta_{S^*}, S^*)}{\left(\frac{1}{n}\right)^n}-\alpha_{n,d} (\mathcal{R}_n(S^*,\theta_{S^*})-\mathcal{R}_n(\theta^*))\right) d\theta_{S^*}\\
&\geq c_0 q(|S^*|)\binom{d}{|S^*|}^{-1} \int_{\|\theta_{S^*}-\hat{\theta}_{S^*}\|_2\leq \frac{1}{n}}\pi_{S^*}(\theta_{S^*}) d\theta_{S^*}\\
&\geq \exp(-c_1s^* \log d)\exp(-\beta_{n,d} s^*).
 \end{aligned}
\end{equation}

\noindent
\textbf{Step 2}: Upper bound $ \Pi_{\rm PE}(|S|\geq s^*+1\,|\,X^n)$.\\
 \noindent
By Theorem 14.20 of~\cite{wainwright_2019}, there exist some constant $(c,c_1,c_2)$ such that it holds with probability at least $1-\exp(-c\log d)$ that 
  \begin{equation}\label{step2eq1}
    \begin{aligned}
& \underset{\theta\in \Theta\atop  \|\theta\|_0\leq s_0}{\sup}|\mathcal{R}_n(\theta)-\mathcal{R}_n(\theta^*)-\mathcal{R}(\theta)+\mathcal{R}(\theta^*)|\leq c_1(\frac{\log d}{n}+\|\theta-\theta^*\|_2\sqrt{\frac{\log d}{n}})
  \end{aligned}
\end{equation}

\noindent Also by Definition of the sparse prior in Section 3.2, there exists a constant $c_4$ such that for any $s\geq s^*+1$, it holds that $q(s)\leq c_4 \exp\left(-\beta_{n,d} (s^*+1)\right)$.
 
\noindent
\begin{enumerate}
\item Under Assumption C.2, by equation~\eqref{step2eq1}, we can further obtain that $ \underset{\theta\in \Theta\atop  \|\theta\|_0\leq s_0}{\sup}\mathcal{R}_n(\theta)-\mathcal{R}_n(\theta^*)\geq -c_2\frac{\log d}{n}$. Then by $\beta_{n,d}=C_0\log d$ and $\alpha_{n,d}\leq C_3 n$, when $C_0$ is large enough, we can get that $ \Pi_{\rm PE}(|S|\geq s^*+1\,|\,X^n) \leq \frac{1}{d}$.
\item Under Assumption C.2',  by equation~\eqref{step2eq1}, we can further obtain that $ \underset{\theta\in \Theta\atop  \|\theta\|_0\leq s_0}{\sup}\mathcal{R}_n(\theta)-\mathcal{R}_n(\theta^*)\geq -c_2\sqrt{\frac{\log d}{n}}$,  then by $\beta_{n,d}\geq C_0 (\log d\vee \alpha_{n,d} \sqrt{\frac{\log d}{n}})$, when $C_0$ is large enough, we can get that $ \Pi_{\rm PE}(|S|\geq s^*+1\,|\,X^n) \leq \exp(-\frac{1}{2}\beta_{n,d})$.
\end{enumerate}

 \noindent
\textbf{Step 3}: Upper bound $\Pi_{\rm PE}(\|\theta-\theta^*\|_2\geq \delta\,|\,X^n)$ with $ \delta\geq c\frac{\log d}{n}$  under Assumption C.2.
 \begin{enumerate}
 \item When $\alpha_{n,d}\geq C_1 n$.  Since by Assumption C.2, there exists a positive constant $c_1$, such that 
 $R(\theta)-R(\theta^*)\geq c_1\|\theta-\theta^*\|_2^2$,  by equation~\eqref{step2eq1}, when $c$ is larger enough, then there exists a positive constant $c_2$  such that  it holds  with probability larger than $1-\frac{1}{n^2}$ that for any $S\subset [d]$ with $|S|\leq s^*$ and $\theta_S\in \Theta_S$ such that $\|(\theta_S,0)-\theta^*\|_2\geq \delta$, it satisfies  that $\exp\big(-\alpha_{n,d} \m (R_n(\theta_S,0)-R_n(\theta^*))\big)\leq \exp(-C_1c_2 n\delta^2)$. Then combined with equation~\eqref{lboudenominator} and the conclusion in Step 2, we can get that when $ \delta\geq c\frac{\log d}{n}$ with a large enough $c$, it holds with probability larger than $1-\frac{1}{n^2}$  that  $\Pi_{\rm PE}(\|\theta-\theta^*\|_2\geq \delta\,|\,X^n)\leq \exp(-\frac{C_1c_1}{2}n\delta^2)\leq \frac{1}{d}$.
 
\item When $\alpha_{n,d}\geq C_2\frac{\log d}{{\min}_{i\in S^*}\, {\theta^*_i}^2}$ and $C_2\frac{\log d}{{\min}_{i\in S^*}\, {\theta^*_i}^2}\leq C_1 n$, then we have ${\min}_{i\in S^*}\, {\theta^*_i}^2\geq \frac{C_2\log d}{C_1n}$. Moreover, we can get when $|S|\leq s^*$ and $S\neq S^*$, there exists a positive constant $c_0$ such that for any $\theta_S\in \Theta_S$, it holds that $R(\theta_S,0)-R(\theta^*)\geq c_0\,{\min}_{i\in S^*}\, {\theta^*_i}^2$, then by equation~\eqref{step2eq1} when $C_2$ is large enough, we can get that  it holds  with probability larger than $1-\frac{1}{n^2}$ that $\Pi_{\rm PE}(|S|\leq s^*, \, S\neq S^*|X^n)\leq \frac{1}{2d}$. Then by Lemma~\ref{lemmath2} and the conclusion in Step 2, we can get that when $c$ is large enough,  $\Pi_{\rm PE}(\|\theta-\theta^*\|_2\geq \delta\,|\,X^n)\leq \frac{1}{d}$.
\end{enumerate}

 
\noindent 
\textbf{Step 4}: Upper bound $ \Pi_{\rm PE}(|S|\leq s^*, \, S\neq S^*|X^n)$ under Assumption C.2'.\\
\noindent
By Assumption C.2', there exists a positive constant $c_2$ such that for any $S\subseteq [d]$ with $|S|\leq s^*$ and $S\neq S^*$,  it holds that $\underset{\theta_S\in \Theta_S}{\inf}\mathcal{R}(\theta_S,0)-\mathcal{R}(\theta^*)\geq c_2 {\min}_{i\in S^*}\, {\theta^*_i}^2$. Moreover, by equation~\eqref{step2eq1}, it holds with probability at least $1-\exp(-c \log d)$ that,
 \begin{equation*}
  \underset{\theta \in \Theta\atop \|\theta\|_0\leq s^*}{\sup} |\mathcal{R}(\theta)-\mathcal{R}_n(\theta)|\lesssim \sqrt{ \frac{\log d}{n}}.
  \end{equation*}
Then  by  $\binom{d}{1}+\cdots+\binom{d}{s^*}\leq p(\frac{ed}{s^*})^{s^*}$, $\log n \leq \log d\leq  C n$ and  ${\min}_{i\in S^*}\, {\theta^*_i}^2\geq c_1 \sqrt{\frac{\log d}{n}}$, when $c_1$ is large enough, it holds with probability at least $1-\frac{1}{n^2}$ that for any $S\subseteq [d]$ with $|S|\leq s^*$ and $S\neq S^*$, it satisfies that
\begin{equation*}
  \underset{\theta_S\in\Theta_S}{\sup}  \exp\left(\log \frac{L(X^n;\theta_S,S)}{(\frac{1}{n})^n}-\alpha_{n,d} (\mathcal{R}_n(\theta_S,0)-\mathcal{R}_n(\theta^*))\right)\leq \exp\left(-\alpha_{n,d} \frac{{\min}_{i\in S^*}\, {\theta^*_i}^2}{2}\right).
    \end{equation*}
Then combined with equation~\eqref{lboudenominator}, there exist some constant $(C_1, C_2)$ such that when $\alpha_{n,d}\geq C_2 \frac{\log d}{{\min}_{i\in S^*}\, {\theta^*_i}^2}$ and $\beta_{n,d} \leq C_1 \alpha_{n,d}\,{\min}_{i\in S^*}\, {\theta^*_i}^2$, it holds with probability at least $1-\frac{1}{n^2}$,
\begin{equation}\label{f9}
       \Pi_{\rm PE}(|S|\leq s^*, \, S\neq S^*|X^n) \leq\exp\left(-\frac{1}{2}\beta_{n,d}\right).
\end{equation}
\noindent Combined with the conclusion in Step 2,  we could then get that it holds with probability at least $1-\frac{1}{n}$ that $\Pi_{\rm PE}(S=S^*|X^n)\geq 1-2\exp(-\frac{1}{2}\beta_{n,d})$.  \\
 
\section{Proof of Technical details}
 
\subsection{Proof of lemma~\ref{l1.2}}
\begin{equation*}
    \Pi_{\rm E}(\theta \in A^c|X^n)=\frac{\int_{A^c} \pi(\theta)\exp\left(\log\frac{L(X^n;\theta)}{(\frac{1}{n})^n}\right)d\theta}{\int \pi(\theta)\exp\left(\log\frac{L(X^n;\theta)}{(\frac{1}{n})^n}\right)d\theta}.
\end{equation*}
\\
\quad\\
\textbf{Step 1}: Lower bound the denominator.
\noindent
\begin{equation*}
    \begin{aligned}
          \int \pi(\theta)\exp\left(\log\frac{L(X^n;\theta)}{(\frac{1}{n})^n}\right)d\theta\geq \int_{B_{\frac{1}{\sqrt{n}}}(\hat{\theta}_1)} \pi(\theta) \exp\left(\log\frac{L(X^n;\theta)}{L(X^n;\hat{\theta}_1)}\right)d\theta
    \end{aligned}
\end{equation*}
\noindent
By equation~\eqref{f13}, there exist constants $(c_0, c)$ such that  it holds with probability at least $1-\frac{c_0}{n^2}$  that when $\|\theta-\hat{\theta}_1\|_2\leq \frac{1}{\sqrt{n}}$,
\begin{equation*}
    \left|\log \frac{L(X^n;\theta)}{L(X^n;\hat{\theta}_1)}\right| \leq c.
\end{equation*}
So, we can get that 
\begin{equation*}
    \int \pi(\theta)\exp\left(\log\frac{L(X^n;\theta)}{(\frac{1}{n})^n}\right)d\theta\geq c_1(\frac{1}{\sqrt{n}})^d,
\end{equation*}
with a positive constant $c_1$.\\
\textbf{Step 2}: Upper bound the numerator.
\begin{equation*}
    \begin{aligned}
     &\int_{A^c} \pi(\theta)\exp\left(\log\frac{L(X^n;\theta)}{(\frac{1}{n})^n}\right)d\theta\\
     &=\int_{A^c} \pi(\theta)\exp\left( \sum_{i=1}^n \log p_i(\theta)-n\log\frac{1}{n}\right)d\theta.
    \end{aligned}
\end{equation*}
When $\theta\in \Theta\cap A^c$, $\|\nabla_{\theta} \mathcal{R}(\theta)\|\geq c$.
\begin{equation*}
    \begin{aligned}
     &\sum_{i=1}^n p_i(\theta)g(X_i,\theta)=0\\
    &\sum_{i=1}^n \left(\frac{1}{n}-p_i(\theta)\right)g(X_i,\theta)=\nabla_{\theta} \mathcal{R}_n(\theta).
    \end{aligned}
\end{equation*}
By equation~\eqref{f12}, it holds with probability at least $1-\frac{1}{n^2}$,
\begin{equation*}
    \underset{\theta\in \Theta}{\sup}\|\nabla_{\theta} \mathcal{R}_n(\theta)-\nabla_{\theta} \mathcal{R}(\theta)\|_2 \lesssim \sqrt{\frac{\log n}{n}}.
\end{equation*}
So, 
\begin{equation*}
 \begin{aligned}
&\sum_{i=1}^n \left(p_i(\theta')-\frac{1}{n}\right)^2 \sum_{i=1}^n \|g(X_i,\theta)\|_2^2 \geq   \frac{c^2}{2}\\
&\sum_{i=1}^n \left(p_i(\theta')-\frac{1}{n}\right)^2 \gtrsim \frac{1}{n}.
 \end{aligned}
 \end{equation*}
 Define $q(p_1, \cdots, p_{n-1})=\sum_{i=1}^{n-1} \log p_i +\log (1-\sum_{i=1}^{n-1} p_i)$. The Hessian matrix of function $q$ at point  $(p_1, \cdots, p_{n-1})$ is 
 \begin{equation*}
     \m H_q|_{(p_1, \cdots, p_{n-1})}= Diag(-\frac{1}{p_1^2},\cdots,-\frac{1}{p_{n-1}^2})-\frac{1}{(1-\sum_{i=1}^{n-1}p_i)^2}\textbf{1}_{(n-1)\times(n-1)}
 \end{equation*}
 Let $p=(p_1, \cdots, p_{n})$ and $p_{-n}=(p_1, \cdots, p_{n-1})$. If  $\|p\|_{\infty} \geq n^{-\frac{2}{3}}$, then 
 \begin{equation*}
 \sum_{i=1}^n \log p_i \leq (n-1) \log \frac{1-n^{-\frac{2}{3}}}{n-1}.
 \end{equation*}
So,
 \begin{equation}\label{f4}
 \begin{aligned}
  -n\log n -\sum_{i=1}^n \log p_i  &\geq -\log n - (n-1) \log\left( (1-n^{-\frac{2}{3}}) \frac{n}{n-1}\right)\\
  &\gtrsim n^{\frac{1}{3}}.
  \end{aligned}
 \end{equation}
 \quad\\
 If $\|p\|_{\infty} \leq n^{-\frac{2}{3}}$, then we have $\sum_{i=1}^{n-1} (p_i-\frac{1}{n})^2\gtrsim \frac{1}{n}$, so by mean value theorem,  
 \begin{equation}
 \begin{aligned}
&q(\frac{1}{n}, \cdots ,\frac{1}{n})-q(p_{-n})\\
&=-\frac{1}{2}(p_{-n}-\frac{1}{n}\textbf{1}_{(n-1)})^T \m H_q|_{(cp_{-n}+(1-c)\frac{1}{n}\textbf{1}_{(n-1)})}(p_{-n}-\frac{1}{n}\textbf{1}_{(n-1)})\\
&\gtrsim n^{\frac{1}{3}}.
  \end{aligned}
 \end{equation}
 So there exists a positive constant $c$, such that it holds with probability at least $1-\frac{2}{n^2}$ that,
 \begin{equation*}\label{f5}
   \int_{A^c} \pi(\theta)\exp\left(\log\frac{L(X^n;\theta)}{(\frac{1}{n})^n}\right)d\theta \leq \exp(-c n^{\frac{1}{3}}).
 \end{equation*}
Then, combined with the lower bound on the denominator, we can get the desired conclusion.

 \subsubsection{Proof of lemma~\ref{l2}}
Fix a vector $ \nu\in \mathbb{S}^{d-1}$, then for any $\theta \in B_r(\tilde{\theta})$, there exists a constant $c \in [0,1]$ depend on $ \nu$ and $\theta$, such that 
\begin{equation*}
    \nabla_{\theta} \mathcal{R}(\theta)^T  \nu=\nabla_{\theta} R(\tilde{\theta})^T  \nu+(\theta-\tilde{\theta})^T \m H_{(c\theta+(1-c)\tilde{\theta})}\, \nu.
\end{equation*}
So, we have 
\begin{equation*}
    \| \nabla_{\theta} \mathcal{R}(\theta)\|_2 \geq \underset{\theta' \in B_r(\tilde{\theta})}{\inf}\left|(\theta-\tilde{\theta})^T \m H_{\theta'}\, \nu\right|.
\end{equation*}
Take the supreme over  $ \nu\in \mathbb{S}^{d-1}$, we can get 
\begin{equation*}
\begin{aligned}
     \| \nabla_{\theta} \mathcal{R}(\theta)\|_2 &\geq \underset{ \nu\in \mathbb{S}^{d-1}}{\sup}\underset{\theta' \in B_r(\tilde{\theta})}{\inf}\left|(\theta-\tilde{\theta})^T \m H_{\theta'}\, \nu\right|\\
     &\geq \underset{\theta' \in B_r(\tilde{\theta})}{\inf}\left|\frac{(\theta-\tilde{\theta})^T \m H_{\theta'}\m H_{\tilde{\theta}}^{-1}(\theta-\tilde{\theta})}{\|\m H_{\tilde{\theta}}^{-1}(\theta-\tilde{\theta})\|}\right|
\end{aligned}
\end{equation*}
Since for any $\theta \in B_r(\tilde{\theta})$, $\m H_\theta^T \m H_\theta\succcurlyeq c I_d$ and $\m H_{\theta}^T \m H_{\tilde{\theta}}^{-1} \succcurlyeq c I_d$, there exists a constant $c_0$ such that
\begin{equation*}
     \| \nabla_{\theta} \mathcal{R}(\theta)\|_2  \geq c_0 \|\theta-\tilde{\theta}\|_2.
\end{equation*}
Since it holds with probability at least $1-\frac{1}{n^2}$ that $\underset{\theta\in \Theta}{\sup} \| \m H_\theta-\mathcal{H}^n_{\theta}\|_{\rm F} \leq c_1 \sqrt{\frac{\log n}{n}}$, we can get 
for any $\theta \in B_r(\tilde{\theta})$, $\mathcal{H}^n_\theta \mathcal{H}^n_\theta\succcurlyeq \frac{c}{2} I_d$ and $\mathcal{H}^n_{\theta} (\mathcal{H}^n_{\tilde{\theta}})^{-1} \succcurlyeq \frac{c}{2} I_d$. Then by $\|\nabla \mathcal{R}_n(\hat{\theta})\|_2=0$, use the same strategy, we can get the conclusion of the second statement.

\subsection{Proof of lemma~\ref{l3}}
Let $b_1=\underset{x \in \mathcal{X} \atop \theta \in B_r(\tilde{\theta})}{\sup}\|g(x,\theta)\|_2$, choose $c_2= \min\left(\sqrt{\frac{a}{8}}, \frac{a}{8b_1}\right)$. Since $\Delta_{\tilde{\theta}} \succcurlyeq aI_{d}$ and $E g(x,\tilde{\theta})=0$, we can find a small enough $\delta_0$, such  that for any $\theta \in B_{\delta_0}(\tilde{\theta})$, $\Delta_{\theta}\succcurlyeq \frac{a}{2} I_d$ and  $\|\mathbb E g(x, \theta)\|_2\leq \frac{a}{8b_1}$.
\\
\quad\\
Then if there exist $\lambda \in \mathbb{S}^{d-1}$ and $\theta \in B_{\delta_0}(\tilde{\theta})$ such that $\m P^*(\lambda^T g(X,\theta) \geq c_2) < c_3$, by 
$\Delta_{\theta}\succcurlyeq \frac{a}{2} I_d$, we can get
\begin{equation*}
\begin{aligned}
\frac{a}{2}&\leq \mathbb E(\lambda^T g(x,\theta))^2\\
&< c_2^2+b_1^2 c_3 +\int_{\lambda^T g(x,\theta)\leq 0} (\lambda^T g(x,\theta))^2 d \m P^*.
\end{aligned}
\end{equation*}
So,
\begin{equation*}
b_1\int_{\lambda^T g(x,\theta)\leq 0} -\lambda^T g(x,\theta) d  \m P^* \geq \int_{\lambda^T g(x,\theta)\leq 0} (\lambda^T g(x,\theta))^2 d \m P^*> \frac{a}{2}-(c_2^2+b_1^2 c_3).
\end{equation*}
Also, 
\begin{equation*}
    \int_{\lambda^T g(x,\theta)\leq 0} -\lambda^T g(x,\theta) d  \m P^*= \int_{\lambda^T g(x,\theta)\geq 0} \lambda^T g(x,\theta) d  \m P^*- \mathbb E(\lambda^T g(x,\theta)).
\end{equation*}
Then, by  $\|E g(x,\theta)\|_2\leq \frac{a}{8b_1}$, we can get 
\begin{equation*}
   b_1 \int_{\lambda^T g(x,\theta)\geq 0} \lambda^T g(x,\theta) d \m P^* +\frac{a}{8}> \frac{a}{2}-(c_2^2+b_1^2 c_3).
\end{equation*}
Then, by 
\begin{equation*}
     \int_{\lambda^T g(x,\theta)\geq 0} \lambda^T g(x,\theta) d  \m P^* <c_2+b_1c_3,
\end{equation*}
we can get 
\begin{equation*}
     c_3>\frac{\frac{3}{8}a-c_2^2-b_1c_2}{2b_1^2}\geq \frac{a}{16 b_1^2}.
\end{equation*}
So, if we choose $c_3= \frac{a}{16 b_1^2}$, we can get  the conclusion of the first statement.
\\
\quad\\
For the second statement, let $\varepsilon=\frac{c_2}{4b_1}$ and $N_{\varepsilon}$ be the minimal $\varepsilon$-covering set of $\mathbb{S}^{d-1}$ with respect to $\ell_2$ distance, then $|N_{\varepsilon}|\leq (\frac{3}{\varepsilon})^d$. \\
 \noindent
Then by Bernstein inequality, there exists a constant $c_4$ such that it hold with probability at least $1-\frac{1}{n^2}$ that 
\begin{equation*}
    \underset{\lambda \in N_{\varepsilon}}{\sup} \left|\m P^*(\lambda^T g(X,\tilde{\theta})\geq c_2)-\frac{1}{n} \sum_{i=1}^n \textbf{1}_{\lambda^Tg(X_i,\tilde{\theta})\geq c_2}\right| \leq c_4 \sqrt{\frac{\log n}{n}}.
\end{equation*}
\quad\\
So,  for any $\lambda \in N_{\varepsilon}$, there are at least $\left(c_3-c_4\sqrt{\frac{\log n}{n}}\right) n$ number of data such that  $\lambda^T g(x,\tilde{\theta})\geq c_2$. Also, for any $\lambda \in \mathbb{S}^{d-1}$, there exists $\tilde{\lambda}\in N_\varepsilon$ such that $\|\lambda-\tilde{\lambda}\|_2\leq \varepsilon$, so we can choose a small enough $\delta_0$, such that for any $\theta\in B_{\delta_0}(\tilde{\theta})$,
\begin{equation*}
    |\lambda^T g(x,\theta)-\tilde{\lambda}^Tg(x,\tilde{\theta})|\leq b_1\varepsilon+\|g(x,\theta)-g(x,\tilde{\theta})\|_2\leq\frac{c_2}{2}.
\end{equation*}
 So for any $1\leq i\leq n$ such that $\tilde{\lambda}^Tg(X_i,\tilde{\theta})\geq c_2$, it holds that $\lambda^T g(X_i,\theta)\geq \frac{c_2}{2}$, we can then get the desired conclusion.

\subsection{Proof of lemma~\ref{l4}}
Consider $\theta \in \left\{\theta\,\big|\, \|\theta-\tilde{\theta}\|_2\leq \frac{2\delta_2(\log n)^{1.5}}{\sqrt{n}}\right\}$, by equation~\eqref{boundlambda}, it holds with probability larger than $1-\frac{1}{n^2}$ that for any $\theta \in \left\{\theta\,\big|\, \|\theta-\tilde{\theta}\|_2\leq \frac{2\delta_2(\log n)^{1.5}}{\sqrt{n}}\right\}$, it satisfies that $\|\lambda(\theta)\|_2\leq \lambda_0$. Since $\lambda(\theta)$ is the solution of 
\begin{equation*}
    \frac{1}{n} \sum_{i=1}^n \exp\left(\lambda(\theta)^Tg(X_i, \theta)\right) g(X_i, \theta)=0
\end{equation*}
we have 
\begin{equation*}
\begin{aligned}
    \lambda^{(1)} (\theta)=&-\left(\frac{1}{n}\sum_{i=1}^n  \exp\left(\lambda(\theta)^Tg(X_i, \theta)\right) g(X_i, \theta)g(X_i, \theta)^T\right)^{-1} \\
    &\cdot \left(\frac{1}{n}\sum_{i=1}^n   \exp\left(\lambda(\theta)^Tg(X_i, \theta)\right) \left(g^{(1)}(X_i,\theta)+g(X_i,\theta)\lambda(\theta)^T g^{(1)}(X_i,\theta)\right)
    \right).\\
    \end{aligned}
\end{equation*}
 For any $\nu\in \mathbb{S}^{d-1}$, let $b_1=\underset{x \in \mathcal{X} \atop  \theta \in \{\theta\,|\, \|\theta-\tilde{\theta}\|_2\leq \frac{2\delta_1(\log n)^{1.5}}{\sqrt{n}}\}}{\sup}\|g(x,\theta)\|_2$,
\begin{equation*}
    \begin{aligned}
       & \nu^T \left(\frac{1}{n}\sum_{i=1}^n  \exp\left(\lambda(\theta)^Tg(X_i, \theta)\right) g(X_i, \theta)g(X_i, \theta)^T\right) \nu\\
       & \geq \exp(-\lambda_0b_1) \frac{1}{n}\sum_{i=1}^n \nu^T g(X_i, \theta)g(X_i, \theta)^T \nu\\
       &=\exp(-\lambda_0b_1)\left(\nu^T \Delta_{\theta} \nu+\nu^T\left(\frac{1}{n}\sum_{i=1}^ng(X_i, \theta)g(X_i, \theta)^T-\Delta_{\theta}\right)\nu\right)
    \end{aligned}
\end{equation*}
Similar as equation~\eqref{f12}, by Dudley's inequality and Bernstein inequality, with probability at least $1-\frac{1}{n^2}$, 
\begin{equation*}
   \underset{\theta \in \left\{\|\theta-\tilde{\theta}\|_2\leq \frac{2\delta_2(\log n)^{1.5}}{\sqrt{n}}\right\}}{\sup} \left\|\frac{1}{n}\sum_{i=1}^ng(X_i, \theta)g(X_i, \theta)^T-\Delta_{\theta}\right\|_{\rm F} \lesssim \sqrt{\frac{\log n}{n}}.
\end{equation*}
Also, by $\Delta_{\tilde{\theta}}\succcurlyeq a I_{d}$, we have for any $\theta \in \left\{\|\theta-\tilde{\theta}\|_2\leq \frac{2\delta_2(\log n)^{1.5}}{\sqrt{n}}\right\}$, $\Delta_{\theta} \succcurlyeq \frac{a}{2} I_{d}$.  Then we can get for any $\nu\in \mathbb{S}^{d-1}$,
\begin{equation*}
\begin{aligned}
      \nu^T \left(\frac{1}{n}\sum_{i=1}^n  \exp\left(\lambda(\theta)^Tg(X_i, \theta)\right) g(X_i, \theta)g(X_i, \theta)^T\right) \nu\geq \exp(-\lambda_0 b_1) \frac{a}{4}
       \end{aligned}
\end{equation*}
So, 
\begin{equation*}
   \left\| \left(\frac{1}{n}\sum_{i=1}^n  \exp(\lambda(\theta)^Tg(X_i, \theta)) g(X_i, \theta)g(X_i, \theta)^T\right)^{-1}\right\|_{op}\leq \exp(\lambda_0b_1) \frac{4}{a}.
\end{equation*}
Then, by 
\begin{equation*}
    \underset{x \in \mathcal{X}\atop \theta \in \Theta}{\sup}\left(\underset{1\leq i\leq d}{\max}\left|\frac{\partial \ell(X,\theta)}{\partial\theta_i}\right|+\underset{1\leq i\leq  d\atop 1\leq j\leq d}{\max}\left|\frac{\partial^2 \ell(X,\theta)}{\partial \theta_i \partial \theta_j}\right|+\underset{1\leq i\leq  d\atop {1\leq j\leq d\atop 1\leq k\leq d}}{\max}\left|\frac{\partial^3 \ell(X,\theta)}{\partial \theta_i \partial \theta_j\partial \theta_k}\right|\right)\leq C,
    \end{equation*}
    and 
    \begin{equation*}
        g(x,\theta)=\nabla \ell(X,\theta),
    \end{equation*}
    we can get the desired conclusion.

\subsection{Proof of Lemma~\ref{lemmaC.3}}
For (b) of Assumption B.2', let $g_j(X_i,\theta)$ denote the $j$th dimension of $g(X_i,\theta)$, for any $1\leq j \leq d$, define the function class $\mathcal{G}_j=\{g_j(\cdot,\theta)-g_j(\cdot,\theta^*),\theta\in \Theta\}$ and its star hull $\bar{\mathcal{G}}_j=\{ag, g\in\mathcal{G}_j\}$. Define 
\begin{equation*}
\mathcal{R}_n(\delta)=\mathbb{E}_X\mathbb{E}_{\varepsilon}\left[\underset{f\in \bar{\mathcal{G}}_j \atop \mathbb{E} f^2 \leq \delta^2 }{\sup}\left|\frac{1}{n}\sum_{i=1}^n \varepsilon_i  f(X_i)\right|\right],
\end{equation*}
where $\{\varepsilon_i\}_{i=1}^n$ are n i.i.d. copies from Rademacher distribution, i.e. $P(\varepsilon_i = 1) = P(\varepsilon_i= -1) = 0.5$. Define the distance between $f,f'\in \bar{\mathcal{G}}_j$,
\begin{equation*}
d_n(f,f')=\sqrt{\sum_{i=1}^n (f(X_i)-f'(X_i))^2}.
\end{equation*}
Then by the uniformly boundness of $\bar{\mathcal{G}}_j$, it follows that  
\begin{equation*}
\begin{aligned}
&\log\mathcal{N}(\bar{\mathcal{G}}_j, d_n, \varepsilon)\\
&\leq \log (\frac{c}{\varepsilon}) + \log \mathcal{N}(\mathcal{G}_j, d_n, \varepsilon)\\
&\leq  \log (\frac{c}{\varepsilon}) + \log \mathcal{N}(\Theta, d^g_n, \varepsilon)\\
&\lesssim \log n + \log (\frac{1}{\varepsilon}) 
 \end{aligned}
\end{equation*}
Then by Dudley inequality~\citep{vershynin_2018} and equation (3.84) of~\cite{wainwright_2019},  it holds that 
\begin{equation*}
\mathcal{R}_n\left(\sqrt{\frac{\log n}{n}}\right)\lesssim \frac{\log n}{n}.
\end{equation*}
Then by Theorem 14.20 of~\cite{wainwright_2019} and $\sqrt{\mathbb{E}\|g(X,\theta)-g(X,\theta^*)\|_2^2}\lesssim \|\theta-\theta^*\|_2^\beta$, there exists a constant $c$ such that it holds with probability at least $1-\frac{1}{n^2}$ that $\underset{\theta\in \Theta} {\sup} \|\frac{1}{n}\sum_{i=1}^n g(X_i,\theta)-\frac{1}{n}\sum_{i=1}^n g(X_i,\theta^*)-\mathbb{E} g(X,\theta)+\mathbb{E} g(X,\theta^*)\|_2 \leq c(\sqrt{\frac{\log n}{n}}\|\theta-\theta^*\|^\beta_2+\frac{\log n}{n})$; We can use the same strategy to prove the statement in (c) of Assumption B.2'. For (a) of  the Assumption B.2', there exists a constant $c$ such that for any $1\leq j,k\leq d$ and $\theta,\theta' \in \Theta$,
 \begin{equation*}
\begin{aligned}
\sqrt{\frac{1}{n}\sum_{i=1}^n (g^h (X_i,\theta)g^k (X_i,\theta)-g^h (X_i,\theta')g^k (X_i,\theta'))^2}\leq c \sqrt{\frac{1}{n}\sum_{i=1}^n \|g(X_i,\theta)-g(X_i,\theta')\|_2^2}.
\end{aligned}
  \end{equation*}
So the statement in (a) of Assumption B.2' is followed by Dudley inequality and Talagrand concentration inequality~\citep{wainwright_2019}.

 \subsection{Proof of Lemma~\ref{le5.1} and Lemma~\ref{le5.2}}
 Let $\mathcal{A}_1$ be the event  $\{\|\frac{1}{n} \sum_{i=1}^n g(X_i,\theta^*)-\mathbb{E} g(X,\theta^*)\|_2\leq c\sqrt{\frac{\log n}{n}}\}\cap \{|\frac{1}{n}\sum_{i=1}^n L(X_i,\theta^*)-\mathbb{E} L(X,\theta^*)| \leq c\sqrt{\frac{\log n}{n}}\}$, then by Assumption B.1, there exists a large enough $c$ such that $\m P^*(\mathcal{A}_1)\geq 1-\frac{1}{n^2}$. Let $\mathcal{A}_2$ be the event that statements (a), (b), (c) in Assumption B.2' hold, then by Lemma~\ref{lemmaC.3}, it holds that $\m P^*(\mathcal{A}_2)\geq 1-\frac{1}{n^2}$. Unless otherwise specified, the following analysis is under event $\mathcal{A}_1\cap \mathcal{A}_2$. For  the statement of  Lemma~\ref{le5.1}, by Assumption B.1 and A.2' and Lemma~\ref{l3}, we can get that there exist some positive constants $(\delta_0, c_2, c_3)$ such that for any $\lambda\in \mathbb{S}^{d-1}$ and $\theta\in B_{\delta_0}(\theta^*)$, it holds that $\m P^*(\lambda^T g(X,\theta)\geq c_2)\geq c_3$. So $\mathbb{E} \max\big(\lambda^T g(X,\theta)-\frac{c_2}{2},0\big)\geq \frac{c_2c_3}{2}$. Moreover, by Assumption B.2 that the $\varepsilon$-covering number of $\Theta$ with respect to $d_n^{g}$ is upper bounded by $(\frac{n}{\varepsilon})^c$,  using Dudley inequality and Talagrand concentration inequality, we can get that there exists a constant $c_4$ such that it hold with probability at least $1-\frac{1}{n^2}$ that 
\begin{equation*}
    \underset{\lambda \in \mathbb{S}^{d-1}\atop \theta\in \Theta}{\sup} \left|\mathbb{E}\max\big(\lambda^T g(X,\theta)-\frac{c_2}{2},0\big)-\frac{1}{n} \sum_{i=1}^n \max\big(\lambda^T g(X_i,\theta)-\frac{c_2}{2},0\big)\right| \leq c_4 \sqrt{\frac{\log n}{n}}.
\end{equation*}
Let $b=\underset{x\in \mathcal{X}, \theta\in B_{\delta_0}(\theta^*)}{\sup} \|g(x,\theta)\|_2$,  it holds with probability at least $1-\frac{1}{n^2}$ that  for any $\lambda\in \mathbb{S}^{d-1}$ and $\theta \in B_{\delta_0}(\theta^*)$, it satisfies that
 \begin{equation*}
 \begin{aligned}
 \frac{c_2c_3}{4}\leq \frac{1}{n} \sum_{i=1}^n \max(\lambda^T g(X_i,\theta)-\frac{c_2}{2},0)\leq (b-\frac{c_2}{2}) \frac{\sum_{i=1}^n \textbf{1}_{\lambda^T g(X_i,\theta)\geq \frac{c_2}{2}}}{n}
 \end{aligned}
 \end{equation*}
 So we can get that there exist some positive constants $(\delta, c_2, c_3)$ such that for any $\lambda\in \mathbb{S}^{d-1}$ and $\theta\in B_{\delta_0}(\theta^*)$, it holds that $\frac{1}{n}\sum_{i=1}^n\textbf{1}_{\lambda^T g(x,\theta)\geq c_2/2}\geq \frac{c_2c_3}{4(b-\frac{c_2}{2})}>0$. So lemma~\ref{le5.1} can be proved using equation~\eqref{boundlambda}.
 \noindent
For the proof of  Lemma~\ref{le5.2}, according to Assumption B.2', it holds that 
\begin{equation*}
   \underset{\theta\in \Theta} {\sup} \|\frac{1}{n}\sum_{i=1}^n g(X_i,\theta)-\frac{1}{n}\sum_{i=1}^n g(X_i,\theta^*)-\mathbb{E} g(X,\theta)+\mathbb{E} g(X,\theta^*)\|_2 \leq c(\sqrt{\frac{\log n}{n}}\|\theta-\theta^*\|^{\beta}_2+\frac{\log n}{n}).
  \end{equation*}
  Also, by Assumption  B.1, we have 
  \begin{equation*}
  \mathbb{E}g(X,\theta)-\mathbb{E} g(X,\theta^*)=\m H_{\theta^*}(\theta-\theta^*)+O(\|\theta-\theta^*\|_2^2),
    \end{equation*}
    \noindent
   so we can get 
     \begin{equation*}
       \underset{\theta\in \Theta} {\sup}\|\frac{1}{n}\sum_{i=1}^n g(X_i,\theta)-\frac{1}{n}\sum_{i=1}^n g(X_i,\theta^*)-\m H_{\theta^*}(\theta-\theta^*)\|_2\lesssim  \|\theta-\theta^*\|_2^2+\sqrt{\frac{\log n}{n}}\|\theta-\theta^*\|^{\beta}_2+\frac{\log n}{n}. 
         \end{equation*}
    So we have 
    \begin{equation*}
\begin{aligned}
&\Big\|\frac{1}{n} \sum_{i=1}^n \exp(\tilde{\lambda}(\theta)^T g(X_i,\theta))g(X_i,\theta)-\frac{1}{n} \sum_{i=1}^n \exp\Big(-\big(\Delta_{\theta^*}^{-1}\frac{1}{n}\sum_{j=1}^n g(x_j,\theta) \big)^T g(X_i,\theta)\Big)g(X_i,\theta)\Big\|_2\\
&\lesssim   \|\theta-\theta^*\|_2^2+\sqrt{\frac{\log n}{n}}\|\theta-\theta^*\|^{\beta}_2+\frac{\log n}{n}. 
  \end{aligned}
  \end{equation*}
  By $\mathbb{E} g(X,\theta^*)=0$, we can get $\|\frac{1}{n}\sum_{i=1}^n g(X_i,\theta)\|\lesssim \|\theta-\theta^*\|_2+\sqrt{\frac{\log n}{n}}$, so
   \begin{equation*}
\begin{aligned}
& \frac{1}{n} \sum_{i=1}^n \exp\Big(-\big(\Delta_{\theta^*}^{-1}\frac{1}{n}\sum_{j=1}^n g(x_j,\theta) \big)^T g(X_i,\theta)\Big)g(X_i,\theta)\\
&=\frac{1}{n}\sum_{i=1}^n g(X_i,\theta)- \frac{1}{n}\sum_{j=1}^n g(x_j,\theta)g(x_j,\theta)^T\Delta_{\theta^*}^{-1} \frac{1}{n}\sum_{i=1}^n g(X_i,\theta)+O( \|\theta-\theta^*\|_2^2+\frac{\log n}{n}).
    \end{aligned}
  \end{equation*}
 Then by
 \begin{equation*}
\| \frac{1}{n}\sum_{j=1}^n g(x_j,\theta)g(x_j,\theta)^T\Delta_{\theta^*}^{-1}-I_{d}\|_2\lesssim \|\theta-\theta^*\|_2+\sqrt{\frac{\log n}{n}},
  \end{equation*}
we can get 
 \begin{equation*}
 \|\frac{1}{n} \sum_{i=1}^n \exp(\tilde{\lambda}(\theta)^T g(X_i,\theta))g(X_i,\theta)\|_2\lesssim   \|\theta-\theta^*\|_2^2+\sqrt{\frac{\log n}{n}}\|\theta-\theta^*\|^{\beta}_2+\frac{\log n}{n}. 
   \end{equation*}
By Lemma~\ref{le5.1}, there exist positive constants $C$ and $r$ such that $\underset{\theta\in B_r(\theta^*)}{\sup} \max\{\|\lambda(\theta)\|_2, \|\tilde{\lambda}(\theta)\|_2 \}\leq C$. By Assumption A.2' and B.1, we can find a small enough  $r_0\leq r$ such that for any $\theta \in B_{r_0}(\theta^*)$, it holds that $\m H_{\theta}\succcurlyeq \frac{a}{2}I_d$ and  $ \Delta_{\theta}\succcurlyeq \frac{b}{2}I_d$, where $a, b>0$. Fix a $\theta \in B_{r_0}(\theta^*)$, define $f(\lambda)=\frac{1}{n}\sum_{i=1}^n \exp(\lambda^T g(X_i,\theta))$, then we have

 \begin{equation*}
\begin{aligned}
 &f^{(1)}(\lambda)=\frac{1}{n}\sum_{i=1}^n \exp(\lambda^T g(X_i,\theta))g(X_i,\theta)\\ 
 &f^{(2)}(\lambda)=\frac{1}{n}\sum_{i=1}^n \exp(\lambda^T g(X_i,\theta))g(X_i,\theta)g(X_i,\theta)^T.
   \end{aligned}
  \end{equation*}
By Assumption A.2', B.1 and B.2', there exists a positive constant $a_1$ such that for any $\|\lambda\|_2\leq C$, it holds that 
  \begin{equation*}
  f^{(2)}(\lambda) \succcurlyeq a_1I_d.
    \end{equation*}
  Moreover, for any  $\|\lambda\|_2\leq C$ and $l\in \mathbb{S}^{d-1}$ there exists  a $\lambda'$ depend on $\lambda$ and $l$ such that  $\|\lambda'\|_2\leq C$ and 
   \begin{equation*}
   f^{(1)}(\lambda)^Tl=  f^{(1)}(\lambda(\theta))^Tl +(\lambda-\lambda(\theta))^T f^{(2)}(\lambda')l.
  \end{equation*}
  So we can get 
  \begin{equation*}
\begin{aligned}
\| f^{(1)}(\tilde{\lambda}(\theta))\|_2 &\geq \underset{l\in \mathbb{S}^{d-1}}{\sup}\underset{\lambda\in \mathbb{R}^{d}\atop \|\lambda\|_2\leq C}{\inf}|(\tilde{\lambda}(\theta)-\lambda(\theta))^T f^{(2)}(\lambda)l|\\
&\geq  \underset{\lambda\in \mathbb{R}^{d}\atop \|\lambda\|_2\leq C}{\inf}\left|\frac{(\tilde{\lambda}(\theta)-\lambda(\theta))^T f^{(2)}(\lambda)(\tilde{\lambda}(\theta)-\lambda(\theta))}{\|\tilde{\lambda}(\theta)-\lambda(\theta)\|_2}\right|\\
&\geq a_1\|\tilde{\lambda}(\theta)-\lambda(\theta)\|_2.
 \end{aligned}
  \end{equation*}
  We can then get  for any $\theta \in B_{r_0}(\theta^*)$,
   \begin{equation*}
   \|\tilde{\lambda}(\theta)-\lambda(\theta)\|_2  \lesssim   \|\theta-\theta^*\|_2^2+\sqrt{\frac{\log n}{n}}\|\theta-\theta^*\|^{\beta}_2+\frac{\log n}{n}. 
  \end{equation*}

\subsection{Proof of Corollary~\ref{coquantile}}
 Recall $\ell(X,\theta)=(Y-\tilde{X}^T\theta)(\tau-\textbf{1}_{Y<\tilde{X}^T\theta})$ and $g(X,\theta)=(\textbf{1}_{Y<\tilde{X}^T\theta}-\tau)\tilde{X}$. We first prove that  the statement in Assumption B.1 is satisfied. 
 Since 
 \begin{equation*}
 \mathcal{R}(\theta)=\mathbb{E}\tau(Y-\tilde{X}^T\theta)-\mathbb{E}_{\tilde{X}} \int_{-\infty}^{\tilde{X}^T\theta} (Y-\tilde{X}^T\theta) p(Y|\tilde{X})dY.
 \end{equation*}
 
 We can get 
 \begin{equation*}
 \nabla \mathcal{R}(\theta)=-\tau \mathbb{E} \tilde{X} +\mathbb{E}_{\tilde{X}}   \int_{-\infty}^{\tilde{X}^T\theta}  p(Y|\tilde{X}) \tilde{X}dY=\mathbb{E} g(X,\theta).
  \end{equation*}
 So we have 
 \begin{equation*}
\m H_{\theta}=\mathbb{E}_{\tilde{X}} p(\tilde{X}^T\theta|\tilde{X}) \tilde{X}\tilde{X}^T
   \end{equation*}
 Then by the assumption that $p(t|\tilde{X})$ is bounded and has bounded derivative w.r.t. $t$ over $t \in \mathbb{R}$ and $\tilde{X}\in \tilde{\mathcal{X}}$ and the assumption that the support of $\tilde{X}$ is compact, we can get that $\mathcal{R}(\theta)$ is bounded and has bounded derivatives w.r.t. $\theta$ up to order three. Moreover, the boundness of $\ell(X,\theta)$ and $g(X,\theta)$ is guaranteed by the compactness of supports of $\tilde{X}$ and $Y$. In addition, there exists a constant $c$ such that for any $\tilde{X}\in \tilde{\mathcal{X}}$ and $\theta,\theta' \in \Theta$ it holds that, 
  \begin{equation}\label{quantile1}
 \begin{aligned}
 & \mathbb{E}_{Y|\tilde{X}} (\textbf{1}_{Y<\tilde{X}^T\theta}-\textbf{1}_{Y<\tilde{X}^T\theta'})^2\\
&  = \mathbb{E}_{Y|\tilde{X}} (\textbf{1}_{\tilde{X}^T\theta'\leq Y<\tilde{X}^T\theta} +\textbf{1}_{\tilde{X}^T\theta\leq Y<\tilde{X}^T\theta'})\\
&\leq c\|\theta-\theta'\|_2.
  \end{aligned}
     \end{equation}
  So there exists a constant $c_1$ such that 
  \begin{equation}. 
 \begin{aligned}
& \|\Delta_{\theta}-\Delta_{\theta^*}\|_{\rm F}\\
&\leq\|\mathbb{E} \tilde{X}\tilde{X}^T (\textbf{1}_{Y<\tilde{X}^T\theta}-\textbf{1}_{Y<\tilde{X}^T\theta^*})\|_{\rm F}\\
&\leq \|\mathbb{E}_{\tilde{X}} \tilde{X}\tilde{X}^T \mathbb{E}_{Y|\tilde{X}} |\textbf{1}_{Y<\tilde{X}^T\theta}-\textbf{1}_{Y<\tilde{X}^T\theta^*}|\|_{\rm F}\\
&\leq c_1\|\theta-\theta^*\|_2.
 \end{aligned}
     \end{equation}
 For the Assumption B.2,   Define 
  \begin{equation*}
        \begin{aligned}
&d_n^g(\theta,\theta')=\sqrt{\frac{1}{n}\sum_{i=1}^n \|g(X_i,\theta)-g(X_i, \theta')\|_2^2},\\
&d_n^{\ell}(\theta,\theta')=\sqrt{\frac{1}{n}\sum_{i=1}^n (\ell(X_i, \theta)-\ell(X_i,\theta'))^2}.
 \end{aligned}
\end{equation*}
   By Lemma 9.12 and Lemma 9.8 of~\cite{Kosorok2008}, we know the function class $\mathcal{F}=\{\textbf{1}_{Y\leq \theta^T\tilde{X}}, \theta\in \Theta\}$ is a VC-class, so by Theorem 8.3.18 of~\cite{vershynin_2018} and the fact that the $\varepsilon$-covering number of $B_1(0)$ with respect to $\ell_2$ norm is upper bounded by $\left(\frac{3}{\varepsilon}\right)^d$, we can get that 
  
   \begin{equation*}
        \begin{aligned}
 &\log \mathcal{N}(\Theta, d_n^g, \varepsilon) \lesssim \log \frac{1}{\varepsilon}\\
 & \log \mathcal{N}(\Theta, d_n^{\ell}, \varepsilon) \lesssim \log \frac{1}{\varepsilon}.
 \end{aligned}
\end{equation*}
Moreover, by equation~\eqref{quantile1}, there exist some constants $(c,c_1)$ such that for any $\theta, \theta' \in \Theta$ it holds that, 
   \begin{equation*}
        \begin{aligned}   
 &\sqrt{ \mathbb{E}(g(X,\theta)-g(X,\theta'))^T(g(X,\theta)-g(X,\theta'))}\\
 &=\sqrt{\mathbb{E}\tilde{X}^T\tilde{X} (\textbf{1}_{Y<\tilde{X}^T\theta}-\textbf{1}_{Y<\tilde{X}^T\theta'})^2}\\
 &\leq c\|\theta-\theta'\|_2^{\frac{1}{2}}
   \end{aligned}
\end{equation*}

\begin{equation*}
        \begin{aligned}   
 &\sqrt{ \mathbb{E}(\ell(X,\theta)-\ell(X,\theta'))^2}\\
 &\leq c_1 \left(\|\theta-\theta'\|_2+\sqrt{\mathbb{E}(\textbf{1}_{Y<\tilde{X}^T\theta}-\textbf{1}_{Y<\tilde{X}^T\theta'})^2}\right)\\
  &\leq c\|\theta-\theta'\|_2^{\frac{1}{2}}
   \end{aligned}
\end{equation*}
Then, we can get the statement of Assumption B.2  with $\beta=\frac{1}{2}$. The desired conclusion is then followed by Theorem 3.

\subsection{Proof of Corollary~\ref{cosvm}}
Recall that  $\ell(X,\theta)=\frac{1}{2}\lambda \theta^T\theta+\textbf{1}_{Y\theta^T\tilde{X}\leq 1}(1-Y\theta^T\tilde{X})$ and $g(X,\theta)=\lambda \theta-Y\textbf{1}_{Y\theta^T\tilde{X}\leq 1} \tilde{X}$. Since
\begin{equation*}
 \mathcal{R}(\theta)= \mathbb{P}(Y=1) \mathbb{E}_{\tilde{X}|Y=1} ((1-\theta^T\tilde{X}) \textbf{1}_{\theta^T\tilde{X} \leq 1})+ \mathbb{P}(Y=-1) \mathbb{E}_{\tilde{X}|Y=-1} ((1+\theta^T\tilde{X}) \textbf{1}_{\theta^T\tilde{X} \geq -1})+ \frac{1}{2}\lambda \theta^T\theta
 \end{equation*}
 We now prove the thirce differentiability of $\mathcal{R}(\theta)$, choose any $\theta=(\theta_1,\cdots, \theta_d)\in \Theta$, w.l.o.g, we can assume $\theta_1, \cdots, \theta_d >0$.  Since  for any $1\leq j\leq d$,
\begin{equation*}
        \begin{aligned}   
        &\mathbb{E}_{\tilde{X}|Y=1} ((1-\theta^T\tilde{X}) \textbf{1}_{\theta^T\tilde{X} \leq 1})\\
        &= \mathbb{E}_{\tilde{X}_{-j}|Y=1}\int_{\tilde{X}_j\leq \frac{1-\sum_{k\neq j}\theta_k\tilde{X}_k}{\theta_j}} (1-\sum_{k\neq j}\theta_k\tilde{X}_k-\theta_j \tilde{X}_j)P^j_{1}(\tilde{X}_j) d\tilde{X}_j
  \end{aligned}
\end{equation*}
We can get 
\begin{equation*}
        \begin{aligned}   
        &\nabla \mathcal{R}(\theta)=\lambda \theta-\mathbb{P}(Y=1) \mathbb{E}_{\tilde{X}|Y=1}\textbf{1}_{\theta^T\tilde{X}\leq 1}\tilde{X}+\mathbb{P}(Y=-1) \mathbb{E}_{\tilde{X}|Y=-1}\textbf{1}_{\theta^T\tilde{X}\geq -1}\tilde{X}\\
        &=-\mathbb{E} Y\textbf{1}_{Y\theta^T\tilde{X}\leq 1}\tilde{X}+\lambda\theta\\
       & =\mathbb{E} g(X,\theta).
          \end{aligned}
\end{equation*}
Moreover, for any $1\leq i \leq d$
\begin{equation*}
    \begin{aligned}   
&\frac{\partial^2 \mathcal{R}(\theta)}{\partial \theta_i^2}=\lambda+\mathbb{P}(Y=1) \mathbb{E}_{\tilde{X}_{-i}|Y=1}\left( \frac{(1-\sum_{k\neq i}^d \theta_k \tilde{X}_k)^2}{\theta_i^3} p^i_1\big( (1-\sum_{k\neq i}^d \theta_k \tilde{X}_k)/\theta_i\big)\right)\\
&+\mathbb{P}(Y=-1) \mathbb{E}_{\tilde{X}_{-i}|Y=-1} \left(\frac{(1+\sum_{k\neq i}^d \theta_k \tilde{X}_k)^2}{\theta_i^3} p^i_{-1}\big ((-1-\sum_{k\neq i}^d \theta_k \tilde{X}_k)/\theta_i\big)\right).
\end{aligned}
\end{equation*}
 when $1\leq i\neq j\leq d$, 
\begin{equation*}
        \begin{aligned}   
&\frac{\partial^2 \mathcal{R}(\theta)}{\partial \theta_i \theta_j}=\mathbb{P}(Y=1) \mathbb{E}_{\tilde{X}_{-i}|Y=1}\left( \frac{\tilde{X}_j(1-\sum_{k\neq i}^d \theta_k \tilde{X}_k)}{\theta_i^2} p^i_1\big( (1-\sum_{k\neq i}^d \theta_k \tilde{X}_k)/\theta_i\big)\right)\\
&+\mathbb{P}(Y=-1) \mathbb{E}_{\tilde{X}_{-i}|Y=-1} \left(\frac{\tilde{X}_j(-1-\sum_{k\neq i}^d \theta_k \tilde{X}_k)}{\theta_i^2} p^i_{-1}\big ((-1-\sum_{k\neq i}^d \theta_k \tilde{X}_k)/\theta_i\big)\right).
\end{aligned}
\end{equation*}
Then by Assumption B.s, it holds that  $\mathcal{R}(\theta)$ is bounded and has bounded derivatives w.r.t. $\theta$ up to order three. Moreover, the boundness of $\ell(X,\theta)$ and $g(X,\theta)$ is guaranteed by the compactness of $\Theta$ and $\tilde{\mathcal{X}}$. In addition, there exist some constants $(c,c_0,c_1,c_2)$ such that 
\begin{equation*}
   \begin{aligned}   
&\|\Delta_{\theta}-\Delta_{\theta'}\|_{\rm F}\\
&\leq\| \mathbb{E} (\tilde{X}\tilde{X}^T |\textbf{1}_{Y\theta^T\tilde{X}\leq 1}-\textbf{1}_{Y\theta'^T\tilde{X}\leq 1}|)\|_F+c_0\| \mathbb{E} (\tilde{X} |\textbf{1}_{Y\theta^T\tilde{X}\leq 1}-\textbf{1}_{Y\theta'^T\tilde{X}\leq 1}|)\|_2+c\|\theta-\theta'\|_2\\
&\leq c_1( \mathbb{E}_{\tilde{X}|Y=1} |\textbf{1}_{\theta^T\tilde{X}\leq 1}-\textbf{1}_{\theta'^T\tilde{X}\leq 1}|+ \mathbb{E}_{\tilde{X}|Y=-1}  |\textbf{1}_{\theta^T\tilde{X}\geq -1}-\textbf{1}_{\theta'^T\tilde{X}\geq -1}|)+c\|\theta-\theta'\|_2\\
&\leq c_2 \|\theta-\theta'\|_2.
 \end{aligned}
\end{equation*}
So the statement in Assumption B.1 holds. For the Assumption B.2,  there exists a constant $c$ such that for any $\theta,\theta' \in \Theta$,
\begin{equation*}
   \begin{aligned} 
  & \sqrt{ \mathbb{E}(g(X,\theta)-g(X,\theta'))^T(g(X,\theta)-g(X,\theta'))}\\
  & =\sqrt{\mathbb{E} ( |\textbf{1}_{Y\theta^T\tilde{X}\leq 1}-\textbf{1}_{Y\theta'^T\tilde{X}\leq 1}|\tilde{X}^T\tilde{X})+\lambda^2 \|\theta-\theta'\|_2^2-2\mathbb{E}(\lambda Y(\textbf{1}_{Y\theta^T\tilde{X}\leq 1}-\textbf{1}_{Y\theta'^T\tilde{X}\leq 1})(\theta-\theta')^T\tilde{X})}\\
   &\leq  c\|\theta-\theta'\|_2^{\frac{1}{2}}.
\end{aligned}
\end{equation*}
Then combined with the fact that  $\mathcal{F}=\{\textbf{1}_{Y\theta^T\tilde{X}\leq 1}, \theta\in \Theta\}$ is a VC-class  and $\ell(X,\theta)$ is uniformly Lipschitz continuous w.r.t $\theta$, similar as the proof of Corollary~\ref{coquantile},  we can get that Assumption B.2 is satisfied with $\beta=\frac{1}{2}$. The desired conclusion is then followed by Theorem 3.

\end{document}